\documentclass[a4paper,aps,prl,reprint,sort&compress,groupedaddress,superscriptaddress,nofootinbib,floatfix,longbibliography]{revtex4-2}
\PassOptionsToPackage{sort&compress}{natbib}
\usepackage{textcomp}
\usepackage[utf8]{inputenc}
\usepackage{color}
\definecolor{note_fontcolor}{rgb}{1, 0, 0}
\usepackage{verbatim}
\usepackage{cprotect}
\usepackage{amsmath}
\usepackage{amssymb}
\usepackage{graphicx}
\PassOptionsToPackage{normalem}{ulem}
\usepackage{ulem}
\usepackage[pdfusetitle,
 bookmarks=true,bookmarksnumbered=false,bookmarksopen=false,
 breaklinks=false,pdfborder={0 0 1},backref=false,colorlinks=false]
 {hyperref}

\usepackage{array}
\usepackage{float}
\usepackage{mathrsfs}
\usepackage{multirow}
\usepackage{varwidth}
\usepackage{tabularx}

\makeatletter

\providecommand{\tabularnewline}{\\}
\newenvironment{cellvarwidth}[1][t]
    {\begin{varwidth}[#1]{\linewidth}}
    {\@finalstrut\@arstrutbox\end{varwidth}}

\usepackage{sistyle}
\usepackage{amsfonts}

\usepackage{bbold} 
\usepackage{dsfont}

\usepackage[TS1,T1]{fontenc}
\usepackage{lmodern}

\DeclareUnicodeCharacter{039B}{lambda} 
\DeclareUnicodeCharacter{2009}{ } 
\usepackage{enumitem}
\newcounter{desccount}

\newcommand{\descref}[1]{\hyperref[#1]{#1}}

\usepackage[super]{nth}                         
\renewcommand{\selectlanguage}[1]{} 
\makeatother

\makeatletter
\def\maketitle{
\@author@finish
\title@column\titleblock@produce
\suppressfloats[t]}
\makeatother

\begin{document}

\title{Uniting Quantum Processing Nodes of Cavity-coupled Ions with Rare-earth Quantum Repeaters Using Single-photon Pulse Shaping Based on Atomic Frequency Comb}

\begin{abstract}
We present an architecture for remotely connecting cavity-coupled trapped ions via a quantum repeater based on rare-earth-doped crystals. The main challenge for its realization lies in interfacing these two physical platforms, which produce photons with a typical temporal mismatch of one or two orders of magnitude. To address this, we propose an efficient protocol that enables custom temporal reshaping of single-photon pulses whilst preserving purity. Our approach is to modify a commonly used memory protocol, called atomic frequency comb, for systems exhibiting inhomogeneous broadening like rare-earth-doped crystals. Our results offer a viable solution for uniting quantum processing nodes with a quantum repeater backbone.
\end{abstract}

\author{P. Cussenot}
\affiliation{Université Paris--Saclay, CEA, CNRS, Institut de physique théorique,
\num{91191} Gif-sur-Yvette, France}
\affiliation{Direction Générale de l'Armement, \num{75015} Paris, France}
\author{B. Grivet}
\affiliation{Université Paris--Saclay, CEA, CNRS, Institut de physique théorique,
\num{91191} Gif-sur-Yvette, France}
\author{B.P. Lanyon}
\affiliation{Institut für Experimentalphysik, Universität Innsbruck, Technikerstraße
25, 6020 Innsbruck, Austria}
\author{T.E. Northup}
\affiliation{Institut für Experimentalphysik, Universität Innsbruck, Technikerstraße
25, 6020 Innsbruck, Austria}
\author{H. de Riedmatten}
\affiliation{ICFO -- Institut de Ciencies Fotoniques, The Barcelona Institute
of Science and Technology, Spain}
\affiliation{ICREA -- Institució Catalana de Recerca i Estudis Avançats, 08015
Barcelona, Spain}
\author{A.S. Sørensen}
\affiliation{Center for Hybrid Quantum Networks (Hy-Q), The Niels Bohr Institute,
University of Copenhagen, Blegdamsvej 17, DK-2100 Copenhagen Ø, Denmark}
\author{N. Sangouard}


\affiliation{Université Paris--Saclay, CEA, CNRS, Institut de physique théorique,
\num{91191} Gif-sur-Yvette, France}
\date{\today}

\maketitle

\paragraph{Introduction --} The realization of a wide network of quantum processors would be a remarkable achievement that might allow us to unlock distributed computing capabilities beyond those of even the most powerful individual future quantum computer \citep{simon_towards_2017}. Trapped ions offer high-fidelity quantum-gate operations
on registers of tens of qubits \citep{friis_observation_2018, moses_race-track_2023}, long
coherence times \citep{wang_single_2021}, and efficient interfacing
with telecom photons \citep{bock_high-fidelity_2018}. They are thus
promising candidates for realizing network nodes with quantum processing capacities \citep{duan_colloquium_2010, bruzewicz_trapped-ion_2019}. Rare-earth doped
materials also exhibit long coherence times \citep{zhong_optically_2015}
and can be interfaced with telecom photons \citep{clausen_quantum_2011,saglamyurek_broadband_2011}.
Furthermore, they possess large inhomogeneous broadening, a characteristic that can be harnessed for both temporal \citep{businger_non-classical_2022}
and spectral \citep{sinclair_spectral_2014} multiplexing, which
makes them inherently suitable for the implementation of long distance
quantum channels by means of quantum repeaters \citep{sangouard_quantum_2011}.
This naturally raises the question of how to interface disparate quantum systems \citep{farrera_generation_2016, morin_deterministic_2019}, here trapped-ion quantum processing nodes with a quantum repeater backbone using rare-earth
doped materials. 

\begin{figure}
\centering 
\includegraphics[width=1\columnwidth]{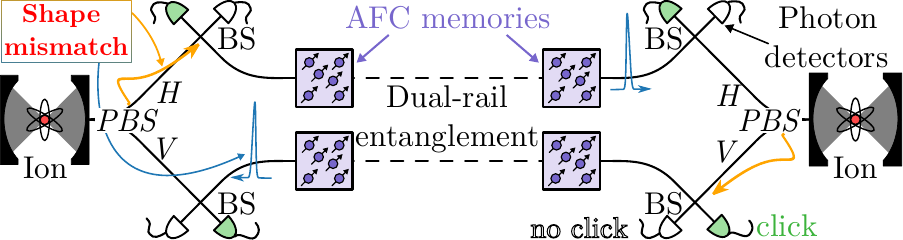} 
\caption{Proposed architecture to connect cavity-coupled ions with two repeater chains, each producing dual-rail entanglement between extreme rare-earth quantum memories (see text for further definitions and explanations).}

\label{Fig1} 
\end{figure}

\noindent
In this Letter, we tackle this question both at the architecture and
physical levels. The architecture we propose uses building blocks that
have already been implemented in distinct experiments, see Fig.~\ref{Fig1}.
Two cavity-integrated trapped-ions nodes \citep{moehring_entanglement_2007,stephenson_high-rate_2020,krutyanskiy_entanglement_2023, saha_high-fidelity_2024} in distant locations produce telecom photons whose polarization
is entangled with the ion energy states \citep{bock_high-fidelity_2018,krutyanskiy_light-matter_2019}.
Two repeater chains made with photon pair sources based on spontaneous
parametric down conversion (SPDC) and memories based on rare-earth
doped crystals \citep{lago-rivera_telecom-heralded_2021,liu_heralded_2021}
are used to distribute dual-rail entanglement \citep{duan_long-distance_2001} between the ions locations. The horizontal (H) and vertical (V) polarizations of photons emitted by the ions are spatially separated by use of polarizing beam splitters (PBS). Each of the two H paths is combined on a beam splitter (BS) with half of a dual-rail entangled state, and similarly for the V paths. A fourfold coincidence detection between photon detectors located after each beam splitter projects the two ions onto an entangled state. This can be seen as an entanglement swapping of ion-photon states
mediated by dual-rail entanglement, which results in ion-ion entanglement, see Appendix~A.
A key condition for the photonic swapping operations to be faithful
is that the fields emitted by the ions and the ones released from the
rare-earth doped crystals should be indistinguishable in all degrees of
freedom. Here, we consider that these two physical platforms produce pure single photons with a temporal mismatch quantified by the square of their waveform overlap, noted $x$, so that the resulting ion-ion fidelity is bounded: $\mathcal{F}_{ion-ion}^{pure}\le(1+x^{2})/2$, see Appendix~A. Fidelities approaching
the $0.5$ separability bound are thus expected for state-of-the-art trapped-ion quantum-network node demonstrations, that employ optical cavities for efficient photon collection and achieve photon waveforms that are tens of microseconds long \citep{meraner_indistinguishable_2020}, and state-of-the-art rare-earth ensembles, typically storing photons produced by SPDC sources that are a few hundreds of nanoseconds long \citep{lago-rivera_telecom-heralded_2021}. While spectral or temporal filtering can be used to restore high fidelities, it comes at the expense of reducing the entanglement distribution
rate by $x^{2}$.

\smallskip{}
\noindent
The solution that we propose relies on a modification of a commonly used
protocol for single-photon storage in rare-earth doped materials \citep{de_riedmatten_solid-state_2008, gundogan_solid_2015,laplane_multiplexed_2016,rakonjac_entanglement_2021,seri_quantum_2017}---the atomic frequency comb (AFC) protocol \citep{afzelius_multimode_2009}. AFC quantum memories are based on a periodic atomic absorption profile
in the form of a comb structure, created on an inhomogeneously broadened
optical transition using e.g. frequency-selective optical pumping.
The absorption of input light pulses whose frequency covers several comb peaks results in echoes
after a fixed-delay storage time, which is given by the inverse of
the comb period. To extend the storage time and read pulses out on demand,
the energy stored in the absorption comb is transferred back and forth
to a long-lived state using two $\pi$-pulses. 
Here, to lengthen the output pulse, we propose partially reading out the energy 
stored on the long-lived state using a readout pulse weaker than
a $\pi$-pulse. Once the light pulse associated with the released energy
is emitted, we proceed with another weak readout pulse, resulting in
the emission of an additional light pulse. The sequence of weak readouts
and light pulse emissions is repeated until all the stored energy is
released. We show that this sequential readout technique can
be efficient when the crystal is embedded in a cavity which
fulfills an impedance matching condition \citep{afzelius_impedance-matched_2010, moiseev_efficient_2010, afzelius_proposal_2013}.
The technique applies to the quantum regime, where the input light is
made of a single photon input, and in this case preserves
the photon purity. It allows an arbitrary temporal shaping with a temporal stretching limited only by the coherence time of the long-lived transition (below, inhomogeneous broadening is neglected on the long-lived transition).

\smallskip{}
\noindent
The relevance of the proposed protocol is highlighted by a feasibility study
in Pr$^{3+}$:Y$_{2}$SiO$_{5}$ where the photon waveform can be made almost indistinguishable from that of a single $^{40}$Ca$^{+}$ ion embedded in a high-finesse cavity. Specifically, we show that the visibility of Hong--Ou--Mandel (HOM)~\citep{hong_measurement_1987} interference between the two photons is limited by the purity of ions' photons for conditions corresponding to a recent experiment with single $^{40}$Ca$^{+}$ ions~\citep{krutyanskiy_entanglement_2023}.  
Our results provide a tangible pathway for uniting $^{40}$Ca$^{+}$ based processors with a quantum repeater backbone using Pr$^{3+}$:Y$_{2}$SiO$_{5}$ memories~\citep{duranti_efficient_2024}.

\medskip{}
\noindent
\paragraph{Modeling cavity-assisted AFC --} Each memory is seen as an ensemble of atoms modeled as $\Lambda$
systems; we label the three levels of each system $|g\rangle,$ $|e\rangle$, and $|s\rangle$, 
see Fig.~\ref{Fig2}. The optical transition $|g\rangle$-$|e\rangle$
is inhomogeneously broadened, and from a frequency-selective optical
pumping, we are left with $N$ atoms whose transition $|g\rangle$-$|e\rangle$
exhibits a periodic comb structure with distribution $n(\omega)$ (period $\Delta$), which is formed
of a series of distinct narrow peaks $w(\omega)$ enveloped
by a function $v(\omega)$ of total width $\Gamma$. The transition $|g\rangle$-$|s\rangle$ exhibits no inhomogeneous broadening and is considered to feature long-lived coherence. The atoms are embedded into a cavity
(amplitude decay rate $\kappa$) which is used to increase the light
absorption efficiency of the transition $|g\rangle$-$|e\rangle$ (collective
atom-light coupling constant is denoted $g\sqrt{N}$). One of the cavity mirrors is
set to couple the single-mode intra-cavity light field to a one dimensional
free space mode as described by the input-output formalism \citep{collett_squeezing_1984,gardiner_input_1985,gardiner_quantum_2004}.
A control field drives the transition $|s\rangle$-$|e\rangle$ with the Rabi frequency $\Omega(t)/2.$
The evolution of the system under an input light pulse to the cavity is given by a set of Heisenberg-Langevin
equations \citep{gorshkov_photon_2007-2,gorshkov_photon_2007}, see Appendices B and C. 
Under assumptions that we specify in Appendix~D (mainly vacuum Markovian reservoirs and single-photon input), the evolution can equivalently be understood with scalar equations. 

\begin{figure}
\centering \includegraphics[width=0.8\columnwidth]{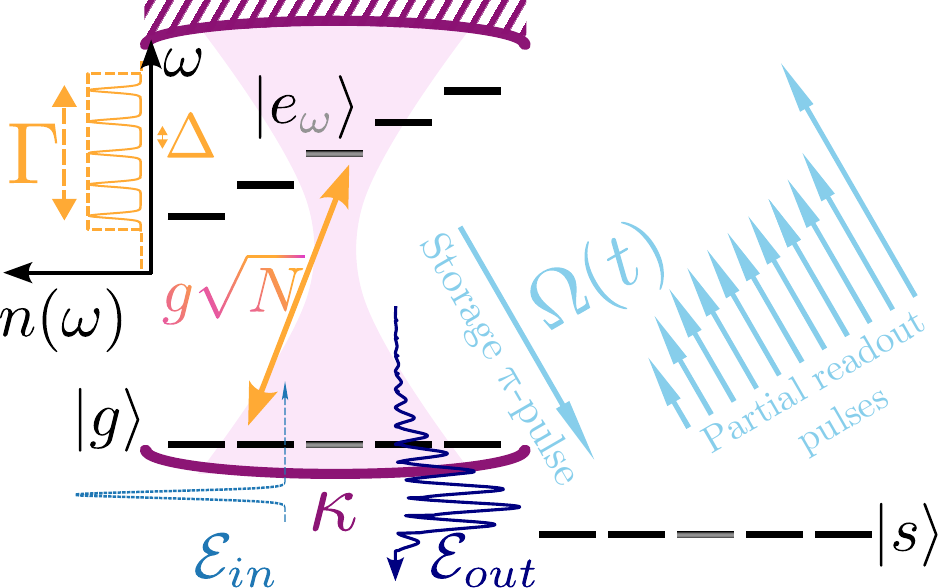} \caption{Relevant energy structure for an AFC protocol considered here with a cavity-enhanced efficiency, see Text for the notations. Temporal shaping of the output waveform is obtained by substituting the readout $\pi$-pulse of the standard AFC protocol with a series of weaker partial control pulses with a sufficiently large temporal separation to allow the partial re-emission of the output field. 
}
\label{Fig2} 
\end{figure}

\smallskip{}
\noindent
It is instructive to solve the dynamics by first considering the
case without the control field. We show in Appendix~B that we can get an explicit expression for the output field (waveform envelope denoted $\mathcal{E}_{out}(t)$) at the first echo time $2\pi/\Delta$ by considering an input single-photon field (waveform envelope labelled $\mathcal{E}_{in}(t)$) with a spectral bandwidth $\delta\omega_{in}$ satisfying $\Gamma, \kappa\gg\delta\omega_{in}>\Delta$, 
{} %

\begin{equation}
\mathcal{E}_{out}(t\approx\frac{2\pi}{\Delta})\approx-\frac{\tilde{w}(\frac{2\pi}{\Delta})}{\tilde{w}(0)}\frac{4\frac{\mathcal{C}}{\mathcal{C}_{opt}}}{\left(1+\frac{\mathcal{C}}{\mathcal{C}_{opt}}\right)^{2}}
\mathcal{E}_{in}(t-\frac{2\pi}{\Delta}).\label{eq:1st-echo}
\end{equation}
Specifically, the Fourier transform $\tilde{n}(t)$ of $n(\omega)$ is approximately a comb of peaks $\tilde{v}(t)$
modulated by $\tilde{w}(t)$. $\mathcal{C} = g^{2}N/{\kappa\Gamma}$ is the cooperativity, and $\mathcal{C}_{opt}$ is a constant which depends on $v(\omega)$. For an atomic distribution given by a
series of delta functions modulated by a large rectangular envelope, we get $\tilde{w}(\frac{2\pi}{\Delta})\approx\tilde{w}(0)$ and $\mathcal{C}_{opt}=1/\pi$. In this case, Eq.~\eqref{eq:1st-echo} tells us that 
the overall
AFC efficiency
$\int_{\frac{2\pi}{\Delta}-\Theta}^{\frac{2\pi}{\Delta}+\Theta}|\mathcal{E}_{out}(t)|^{2}dt/\int_{-\infty}^{+\infty}|\mathcal{E}_{in}(t)|^{2}dt$,
with bounds $\Theta$ encompassing only the first AFC echo, is given by
$\frac{16\left(\mathcal{C}/\mathcal{C}_{opt}\right)^{2}}{\left(1+\mathcal{C}/\mathcal{C}_{opt}\right)^{4}}.$
It equals $1$ under the impedance matching condition $\mathcal{C}=\mathcal{C}_{opt}$.

\smallskip{}
\noindent
We now add the control field into the analysis. When using the control field to implement a pair of $\pi$-pulses with a temporal duration much smaller than $\delta\omega_{in}^{-1}$ for an on-demand storage and readout of the excitation, the relation between the output and input fields is unchanged except for
a loss factor $e^{-\gamma_{S}T_{\text{stor}}}$ coming from the
decoherence (quantified by a Langevin noise with rate $\gamma_{S}$) during the storage time $T_{\text{stor}}$, i.e., the temporal delay between the two $\pi$-pulses.

\medskip{}
\noindent
\paragraph{Single-photon piecewise pulse shaper based on AFC --}

The proposed approach for stretching the output field starts with applying a storage $\pi$-pulse right before the AFC echo. We then use a series of $N_{shape}$ rectangular readout pulses separated by the input-echo time width $\sim$$\delta\omega_{in}^{-1}$, each of which partially transfers the excitation from $|s\rangle$ to $|e\rangle$. The pulses all
have a duration $\tau$ much smaller than $\delta\omega_{in}^{-1}$ but distinct areas. The area of pulse $j$ ($1\leq j\leq N_{shape}$) is labelled $r_{j}$ and is such that $\forall j\int\Omega_{j}(t)dt=r_{j}\pi=:2~\mathrm{arcsin}(q_{j})$. When the decoherence on the $|g\rangle$-$|s\rangle$ transition is negligible, we expect that the output field is given by a sequence of bins, all filled by the shape of the input pulse but with an intensity weight given by $p_{j}^{2}=q_{j}^{2}\prod_{k<j}(1-q_{k}^{2})$. If the $r_{j}$ are chosen such that $\sum_{j=1}^{N_{shape}}p_{j}^{2}=1$,
and if $\tilde{w}(\frac{2\pi}{\Delta})=\tilde{w}(0)$ and $\mathcal{C}/\mathcal{C}_{opt}=1$,
we expect unit overall efficiency of the $N_{shape}$-factor stretching procedure. Refer to Fig.~\ref{Fig3} introduced in the next Section for a numerical illustration.

\smallskip{}
\noindent
Crucially, we find a recipe to choose the amplitudes
and phases of the readout pulses so as to shape the waveform of the output photon and maximize the overlap with any targeted waveform, beyond a simple stretching of the input. In particular, $p_j$ is chosen proportional to the overlap between the input translated to bin $j$ and the target restricted to that bin (see Appendix~E for more details).
This waveform shaping technique preserves photon purity (see Appendix~F).

\smallskip{}
\noindent
Interestingly, the technique can easily be adapted to preserve the multimode capacity of the AFC protocol, a feature facilitating the realization of quantum repeaters with high entanglement distribution rates~\cite{simon_quantum_2007}. Consider an input field which is a series of time bins, all empty except one that contains a single photon, with the exact temporal location of the photon revealed after the storage $\pi$-pulse. In this case, two synchronisation $\pi$-pulses are added between the storage pulse and the readout pulse series so that the echo of the time bin filled with the photon is emitted immediately after the readout pulses. This avoids unwanted delays between the echos emitted after each readout pulse and preserves the efficiency and versatility of the waveform shaping. 

The synchronization pulses turn out to be also useful in the standard case of an input field defined as a single time bin filled with a single photon. They can indeed be used to crop the echos and shorten the temporal separation between the readout pulses. This enhances the overlap with a targeted waveform without much detriment to the efficiency. We call this the cropped-echo technique (see Appendix E).

\begin{figure}
\centering \includegraphics[width=1.\columnwidth]{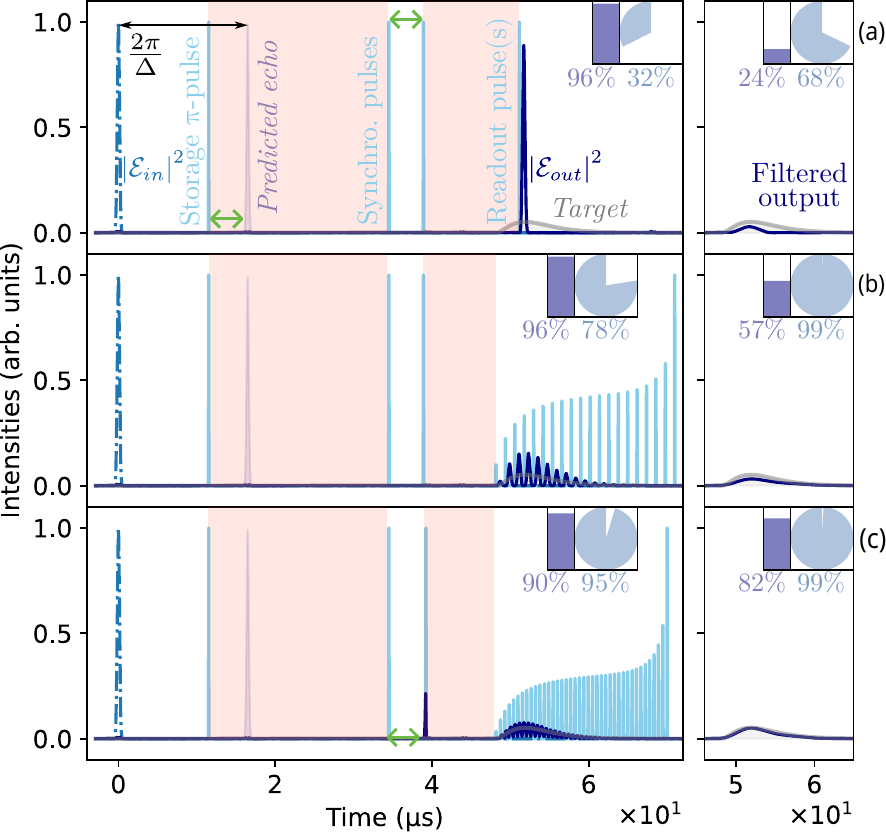}
\cprotect\caption{Simulation of cavity-assisted AFC protocols implemented in Pr$^{3+}$:Y$_{2}$SiO$_{5}$. \cprotect\textbf{Left panels} show the input field intensity $|\mathcal{E}_{in}(t)|^2$ (dash-dotted line), the storage $\pi$-pulse (light blue line) applied before the AFC rephasing time $2\pi/\Delta$ (grayed out), readout pulses (light blue lines), and output field intensity $\left|\mathcal{E}_{out}\right|^{2}$ (dark blue), synchronized and
superimposed on a target mode wavefunction (light gray). In the case that the input field occupies an unknown time bin revealed after the storage pulse, two synchonization pulses (light blue lines) are added with a temporal separation (indicated by green double arrows) corresponding to the time delay between the storage pulse and the AFC rephasing time. Input and output intensities are renormalized w.r.t.
the input maximal value, while control pulses are renormalized w.r.t. their maximal Rabi frequency.
\cprotect\textbf{Right panels} display filtered output signals,
with a sharp cutoff below/above $\pm2\pi\times0.15$~MHz. Insets show the AFC efficiency of the output main component (blue rectangular indicators) and the conditional overlap between the output and the target waveform (circular indicators), without and with filtration. The numerical values are also provided.

\cprotect\textbf{(a)} Standard cavity-assisted AFC protocol with a single
readout pulse. \cprotect\textbf{(b)} Waveform shaping with a sequence of readout pulses separated by the echo duration. \cprotect\textbf{(c)} Cropped-echo readout sequence, where the delay between the synchronization pulses is slightly increased to crop the subsequent echo (see the small output tail before the second synchronization pulse) and the delay between readout pulses is decreased so as to improve overlap with the target.
}
\label{Fig3} 
\end{figure}

\medskip{}
\noindent
\paragraph{Numerical results --}
For concreteness, we present results of realistic numerical simulations. We focus on a cavity-assisted protocol implemented with a Pr$^{3+}$:Y$_{2}$SiO$_{5}$ crystal for which an AFC efficiency reaching up to 62\% has recently been reported \citep{duranti_efficient_2024}.
In particular, the hyperfine
transitions $\pm1/2$ -- $\pm3/2$ and $\pm3/2$ -- $\pm3/2$ of the $^{3}$H$_{4}$ -- $^{1}$D$_{2}$ line are considered for the $|g\rangle$-$|e\rangle$ and $|e\rangle$-$|s\rangle$ transitions respectively. 
For the AFC, we focus on a rectangular distribution $v$ (width
$\Gamma=2\pi\times4$~MHz) with $67$ Gaussian teeth $w$ (width $\gamma_{tooth}=2\pi\times1$ kHz), separated by $\Delta=2\pi\times61$~kHz, and free space mean optical depth $\tilde{d}=0.48$. This choice fulfils the impedance matching condition for a cavity amplitude decay rate $\kappa=2\pi\times55$~MHz (cavity
finesse $\mathcal{F}_{cav}=\pi c/(2L_{cav}\kappa)= 6.6$ for cavity length $L_{cav}=208$~mm, mirrors with input and output reflectivity $R_{in}=0.4$
and $R_{out}=1$): $g\sqrt{N}=2\pi\times8.4$~MHz within the cavity.

We consider a Gaussian input photon with $330$~ns intensity FWHM, and take control pulses almost covering the corresponding 
$1.3$~MHz photon bandwidth FWHM, hence $\tau=0.07$~µs (from the dipole
moment of the optical transition $\pm3/2$ -- $\pm3/2$,
and assuming a beam diameter of $50$~µm, this corresponds to $100$~mW $\pi$-pulse peak power). 

\smallskip{}
\noindent
The scalar equations discussed above are solved using a numerical integrator, see Appendices~G and H for additional details on the parameters and numerical methods. The result is shown in Fig.~\ref{Fig3}(a), where we observe an overall AFC efficiency of 96\% after a $1.6$~µs rephasing time, limited by the AFC (99\% absorption) and the $\pi$-pulses bandwidths. 
We then replace the single readout
$\pi$-pulse by $20$ pulses with increasing Rabi frequencies to achieve optimal overlap with a waveform corresponding to the (pure part of a) photon of a cavity-coupled-ion (see below), cf. Fig.~\ref{Fig3}(b). The efficiency remains unchanged with respect to (w.r.t.)
the AFC protocol without pulse shaping but the conditional overlap improves from 32\% to 78\%. Fig.~\ref{Fig3}(c) shows that the cropped-echo technique removes detrimental gaps
in the piecewise output. The efficiency is almost preserved (90\%) and the overlap reaches 95\%. Filtering of output photons is also considered in each case to further improve the overlaps with a limited efficiency reduction, see caption of Fig.~\ref{Fig3}.

\medskip{}
\noindent
\paragraph{Photon waveform matching trapped ion's emission --} The target waveform was chosen as that of a photon produced by a cavity-coupled ion. 

Specifically, we consider a single $^{40}$Ca$^{+}$ ion, trapped at the waist of an optical cavity, generating single photons via a cavity-mediated
Raman transition \citep{keller_continuous_2004}. We employ here the theoretical model that was used in Ref.~\citep{krutyanskiy_entanglement_2023} to accurately reproduce the photon field observed in their experiments. The photon waveform obtained from this model is asymmetric with a full width half maximum (FWHM) of $\sim$$11$~µs, a fast rising front and longer decreasing tail, see Fig.~\ref{Fig4}(a). The quality of these photons can be assessed by HOM-like interference, which consists of sending two photons into separate input ports of a beam splitter and recording coincidences on photon detectors placed at the beam splitter outputs (see Appendix~I). In the case of perfectly pure and indistinguishable waveforms, the photons bunch, and there
is no coincidence detection: the interference visibility reaches $1$. One of the primary sources of imperfections in Ref.~\citep{krutyanskiy_entanglement_2023} was identified as spontaneous emission. Removing spontaneous emission in the simulation leads to pure photons with a FWHM of $\sim$$6$~µs, see Fig.~\ref{Fig4} a).

\begin{figure}
\centering{}\centering \includegraphics[width=1.\columnwidth]{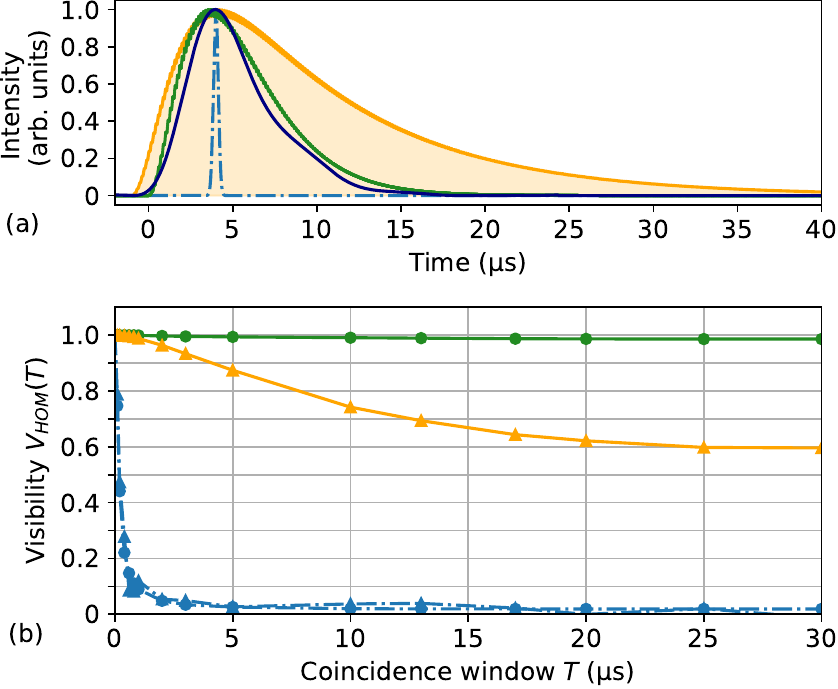}
\cprotect\caption{ \cprotect\textbf{(a)}~Photon waveform (renormalized field intensities w.r.t. their maximum) emitted by a single $^{40}$Ca$^{+}$ ion embedded in a cavity~\citep{krutyanskiy_entanglement_2023} (orange) and in the ideal case where unwanted ion spontaneous emission is removed (green). The waveform associated with the photon used as input of the AFC memory (dash-dotted light blue) and the shaped waveform output filtered to discard the arches generated by the shaping (dark blue) are also shown. Synchronization (time offsets) between the waveforms is set to achieve the best possible
visibilities. \protect 
\cprotect\textbf{(b)}~HOM visibility as a function of the acceptance coincidence window between full photons emitted by a single ion 
and an AFC memory without (blue dash-dotted line with triangles) and with (orange solid line with triangles) waveform shaping. For comparison, the visibilities that would be obtained between a pure ion photon and an AFC photon with (green solid line with bullets) and without shaping (blue dash-dotted line with triangles) are also given.
\protect 
}
\label{Fig4} 
\end{figure}

\smallskip{}
\noindent
We are here interested in the HOM interference between a photon emitted by a single $^{40}$Ca$^{+}$ ion and another coming from an AFC protocol realized in Pr$^{3+}$:Y$_{2}$SiO$_{5}$. The AFC parameters and the input for the AFC memory are as mentioned above, with the ion-emitted $\sim$$6$~µs--waveform as a shaping target.

The resulting filtered AFC photon waveform is shown in Fig.~\ref{Fig4}~(a). Fig.~\ref{Fig4}~(b) shows the interference visibility $V_{HOM}$ as a function of the acceptance coincidence window $T$ without and with the shaping protocol including a filtration of the output photon, as in Fig.~\ref{Fig3} right panels. The visibility that would be obtained in the ideal case without spontaneous emission during the ion's photon emission shows that the visibility in the shaped case is mostly limited by the impurity of the ion's photon. We add that taking the wider mean waveform as a shaping target did not improve the results.

\medskip
\paragraph{Conclusion --}
We have demonstrated how to expand and shape the waveform of single photons by modifying the AFC protocol, thereby addressing the general problem of how to interface two physical platforms that interact with light on very different timescales. We conclude by quantifying the advantage of this waveform shaping for the long distance entanglement distribution channel shown in Fig.~\ref{Fig1}. We consider the case where the end memories of the two long-distance repeater chains are each loaded with dual-rail entanglement. As detailed in Appendix~A, the probability of getting the four clicks heralding ion-ion entanglement is given by $\mathds{P}_{4cl}=(\eta_{det}^{4}\eta_{mem}^{2}\eta_{ion}^{2})/8$, with $\eta_{det}$ the photon detector efficiency, $\eta_{mem}$ and $\eta_{ion}$ the overall memory and ion efficiencies, which include photon frequency conversion to a shared frequency. The conditional ion-ion state fidelity is given by $\mathcal{F}_{ion-ion}^{mixed}=\frac{1+{V_{HOM}^{\infty}}^2}{2}$, where $\infty$ means that we are considering an infinitely large acceptance window $T$ (no post-selection of detection times) and "mixed" refers to the full ion photon. Assuming proven values $\eta_{det}=0.9$, $\eta_{mem}=0.5$ and $\eta_{ion}=0.1$, the basic case without shaping leads to a heralding probability $\mathds{P}_{4cl}=2.2\times10^{-3}$ and a fidelity $\mathcal{F}_{ion-ion}^{mixed}=0.50$, also achievable with separable states. When the AFC photons are filtered (sharp cutoff of $\pm 2\pi\times0.15$~MHz) the fidelity increases to $\mathcal{F}_{ion-ion}^{mixed}=0.52$, but the heralding probability is reduced to $\mathds{P}_{4cl}=1.3\times10^{-4}$. When the AFC photon waveform is shaped and filtered, $\mathds{P}_{4cl}=1.6\times10^{-3}$ almost reaches the value of the basic case. Moreover, significantly, the fidelity is now well above the separability bound: $\mathcal{F}_{ion-ion}^{mixed}=0.68$. The primary limitation arises from the impurity of the ion-photon state, which could be mitigated by enhancing the cavity cooperativity~\citep{schupp_interface_2021}. Nonetheless, our analysis indicates that entanglement between two ions mediated by rare-earth quantum repeaters is achievable with the current performance parameters.

\medskip
\paragraph{Acknowledgments --}
We thank M.~Afzelius, J.-D.~Bancal, E.~Gouzien, L.~Feldmann, S.~Grandi, G.~Misguich, C.~Lanore, P.~Sekatski and S.~Wengerowsky for insightful discussions. 
This work received funding from the European Union's Horizon Europe research and innovation programme under grant agreement No. 101102140 and project name 'QIA-Phase 1'. The opinions expressed in this document reflect only the author's view and reflect in no way the European Commission's opinions. The European Commission is not responsible for any use that may be made of the information it contains. P.~Cussenot acknowledges funding from the French Direction Générale de l'Armement (DGA).\\

N.S., H.d.R., T.N., and B.L. proposed the network architecture. P.C., N.S., and A.S. introduced the waveform shaping technique. P.C. and
N.S. performed the theoretical analysis with the help of B.G. for the
network analysis. P.C. and N.S. wrote the manuscript, with contribution
of B.G., and with inputs from all authors. The project was
supervised by N.S., B.L., T.N., H.d.R., and A.S..%

\bibliography{ReferencesArticleVBB.bib}

\clearpage
\newpage

\newpage
\setcounter{figure}{0}
\renewcommand{\thefigure}{S\arabic{figure}} 

\appendix

\title{Appendices: Uniting Quantum Processing Nodes of Cavity-coupled
Ions with Rare-earth Quantum Repeaters Using Single-photon Pulse Shaping
Based on Atomic Frequency Comb}

\date{\today}
\begin{abstract}
In these Appendices,
some calculation details to support the main text results are provided, as well as explanations about the different physical platforms used for the network. First, we detail the suggested quantum network architecture for long-distance entanglement distribution and assess the fidelities and rates that can be achieved thanks to our proposal. Second, we review the physics of single-photon storage in a cavity-assisted AFC quantum memory. Third, we explain our photon-shaping protocol in details and show that it is compatible with the emission of a pure photon. Finally, we provide insight into photon emission by a single trapped ion and recall some results about Hong-Ou-Mandel interferences as a distinguishability witness. Along the way, some technical points are discussed to support the former sections. Translation of the parameters of our model into experimental values is also detailed.
\end{abstract}
\maketitle
\tableofcontents{}

\newpage
\clearpage

\begin{widetext}

\begin{figure*}
\centering \includegraphics[width=0.9\textwidth]{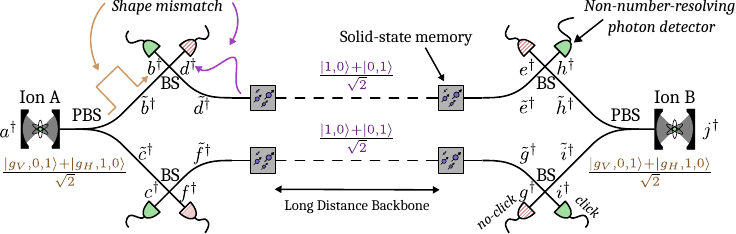}
\caption{Proposed network architecture, with notations for the different modes
involved.\label{fig:net@base}}
\end{figure*}

\end{widetext}

\section{Quantum network architecture and fidelity estimation}

\label{sec:Quantum-network-architecture}

In this Appendix, we start with an analysis of the quantum network
architecture that is given in Fig.~1 of the main text. We explain
how entanglement between the two extreme nodes can be generated conditioned
on particular detection events and provide explicit formulas for
the probability of generation and for the fidelity of the resulting
entangled state.

\subsection{Presentation of the network protocol}

\label{sec:net@arc} 

The network architecture is depicted again in Fig.~\ref{fig:net@base}.
It consists of two long-distance repeater chains \citep{sangouard_quantum_2011}, each generating entanglement between two solid-state quantum memories;
four balanced beam splitters (BS); eight non-photon-number-resolving
photodetectors; two polarising beam splitters (PBS); and two ion nodes
($A$ and $B$) \citep{krutyanskiy_entanglement_2023}, thought
to be located far apart from one another. In the following analysis,
we will assume that the long-distance repeater chains work as ideal
entanglement generators. That is, we start with loaded long-distance
repeater chains, each producing a pure Bell state $|\Psi^{+}\rangle=\frac{|1,0\rangle+|0,1\rangle}{\sqrt{2}}$,
where $|1,0\rangle$ is the state for which the left memory stores
an excitation while the right one does not, and conversely for $|0,1\rangle$. 

We introduce the vacuum state
\begin{eqnarray*}
|\bar{0}\rangle & := & \quad|g\rangle_{A}\otimes|0\rangle_{\tilde{b}}\otimes|0\rangle_{\tilde{c}}\otimes|0\rangle_{\tilde{d}}\otimes|0\rangle_{\tilde{f}}\\
 &  & \otimes|0\rangle_{\tilde{e}}\otimes|0\rangle_{\tilde{g}}\otimes|0\rangle_{\tilde{h}}\otimes|0\rangle_{\tilde{i}}\otimes|g\rangle_{B},
\end{eqnarray*}

\noindent where letters $(\tilde{b},\tilde{c},\tilde{d},\tilde{e},\tilde{f},\tilde{g},\tilde{h},\tilde{i})$
refer to photon modes of the BS input ports (and PBS output ports).
For each of these modes, for instance $\tilde{b}$, the creation of
a photon is denoted by $\tilde{b}^{\dagger}|0\rangle_{\tilde{b}}\equiv\tilde{b}^{\dagger}|\bar{0}\rangle$.
The driving from state $|g\rangle$ of any of the ions leads to the
emission of a photon of polarisation $V$ or $H$, with the ion decaying
either in the $|g_{V}\rangle$ or the $|g_{H}\rangle$ state. For
convenience, we introduce mode--like operators $a_{H},a_{V}$ and
$j_{H},j_{V}$ such that $a_{H}^{\dagger}\left|g\right\rangle _{A}=|g_{H}\rangle_{A}$,
$a_{V}^{\dagger}\left|g\right\rangle _{A}=|g_{V}\rangle_{A}$, $j_{H}^{\dagger}\left|g\right\rangle _{B}=|g_{H}\rangle_{B}$,
and $j_{V}^{\dagger}\left|g\right\rangle _{B}=|g_{V}\rangle_{B}$.\\

\paragraph{Initialisation}

We start by preparing each ion in a superposition of $|g_{V}\rangle$
and $|g_{H}\rangle$ and collecting the emitted photons, as well as
independently loading the repeaters with Bell states and triggering
emission from the memories. More precisely, one starts from the state
\begin{eqnarray}
|\psi_{0}\rangle & := & \quad\left[\frac{a_{H}^{\dagger}\tilde{b}^{\dagger}+a_{V}^{\dagger}\tilde{c}^{\dagger}}{\sqrt{2}}\right]\left[\frac{\tilde{d}^{\dagger}+\tilde{e}^{\dagger}}{\sqrt{2}}\right]\nonumber \\
 &  & \times\left[\frac{\tilde{f}^{\dagger}+\tilde{g}^{\dagger}}{\sqrt{2}}\right]\left[\frac{\tilde{h}^{\dagger}j_{H}^{\dagger}+\tilde{i}^{\dagger}j_{V}^{\dagger}}{\sqrt{2}}\right]|\bar{0}\rangle.\label{eq:psi_0_bellrepeater}
\end{eqnarray}

\noindent As such, the network is loaded with four excitations (photons)
in total: one in $\tilde{b}$ or $\tilde{c}$, one in $\tilde{d}$
or $\tilde{e}$, one in $\tilde{f}$ or $\tilde{g}$, and one in $\tilde{h}$
or $\tilde{i}$.\\

\paragraph{Beam Splitters}

The presence of the four beam splitters leads to interference between
pairs of modes $(\tilde{b},\tilde{d})$, $(\tilde{c},\tilde{f})$,
$(\tilde{h},\tilde{e})$, and $(\tilde{i},\tilde{g})$. For now, we
assume that the interference is perfect\footnote{We relax this assumption in Sec.~\ref{subsec:Imperfect-mode-overlap}.},
so that we introduce modes $(b,d)$, $(c,f)$, $(h,e)$, and $(i,g)$,
such that the unitary action of the beam splitters 
corresponds to the mapping, for instance
\begin{equation}
\begin{cases}
\tilde{b}\rightarrow & b=\frac{\tilde{b}+\tilde{d}}{\sqrt{2}}\\
\tilde{d}\rightarrow & d=\frac{\tilde{b}-\tilde{d}}{\sqrt{2}}
\end{cases}.\label{eq:BS_unitary}
\end{equation}

\noindent The BS unitary action transforms $|\psi_{0}\rangle$ into
\begin{eqnarray}
|\psi\rangle & = & \frac{1}{4}\left[a_{H}\cdot\frac{b+d}{\sqrt{2}}+a_{V}\cdot\frac{c+f}{\sqrt{2}}\right]^{\dagger}\left[\frac{b-d}{\sqrt{2}}+\frac{h-e}{\sqrt{2}}\right]^{\dagger}\label{eq:state_before_detection}\\
 &  & \times\left[\frac{c-f}{\sqrt{2}}+\frac{i-g}{\sqrt{2}}\right]^{\dagger}\left[j_{H}\cdot\frac{h+e}{\sqrt{2}}+j_{V}\cdot\frac{i+g}{\sqrt{2}}\right]^{\dagger}|\bar{0}\rangle.\nonumber 
\end{eqnarray}

\paragraph{Detection and heralding}

\begin{figure}
\begin{centering}
\includegraphics[width=1\columnwidth]{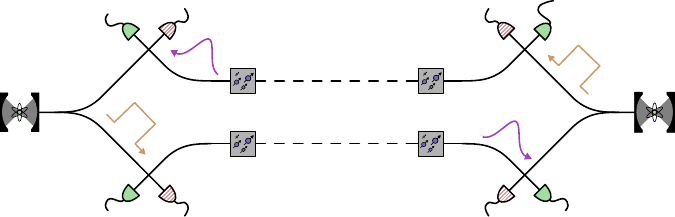}
\par\end{centering}
\begin{centering}
\includegraphics[width=1\columnwidth]{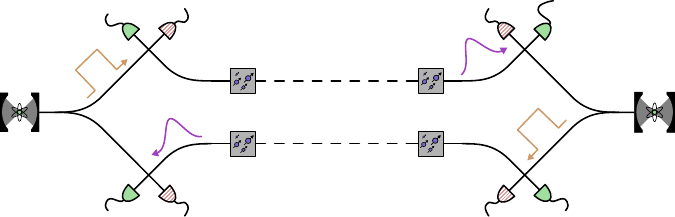}
\par\end{centering}
\caption{Detection diagrams for four-click scenarios. In the first case, the
top-left click comes from the photon being in $\tilde{d}$, which in turn
requires the top-right click to stem from a photon in $\tilde{h}$, the bottom-right
click from $\tilde{g}$, and the bottom-left from $\tilde{c}$. A possible set of detectors that click is in green, with the other detectors being striped in red. \label{fig:Detection-diagram}}
\end{figure}

The heralding step consists of fou\textit{\emph{r single-click}} measurements,
one after each beam splitter. Since the network contains exactly four
photons delocalised across the optical modes, there are two possible
scenarios, depicted in Fig.~\ref{fig:Detection-diagram}. Clicks
on $b$, $c$, $h$, and $i$ leave the ions in $|g_{V}\rangle_{A}\otimes|g_{H}\rangle_{B}$
or in $|g_{H}\rangle_{A}\otimes|g_{V}\rangle_{B}$ depending on whether
detector clicks are produced by memories' or ions' photons. Due to
the lack of which-path information, one ends up with the ion-encoded
Bell state
\begin{equation}
|\Psi_{ion-ion}^{+}\rangle=\frac{|g_{V}\rangle_{A}\otimes|g_{H}\rangle_{B}+|g_{H}\rangle_{A}\otimes|g_{V}\rangle_{B}}{\sqrt{2}}.\label{eq:ion-ion-Bell-state}
\end{equation}

\noindent The case where two photons (here $\tilde{b}$ and $\tilde{d}$)
bunch and lead to the same single-click event is discarded by restricting to the heralding of four clicks and not fewer. Note that there are $16$
combinations of detection events with one click after each BS, which
lead to the same state \ref{eq:ion-ion-Bell-state} up to a local
correction (change of the relative phase in the superposition).

\subsection{Entanglement generation probability and fidelity}

\label{sec:net@pre}

We now tackle the network analysis more in depth so as to compute
the probability of getting the four clicks and the fidelity of the
heralded ion-ion state w.r.t. state~\ref{eq:ion-ion-Bell-state}, in
the case where losses happen and devices do not have a unit efficiency.
What is more, we now allow the input modes of the beam splitters to
represent partially overlapping photon waveforms \citep{fabre_modes_2020}.\\

\subsubsection{Imperfect mode overlap}

\label{subsec:Imperfect-mode-overlap}

Indeed, the transformation~\ref{eq:BS_unitary} implicitly assumes
that the two incident modes $\tilde{b}$ and $\tilde{d}$ are indistinguishable
\citep{legero_characterization_2006}. In our context, we remind the reader that
the photons emitted by the ions and the memories may have different
waveforms (see Appendices~\ref{sec:Ion-Photon-state-and} and \ref{sec:Modelling-a-cavity-asisted});
this may result in very different spatio-temporal shapes. Following
Ref.~\citep{craddock_quantum_2019}, we will take modes $\tilde{d}$, $\tilde{e}$,
$\tilde{f}$, and $\tilde{g}$ as references for each of the
beam splitters (modes coming from the memories), and decompose modes
(coming from the ions) $\tilde{b}$, $\tilde{h}$, $\tilde{c}$ and
$\tilde{i}$ into some interfering components, for which we keep the
same notations, and some non-interfering orthogonal components $\tilde{b}_{\perp}$,
$\tilde{h}_{\perp}$, $\tilde{c}_{\perp}$, and $\tilde{i}_{\perp}$.
For instance, we write
\begin{equation*}
\tilde{b}^{\dagger}\rightarrow\sqrt{x_{A}}\tilde{b}^{\dagger}+\sqrt{1-x_{A}}\tilde{b}_{\perp}^{\dagger},
\end{equation*}
where $x_{A}\in[0,1]$ quantifies the degree of waveform overlap (scalar
product) with mode $\tilde{d}$ ($x_{A}=1$ when the photons have
identical shapes and fully interfere, and $x_{A}=0$ when the modes
are orthogonal and do not interfere), with $\langle\bar{0}|\tilde{b}\tilde{b}_{\perp}^{\dagger}|\bar{0}\rangle=\langle\bar{0}|[\tilde{b},\tilde{b}_{\perp}^{\dagger}]|\bar{0}\rangle=0$.
For simplicity, we take a symmetric situation where the value of $x_{A}$
is the same for the overlaps ($\tilde{b},\tilde{d}$) and ($\tilde{c},\tilde{f}$),
while $x_{B}$ is defined for the overlaps ($\tilde{e},\tilde{h}$)
and ($\tilde{g},\tilde{i}$). Then the unitary transformation due
to the beam splitters acts independently on orthogonal modes, so that
the state \ref{eq:state_before_detection} before detection is now given by 

\begin{widetext}

\begin{equation}
\begin{aligned}\\|\psi\rangle= & \frac{1}{4}\left[\sqrt{x_{A}}\left(a_{H}\cdot\frac{b+d}{\sqrt{2}}+a_{V}\cdot\frac{c+f}{\sqrt{2}}\right)+\sqrt{1-x_{A}}\left(a_{H}\cdot\frac{b_{\perp}+d_{\perp}}{\sqrt{2}}+a_{V}\cdot\frac{c_{\perp}+f_{\perp}}{\sqrt{2}}\right)\right]^{\dagger}\times\left[\frac{b-d}{\sqrt{2}}+\frac{h-e}{\sqrt{2}}\right]^{\dagger}\\
 & \times\left[\frac{c-f}{\sqrt{2}}+\frac{i-g}{\sqrt{2}}\right]^{\dagger}\times\left[\sqrt{x_{B}}\left(j_{H}\cdot\frac{h+e}{\sqrt{2}}+j_{V}\cdot\frac{i+g}{\sqrt{2}}\right)+\sqrt{1-x_{B}}\left(j_{H}\cdot\frac{h_{\perp}+e_{\perp}}{\sqrt{2}}+j_{V}\cdot\frac{i_{\perp}+g_{\perp}}{\sqrt{2}}\right)\right]^{\dagger}|\underline{0}\rangle,
\end{aligned}
\label{eq:net@pur@BS}
\end{equation}

\end{widetext}

\noindent where we introduced $d_{\perp}$, $e_{\perp}$, $f_{\perp}$,
and $g_{\perp}$ (as well as $b_{\perp}$, $h_{\perp}$, $c_{\perp}$,
and $i_{\perp}$without the tilde superscript) as the orthogonal
output modes after the beam splitters (for orthogonal input modes, the
BS relation is for instance given by $\tilde{b}_{\perp}\rightarrow\frac{b_{\perp}+d_{\perp}}{\sqrt{2}}$,
and no input of the type $\tilde{d}_{\perp}$ is to be considered).

\subsubsection{Detection POVMs}

We then build the Positive Operator-Valued Measure operator (POVM)
associated to the non-photon-resolving photodetectors. If one provides
an orthogonal mode basis \citep{fabre_modes_2020} for the light arriving on detector $b$, that we note as a set of  $\beta_{k}$ with mode index $k$, then the POVM for a no-click
event on $b$ is defined as
\begin{equation}
\Pi_{nc}^{b}:=(1-\eta_{det})^{\sum_{k}\beta_{k}^{\dagger}\beta_{k}}=\prod_{k}(1-\eta_{det})^{\beta_{k}^{\dagger}\beta_{k}},\label{eq:POVM_non_clic}
\end{equation}

\noindent where $\eta_{det}$ is the efficiency of the detector and
$[\beta_{k},\beta_{k'\neq k}]=0$, $[\beta_{k},\beta_{k'\neq k}^{\dagger}]=0$.
Using the Baker-Campbell-Hausdorff formula, one can see that $\Pi_{nc}^{b}$
is diagonal in the Fock basis for modes $\beta_{k}$, and the probability
of not detecting a number state with $M_{k}$ photons in the mode
$\beta_{k}$ is $\frac{1}{M_{k}!}\langle0|\beta_{k}^{M_{k}}\Pi_{nc}^{b}(\beta_{k}^{M_{k}})^{\dagger}|0\rangle=(1-\eta_{det})^{M_{k}}$.
In particular, $\langle0|\Pi_{nc}^{b}|0\rangle=1$. Then if we decompose
the output $b$ into the basis $\{\beta_{k}\}_{k}$ by writing $b^{\dagger}=\sum_{k}\gamma_{k}\beta_{k}^{\dagger}$,
we note that
\[
\begin{aligned}\Pi_{nc}^{b}b^{\dagger}|0\rangle & =(1-\eta_{det})^{\sum_{k}\beta_{k}^{\dagger}\beta_{k}}b^{\dagger}|0\rangle\\
 & =(1-\eta_{det})^{\sum_{k}\beta_{k}^{\dagger}\beta_{k}}\left(\sum_{k}\gamma_{k}\beta_{k}^{\dagger}\right)|0\rangle\\
 & =\sum_{k}\gamma_{k}\cdot(1-\eta_{det})\beta_{k}^{\dagger}|0\rangle\\
 & =(1-\eta_{det})b^{\dagger}|0\rangle.
\end{aligned}
\]

\noindent An identical derivation leads to $\Pi_{nc}^{b}b_{\perp}^{\dagger}|0\rangle=(1-\eta_{det})b_{\perp}^{\dagger}|0\rangle$.
This shows that $\Pi_{nc}^{b}$ is a suitable POVM for the non-click
event at the detector whatever the incoming mode (like $b$, $b_{\perp}$
and others)\footnote{This of course assumes that the experiment is run so that all modes
are detected with the same efficiency.}. Similarly, the complementary POVM for a click event is introduced as 
\[
\Pi_{c}^{b}=\mathds{1}-\Pi_{nc}^{b}=\mathds{1}-(1-\eta_{det})^{\sum_{k}\beta_{k}^{\dagger}\beta_{k}}.
\]

\noindent Note that 
\begin{eqnarray}
\Pi_{c}^{b}|0\rangle & = & |0\rangle-|0\rangle=0\nonumber \\
\Pi_{c}^{b}b_{\perp}^{\dagger}|0\rangle & = & b_{\perp}^{\dagger}\Pi_{c}^{b}|0\rangle=0.\label{eq:0_click}
\end{eqnarray}
Likewise, POVMs $\Pi_{nc}^{d}$, $\Pi_{c}^{d}$, $\Pi_{nc}^{c}$,
$\Pi_{c}^{c}$, $\Pi_{nc}^{f}$, $\Pi_{c}^{f}$, $\Pi_{nc}^{e}$,
$\Pi_{c}^{e}$, $\Pi_{nc}^{h}$, $\Pi_{c}^{h}$, $\Pi_{nc}^{g}$,
$\Pi_{c}^{g}$, $\Pi_{nc}^{i}$, $\Pi_{c}^{i}$ are introduced for
the other photodetectors.

\subsubsection{Four-click event}

There are $2^{4}=16$ ways for exactly one detector per beam splitter
pair to click simultaneously as a four-click event that we look for.
For clarity, we focus on one way, which corresponds to the situation
on the top of Fig.~\ref{fig:Detection-diagram} (detectors that click
are in green, namely $b$, $c$, $h$, and $i$). For this situation,
the four-click event POVM is given by

\[
\Pi_{4cl,1}=(\Pi_{c}^{b}\times\Pi_{nc}^{d})\times(\Pi_{c}^{c}\times\Pi_{nc}^{f})\times(\Pi_{c}^{h}\times\Pi_{nc}^{e})\times(\Pi_{c}^{i}\times\Pi_{nc}^{g}),
\]

\noindent where all factors commute with one another. Starting from
Eq.~\ref{eq:net@pur@BS}, the heralded state after detection is then

\[
\rho_{ion-ion}=\frac{1}{\mathds{P}_{4cl,1}}\mathrm{Tr}_{repeaters}\left(\Pi_{4c,1}|\psi\rangle\langle\psi|\right),
\]

\noindent where $\mathrm{Tr}_{repeaters}$ denotes the partial trace
over the inner modes (i.e., all modes but $a$ and $j$), and the normalisation
constant 

\[
\mathds{P}_{4cl,1}=\mathrm{Tr}\left(\Pi_{4c,1}|\Psi\rangle\langle\Psi|\right)
\]

\noindent is the heralding probability of that specific four-click
event. 

Expanding the expression of $\Pi_{4c,1}|\Psi\rangle\langle\Psi|$
yields many zero terms: the only non-zero terms are those for which
the detection modes ($b$, $c$, $h$, and $i$ or their orthogonal
counterparts) are loaded with four excitations in $|\Psi\rangle$
(see Eq.~\ref{eq:0_click}, and recall that $\Pi_{c}^{b}$ is a click
due to a photon arriving either through $b$ or through $b_{\perp}$).
Typically, $\Pi_{4c,1}b^{\dagger}c^{\dagger}h_{\perp}^{\dagger}i^{\dagger}|\bar{0}\rangle=\eta_{det}^{4}b^{\dagger}c^{\dagger}h^{\dagger}i^{\dagger}|\bar{0}\rangle$
but $\Pi_{4c,1}a_{V}^{\dagger}c_{\perp}^{\dagger}b^{\dagger}f^{\dagger}i_{\perp}^{\dagger}j_{V}^{\dagger}|\bar{0}\rangle=0$.
Picking relevant creation operators yields $8$ non-zero terms only,
which are provided in Table~\ref{tab:net@pur@terms-1} together with
the related coefficient in the expansion of $|\psi\rangle$ and the state $|g_{H}\rangle$ or $|g_{V}\rangle$ to which the ions decayed. The expansion is also weighted by a $\eta_{det}^{4}$ coefficient,
corresponding to the success probability of $4$ detections.

\begin{table*}[th]
\centering \fontsize{10pt}{14pt}\selectfont %
\begin{tabular}{ccccccc}
\hline 
\multirow{1}{*}{} & \nth{1} factor & \nth{2} factor & \nth{3} factor & \nth{4} factor & \multirow{1}{*}{Coefficient} & \multirow{1}{*}{Ion modes involved}\tabularnewline
\hline 
Among & $b,c,b_{\perp},c_{\perp}$ & $b,h$ & $c,i$ & $h,i,h_{b},i_{\perp}$ &  & \tabularnewline
\hline 
\multirow{8}{*}{} & $b$ & $h$ & $c$ & $i$ & $\frac{1}{16}\cdot\sqrt{x_{A}x_{B}}$ & \multirow{4}{*}{$a_{H}$ and $j_{V}$}\tabularnewline
\cline{2-6}
 & $b$ & $h$ & $c$ & $i_{\perp}$ & $\frac{1}{16}\cdot\sqrt{x_{A}}\cdot\sqrt{1-x_{B}}$ & \tabularnewline
\cline{2-6}
 & $b_{\perp}$ & $h$ & $c$ & $i$ & $\frac{1}{16}\cdot\sqrt{1-x_{A}}\cdot\sqrt{x_{B}}$ & \tabularnewline
\cline{2-6}
 & $b_{\perp}$ & $h$ & $c$ & $i_{\perp}$ & $\frac{1}{16}\cdot\sqrt{1-x_{A}}\sqrt{1-x_{B}}$ & \tabularnewline
\cline{2-7}
 & $c$ & $b$ & $i$ & $h$ & $\frac{1}{16}\cdot\sqrt{x_{A}x_{B}}$ & \multirow{4}{*}{$a_{V}$ and $j_{H}$}\tabularnewline
\cline{2-6}
 & $c$ & $b$ & $i$ & $h_{\perp}$ & $\frac{1}{16}\cdot\sqrt{x_{A}}\cdot\sqrt{1-x_{B}}$ & \tabularnewline
\cline{2-6}
 & $c_{\perp}$ & $b$ & $i$ & $h$ & $\frac{1}{16}\cdot\sqrt{1-x_{A}}\cdot\sqrt{x_{B}}$ & \tabularnewline
\cline{2-6}
 & $c_{\perp}$ & $b$ & $i$ & $h_{\perp}$ & $\frac{1}{16}\cdot\sqrt{1-x_{A}}\sqrt{1-x_{B}}$ & \tabularnewline
\hline 
\end{tabular}\caption{Summary of all the 8 combinations built from the modes $b,c,h,i$
or their respective orthogonal modes, representing the non-zero terms
of $\Pi_{4c,1}|\Psi\rangle$. Note that the first and the fifth combinations
involve the same inner network nodes $b,h,c,i$. \label{tab:net@pur@terms-1}}
\end{table*}

Taking the partial trace w.r.t. all mode operators except $a_{H}$, $a_{V}$,
$j_{H}$, and $j_{V}$ yields the decomposition

\begin{widetext}

\begin{eqnarray*}
\mathrm{Tr}_{repeaters}\left(\Pi_{4c,1}|\psi\rangle\langle\psi|\right)= & \eta_{det}^{4} & \left[\frac{x_{A}x_{B}}{256}|\Psi_{ion-ion}^{+}\rangle\langle\Psi_{ion-ion}^{+}|\right.\\
 &  & +\left(\frac{x_{A}(1-x_{B})}{256}+\frac{(1-x_{A})x_{B}}{256}+\frac{(1-x_{A})(1-x_{B})}{256}\right)|g_{H}\rangle_{A}\otimes|g_{V}\rangle_{B}\langle g_{H}|_{A}\otimes\langle g_{V}|_{B}\\
 &  & \left.+\left(\frac{x_{A}(1-x_{B})}{256}+\frac{(1-x_{A})x_{B}}{256}+\frac{(1-x_{A})(1-x_{B})}{256}\right)|g_{V}\rangle_{A}\otimes|g_{H}\rangle_{B}\langle g_{V}|_{A}\otimes\langle g_{H}|_{B}\right]\\
= & \eta_{det}^{4} & \left[\frac{1}{256}|g_{V}\rangle_{A}\otimes|g_{H}\rangle_{B}\langle g_{V}|_{A}\otimes\langle g_{H}|_{B}+|g_{H}\rangle_{A}\otimes|g_{V}\rangle_{B}\langle g_{H}|_{A}\otimes\langle g_{V}|_{B}\right.\\
 &  & \left.+\frac{x_{A}x_{B}}{256}|g_{V}\rangle_{A}\otimes|g_{H}\rangle_{B}\langle g_{H}|_{A}\otimes\langle g_{V}|_{B}+|g_{H}\rangle_{A}\otimes|g_{V}\rangle_{B}\langle g_{V}|_{A}\otimes\langle g_{H}|_{B}\right].
\end{eqnarray*}

\end{widetext}

\noindent The four-click probability of our particular event is
\begin{equation*}
\mathds{P}_{4cl,1}=\frac{\eta_{det}^{4}}{128},
\end{equation*}
\noindent and we finally obtain the density matrix
\begin{eqnarray}
\rho_{ion-ion}= &  & \frac{1}{2}\left(1+x_{A}x_{B}\right)|\Psi_{ion-ion}^{+}\rangle\langle\Psi_{ion-ion}^{+}|\label{eq:four_click_heralded_state}\\
 & + & \frac{1}{2}\left(1-x_{A}x_{B}\right)|\Psi_{ion-ion}^{+}\rangle\langle\Psi_{ion-ion}^{+}|.\nonumber 
\end{eqnarray}

\noindent The fidelity w.r.t. the expected target state $|\Psi_{ion-ion}^{+}\rangle$
(see Eq.~\ref{eq:ion-ion-Bell-state}) is thus

\begin{equation}
\mathcal{F}_{ion-ion}=\frac{1}{2}\left(1+x_{A}x_{B}\right).\label{eq:fidelity_ion_ion_pure}
\end{equation}

\noindent Furthermore, the total four-click probability is

\begin{equation}
\mathds{P}_{4cl}=16\times\mathds{P}_{4cl,1}=\frac{\eta_{det}^{4}}{8},\label{eq:four_click_total_probability}
\end{equation}

\noindent which is not suprising since $2$ out of $16$ path combinations
for the four excitations lead to a four-click event.

\subsubsection{Imperfect situation}

So far, we assumed a perfect situation where both the memories and
the ions emit photons with unit efficiency, and no propagation loss
occur. We can now introduce $\eta_{ion}$ and $\eta_{rep}$ as the
efficiencies, including fiber losses, for the ions and solid-state memories
(assumed symmetric for the left and right sides of the network architecture). Tracing out the loss channels
leads to the initialised state $\rho_{0}=\begin{aligned}\rho_{ion}^{A}\end{aligned}
\otimes\rho_{rep}^{up}\otimes\rho_{rep}^{down}\otimes\begin{aligned}\rho_{ion}^{B}\end{aligned}$,  with
\begin{eqnarray*}
\begin{aligned}\rho_{ion}^{A}\end{aligned}
 & = & \quad\left(1-\eta_{ion}\right)|0\rangle\langle0|\\
 &  & +\eta_{ion}\left[\frac{a_{H}^{\dagger}\tilde{b}^{\dagger}+a_{V}^{\dagger}\tilde{c}^{\dagger}}{\sqrt{2}}\right]|0\rangle\langle0|\left[\frac{a_{H}\tilde{b}+a_{V}\tilde{c}}{\sqrt{2}}\right],\\
\rho_{rep}^{up} & = & \quad\left(1-\eta_{rep}\right)|0\rangle\langle0|\\
 &  & +\eta_{rep}\left[\frac{\tilde{d}^{\dagger}+\tilde{e}^{\dagger}}{\sqrt{2}}\right]|0\rangle\langle0|\left[\frac{\tilde{d}+\tilde{e}}{\sqrt{2}}\right],\\
\rho_{rep}^{down} & = & \quad\left(1-\eta_{rep}\right)|0\rangle\langle0|\\
 &  & +\eta_{rep}\left[\frac{\tilde{f}^{\dagger}+\tilde{g}^{\dagger}}{\sqrt{2}}\right]|0\rangle\langle0|\left[\frac{\tilde{f}+\tilde{g}}{\sqrt{2}}\right],\\
\begin{aligned}\rho_{ion}^{B}\end{aligned}
 & = & \quad\left(1-\eta_{ion}\right)|0\rangle\langle0|\\
 &  & +\eta_{ion}\left[\frac{\tilde{h}^{\dagger}j_{H}^{\dagger}+\tilde{i}^{\dagger}j_{V}^{\dagger}}{\sqrt{2}}\right]|0\rangle\langle0|\left[\frac{\tilde{h}j_{H}+\tilde{i}j_{V}}{\sqrt{2}}\right].
\end{eqnarray*}

\noindent Using previous derivations to identify the relevant four-excitation
components after the beam splitters' unitary action leads to the total
four-click probability
\[
\mathds{P}_{4cl}=\frac{\eta_{\text{det}}^{4}\eta_{\text{ion}}^{2}\eta_{\text{rep}}^{2}}{8},
\]

\noindent which is not surprising since a four-click event requires
$2$ photons coming from the ions and $2$ photons coming from the
memories. The heralded state and its fidelity are unaffected.

\subsection{Mixture}

\label{sec:net@mix}

We then point out that the initial ion-photon states may not be
pure for realistic experiments. As detailed in Appendix~\ref{sec:Ion-Photon-state-and}
(which focuses on experiments \citep{meraner_indistinguishable_2020,krutyanskiy_telecom-wavelength_2023}),
the initial ion-photon state is rather described by a continuous mixture
of pure components that we write here

\begin{widetext}

\[
\begin{aligned}\rho_{ion}^{A}\end{aligned}
=\left(1-\eta_{ion}\right)|0\rangle\langle0|+\int\bar{P}_{A}(s)\left[\frac{a_{H}^{\dagger}\tilde{b}^{\dagger}(s)+a_{V}^{\dagger}\tilde{c}^{\dagger}(s)}{\sqrt{2}}\right]|0\rangle\langle0|\left[\frac{a_{H}\tilde{b}(s)+a_{V}\tilde{c}(s)}{\sqrt{2}}\right]\mathrm{ds}
\]

\end{widetext}

\noindent with a probability distribution $\bar{P}_{A}(s)$ such that
$\int\bar{P}_{A}(s)\mathrm{d}s=\eta_{ion}$ and a collection of optical
modes $\{\tilde{b}(s),\tilde{c}(s)\}$ (not orthogonal to one another)
that depend on the value of $s$ (while the related ion state is supposed
independent of $s$). A similar state $\rho_{ion}^{B}$ is introduced
with a distribution $\bar{P}_{B}(s)$ and modes $\{\tilde{h}(s),\tilde{i}(s)\}$.

The linearity of the equations (that crucially relies on the fact
that we introduced some detection POVMs that are independent of the
incoming modes) then leads to the heralded state fidelity w.r.t. the
Bell state 

\begin{equation}
\mathcal{F}_{ion-ion}^{mixed}=\frac{1}{2}\left(1+\int\bar{P}_{A}(s)x_{A}(s)\mathrm{ds}\int\bar{P}_{B}(s')x_{B}(s')\mathrm{ds'}\right),\label{eq:fidelity_mixedion}
\end{equation}

\noindent where $x_{A}(s)$ ($x_{B}(s)$) is the overlap between the
modes $\tilde{b}(s)$ or $\tilde{c}(s)$ ($\tilde{h}(s)$ or $\tilde{i}(s)$)
and the mode $\tilde{d}$ or $\tilde{f}$ ($\tilde{e}$ or $\tilde{g}$).
The total four-click probability is kept unchanged:

\begin{equation}
\mathds{P}_{4cl}=\frac{\eta_{\text{det}}^{4}\int\bar{P}_{A}(s)\mathrm{d}s\int\bar{P}_{B}(s)\mathrm{d}s\eta_{\text{rep}}^{2}}{8}=\frac{\eta_{\text{det}}^{4}\eta_{\text{ion}}^{2}\eta_{\text{rep}}^{2}}{8}.\label{eq:total4c_probability}
\end{equation}

In Appendix~\ref{sec:Ion-Photon-state-and}, we will show that the
fidelities \ref{eq:fidelity_ion_ion_pure} and \ref{eq:fidelity_mixedion}
can be related to the visibilities of Hong Ou Mandel--like experiments
between the photons of the memories and of the ions (see Eqs.~\ref{eq:visi_mixed_mixed},
\ref{eq:visi_mixed_novacuum}, \ref{eq:visi_pure_mixed}, and \ref{eq:visi_pure_pure}).

Furthermore, we point out that a clear way to enhance the heralded state
fidelity is to increase the overlap between the memory photon mode
and at least one ion photon mode. However, this enhancement in overlap
should not be too detrimental to the emission efficiencies and thus
the value of entanglement generation rate of the repeater. This is
the goal of our shaping protocol, which we will detail in Appendix~\ref{sec:Protocol-for-photon-shaping}.
Numerical estimations of the overlaps are provided.

\subsection{Experimental efficiencies}

Finally, we provide experimentally relevant values for the efficiencies
$\eta_{det}$, $\eta_{ion}$ and $\eta_{mem}$. For the memory emission,
we write $\eta_{mem}=\eta_{AFC,shaping}\times\eta_{QFC}$ where we
include the efficiency of the AFC shaping process $\eta_{AFC,shaping}$,
which we take to be $82$~\% (see main text), and some end-to-end conversion to telecom wavelength with efficiency $\eta_{QFC}=60$~\%
based on Refs.~\citep{van_leent_long-distance_2020,geus_low-noise_2024}\footnote{The frequency conversion to a frequency shared with the ions' photons is needed for indistinguishability, see Appendix~\ref{sec:Ion-Photon-state-and}.}.
For the ions' emission, we assume an overall efficiency $\eta_{ion}=10$~\%
including the wavelength conversion \citep{krutyanskiy_multimode_2024}.
For non-photon-number-resolving photodetectors, we take $\eta_{det}=90$~\%.

Note that the repeaters might start with some infidelity w.r.t.
the Bell states involved in state \ref{eq:psi_0_bellrepeater}, which
would result in the heralded fidelities being only bounded from above by
the values given in \ref{eq:fidelity_ion_ion_pure} and \ref{eq:fidelity_mixedion}.

\clearpage
\newpage

\section{Modelling a cavity-assisted AFC quantum memory}

\label{sec:Modelling-a-cavity-asisted}

In this Appendix, we build upon previous derivations from the literature
to provide an extended presentation of a model of an AFC quantum memory
at the single photon level. The conventions chosen for the different
parameters as compared with other articles, and suggested experimental
values extracted from the literature will be provided in Appendix
\ref{subsec:Experimental-parameters}.

\subsection{Inhomogeneous ensemble of $\Lambda$-systems}

We first set out the physical model of a cavity-assisted AFC quantum
memory, which consists of a collection of doping ions embedded in
a crystal structure within an optical cavity \citep{afzelius_impedance-matched_2010,moiseev_efficient_2010}.
We model it as an inhomogeneously broadened ensemble of $\Lambda$-systems
interacting with a single cavity mode which is coupled to input and
output fields in the environment. We build the equations for quantum
operators involved in the description of the full $\Lambda$-structure,
and then obtain a set of scalar equations. 

\subsubsection{Set of equations for quantum operators}

The system of interest comprises a cavity, an assembly of $N$ atoms,
a collection of modes of the environment that will collect input and
output light, and the rest of the environment, made of different baths.\\

\paragraph{Hamiltonian part of the equations}

The Hamiltonian Hilbert space is initially defined as the tensor product
of all Hilbert spaces introduced to describe individual components
of the system (without the environment in a first place), as

\begin{eqnarray}
\mathcal{H}_{mem} & = & \mathcal{H}_{cav}\otimes\bigotimes_{j=1}^{N}\mathcal{H}_{j}.\label{eq:HilbertSpace}
\end{eqnarray}

\noindent To represent the interactions between all the components,
we follow Refs.~\citep{gorshkov_photon_2007-1,gorshkov_photon_2007-2}
and introduce the simplified Hamiltonian

\begin{eqnarray}
\hat{H} & = & \hat{H}_{cav}+\hat{H}_{at}+\hat{H}_{int}+\hat{H}_{trans},\label{eq:Hamiltonian_ref}
\end{eqnarray}

\noindent with

\begin{eqnarray*}
\hat{H}_{cav} & = & \hbar\omega_{c}\hat{a}^{\dagger}\hat{a},\\
\hat{H}_{at} & = & \hbar\sum_{k}\left(\Delta_{k}\hat{\sigma}_{ee}^{(k)}-\omega_{s}\hat{\sigma}_{ss}^{(k)}\right),\\
\hat{H}_{int} & = & -\hbar g\sum_{k}\left(\hat{a}\hat{\sigma}_{eg}^{(k)}+\hat{a}^{\dagger}\hat{\sigma}_{ge}^{(k)}\right),\\
\hat{H}_{trans} & = & -\frac{\hbar}{2}\sum_{k}\left(\Omega(t)e^{-i\omega_{r}t}\hat{\sigma}_{es}^{(k)}+\Omega^{*}(t)e^{+i\omega_{r}t}\hat{\sigma}_{se}^{(k)}\right).
\end{eqnarray*}
In this Hamiltonian, $\hat{H}_{cav}$ and $\hat{H}_{at}$ describe
the free evolution of a single mode of the cavity field and the atoms,
$\hat{H}_{int}$ models the interaction between the cavity mode and
the atoms in a two-level approximation, and $\hat{H}_{trans}$ is
the semi-classical Rabi Hamiltonian for the driving of the third levels
by a strong classical pulse. 

\begin{figure}
\begin{centering}
\includegraphics[width=0.8\columnwidth]{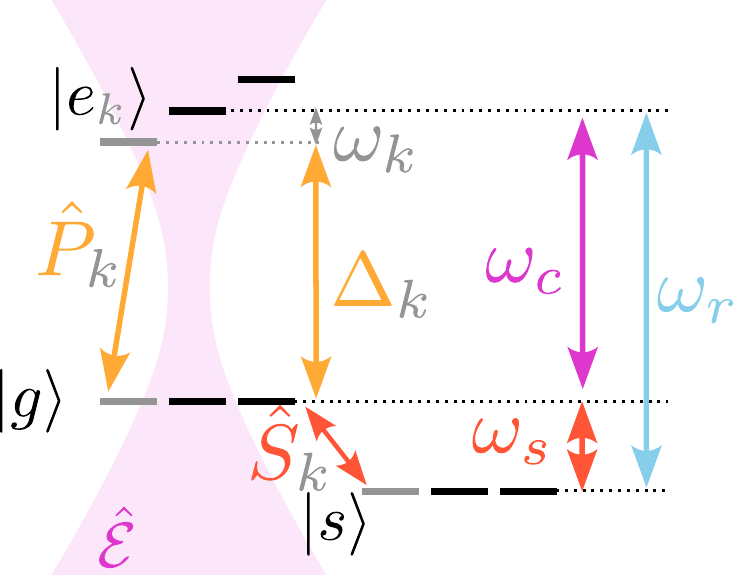}
\par\end{centering}
\caption{Level structure and notations in the rotating frame.\label{fig:Level-structure.} }
\end{figure}

\noindent Furthermore,
\begin{itemize}
\item $\hat{a}$ is the annihilation operator for the mode (frequency $\omega_{c}$)
of interest of the electric field inside the cavity at the position
where the atoms are considered, namely $\hat{\boldsymbol{E}}(t)=\boldsymbol{\epsilon}\sqrt{\frac{\hbar\omega_{c}}{2\varepsilon_{0}V}}\left[\hat{a}(t)+\hat{a}^{\dagger}(t)\right]$.
Recall that, in the free field evolution only, $\hat{a}(t)=e^{-i\omega_{c}t}\hat{a}(0)$
due to Hamiltonian evolution under $\hbar\omega_{c}\hat{a}^{\dagger}\hat{a}$:
the exponential term then gives the oscillations of the field in the
Heisenberg picture. The polarisation unit vector $\boldsymbol{\epsilon}$
is chosen so as to correctly address the atomic $|g\rangle\rightarrow|e\rangle$
transition under angular momentum conservation. $V$ is the quantisation
volume for the cavity mode, typically 
\begin{equation}
V=AL_{cav}\label{eq:quantisation_volume}
\end{equation}
 with $A$ the mode cross section and $L_{cav}$ the length of the
cavity\footnote{In some references, the crystal length would be taken instead, as
discussed in Sec.~\ref{eq:g2N_cav}.}. The operator $\hat{a}$ abides by the commutation relation 
\begin{equation}
[\hat{a}(0),\hat{a}^{\dagger}(0)]=\mathds{I},\label{eq:a(0)_commutation}
\end{equation}
which is conserved for any unitary evolution. 
\item $k$ indices in summations refer to different atomic frequency classes.
In each class, $N_{k}$ atoms share the same $|g\rangle\rightarrow|e\rangle$
transition frequency $\Delta_{k}$. It is assumed that all classes
share the same $|g\rangle\rightarrow|s\rangle$ transition frequency
$\omega_{s}$, which does not depend on $k$. For every class, a collective
coherence operator is introduced as the sum of atomic coherences:
\begin{itemize}
\item $\hat{\sigma}_{eg}^{(k)}:=\sum_{j=1}^{N_{k}}|e\rangle_{jj}\langle g|=\left(\hat{\sigma}_{ge}^{(k)}\right)^{\dagger}$
is the internal state operator for the $|g\rangle\rightarrow|e\rangle$
optical transition of atomic class $k$;
\item $\hat{\sigma}_{sg}^{(k)}:=\sum_{j=1}^{N_{k}}|s\rangle_{jj}\langle g|=\left(\hat{\sigma}_{gs}^{(k)}\right)^{\dagger}$
is the internal state operator for the $|g\rangle\rightarrow|s\rangle$
spin transition of atomic class $k$;
\item More generally, $\hat{\sigma}_{\alpha\beta}^{(k)}:=\sum_{j=1}^{N_{k}}|\alpha\rangle_{jj}\langle\beta|=\left(\hat{\sigma}_{\beta\alpha}^{(k)}\right)^{\dagger}$
refers to the internal state operator for the $|\alpha\rangle\rightarrow|\beta\rangle$
level transition of atomic class $k$. When $\alpha=\beta$, $\hat{\sigma}_{\alpha\alpha}$
refers to the population of energy level $\alpha$ so that for all
$k$ $\hat{\sigma}_{gg}^{(k)}+\hat{\sigma}_{ee}^{(k)}+\hat{\sigma}_{ss}^{(k)}=N_{k}\mathds{1}_{\mathcal{H}_{k}}$.
In addition, one can check that the collective operators keep similar
commutation relations, namely $\left[\hat{\sigma}_{\alpha\beta}^{(k)}(0),\hat{\sigma}_{\mu\nu}^{(j)}(0)\right]=\delta_{kj}\left(\delta_{\beta\mu}\hat{\sigma}_{\alpha\nu}^{(k)}(0)-\delta_{\nu\alpha}\hat{\sigma}_{\mu\beta}^{(k)}(0)\right)$,
which are conserved throughout any unitary evolution ($\delta_{\alpha\beta}$
are Kronecker deltas here);
\item The total number of atoms in the system is $N:=\sum_{k}N_{k}$, and
for every $k$ we define 
\begin{equation}
n_{k}:=\frac{N_{k}}{N}\label{eq:def_pk}
\end{equation}
 the density of atoms with frequency class $k$ ($\sum_{k}n_{k}=1$);
\item Typically, $N$ is much bigger than the number of excitations in the
system, namely $N\gg1$, so that one can introduce the low population
hypothesis (see \ref{enu:Low-excitation-number} below).
\end{itemize}
\item $g$ is the coupling constant between the cavity mode field and the
atoms for the $|g\rangle\rightarrow|e\rangle$ optical transition,
assumed to be homogeneous accross classes. Working with the dipole
approximation, we set $\hat{\boldsymbol{d}}_{j}=\langle g|_{j}\hat{\boldsymbol{d}}_{j}|e\rangle_{j}|g\rangle_{jj}\langle e|+\langle e|_{j}\hat{\boldsymbol{d}}_{j}|g\rangle_{j}|e\rangle_{jj}\langle g|=\langle g|_{j}\hat{\boldsymbol{d}}_{j}|e\rangle_{j}\left[|g\rangle_{jj}\langle e|+\left(|g\rangle_{jj}\langle e|\right)^{\dagger}\right]$
for the dipole moment vector operator of atom $j$, with a convenient
phase so that the conjugated matrix elements appear real\footnote{See Sec.~5.1.1 of Ref.~\citep{steck_quantum_nodate-1}.}.
Then the dipole-field interaction for the $|g\rangle\rightarrow|e\rangle$
transition is given by $-\hat{\boldsymbol{d}}_{j}\cdot\hat{\boldsymbol{E}}=-\langle g|_{j}\hat{\boldsymbol{d}}_{j}\cdot\boldsymbol{\epsilon}|e\rangle_{j}\sqrt{\frac{\hbar\omega_{c}}{2\varepsilon_{0}V}}\left[|g\rangle_{jj}\langle e|+\left(|g\rangle_{jj}\langle e|\right)^{\dagger}\right]\left[\hat{a}+\hat{a}^{\dagger}\right]$.
With the rotating-wave approximation (RWA), terms involving $|g\rangle_{jj}\langle e|\hat{a}$
and $\left(|g\rangle_{jj}\langle e|\right)^{\dagger}\hat{a}^{\dagger}$
are discarded. We obtain $-\hat{\boldsymbol{d}}_{j}\cdot\hat{\boldsymbol{E}}\approx-\langle g|_{j}\hat{\boldsymbol{d}}_{j}\cdot\boldsymbol{\epsilon}|e\rangle_{j}\sqrt{\frac{\hbar\omega_{c}}{2\varepsilon_{0}V}}\left[|g\rangle_{jj}\langle e|\hat{a}^{\dagger}+\left(|g\rangle_{jj}\langle e|\right)^{\dagger}\hat{a}\right]=:-\hbar g\left[|g\rangle_{jj}\langle e|\hat{a}^{\dagger}+\left(|g\rangle_{jj}\langle e|\right)^{\dagger}\hat{a}\right]$
where we take (approximately for any $j$)
\begin{equation}
g:=\langle g|_{j}\hat{\boldsymbol{d}}_{j}\cdot\boldsymbol{\epsilon}|e\rangle_{j}\sqrt{\frac{\omega_{c}}{2\hbar\varepsilon_{0}V}}.\label{eq:g_cav_def}
\end{equation}
\item $\Omega$ is the Rabi frequency of the driving field for the $|s\rangle\rightarrow|e\rangle$
transition. For this, we introduce the classical driving field of
frequency $\omega_{r}$, $\boldsymbol{E}_{r}(t)=\boldsymbol{\epsilon}_{r}\frac{\left[E_{r}(t)e^{-i\omega_{r}t}+E_{r}^{*}(t)e^{+i\omega_{r}t}\right]}{2}$
, where $E_{r}(t)$ gives the slowly-varying modulation shape of the
driving pulse (typically chosen real and rectangular-shaped in the following)
and $\boldsymbol{\epsilon}_{r}$ is the polarisation unit vector.
With the rotating-wave and dipole approximations, we introduce the
dipole moment vector $\hat{\boldsymbol{b}}_{j}=\langle s|_{j}\hat{\boldsymbol{b}}_{j}|e\rangle_{j}|s\rangle_{jj}\langle e|+\langle e|_{j}\hat{\boldsymbol{b}}_{j}|s\rangle_{j}|e\rangle_{jj}\langle s|=\langle s|_{j}\hat{\boldsymbol{b}}_{j}|e\rangle_{j}\left[|s\rangle_{jj}\langle e|+\left(|s\rangle_{jj}\langle e|\right)^{\dagger}\right]$,
and the dipole-field interation for the $|s\rangle\rightarrow|e\rangle$
transition is given by $-\hat{\boldsymbol{b}}_{j}\cdot\boldsymbol{E}_{r}\approx-\frac{\langle s|_{j}\hat{\boldsymbol{b}}_{j}\cdot\boldsymbol{\epsilon}_{r}|e\rangle_{j}}{2}\left[E_{r}^{*}(t)e^{+i\omega_{r}t}|s\rangle_{jj}\langle e|+h.c.\right]=:-\frac{\hbar}{2}\left[\Omega^{*}(t)e^{+i\omega_{r}t}|s\rangle_{jj}\langle e|+\Omega(t)e^{-i\omega_{r}t}\left(|s\rangle_{jj}\langle e|\right)^{\dagger}\right]$
with for any $j$
\begin{equation}
\Omega(t)=\frac{1}{\hbar}\langle g|_{j}\hat{\boldsymbol{b}}_{j}\cdot\boldsymbol{\epsilon}_{r}|s\rangle_{j}E_{r}(t).\label{eq:Rabi_frequency}
\end{equation}
Contrary to Ref.~\citep{gorshkov_photon_2007}, we do not include
any factor $2$ into the definition of $\Omega$ so that a factor
$\frac{1}{2}$ will appear in our equations.
\item Finally, our notations do not distinguish between operators acting
on a single Hilbert space or on $\mathcal{H}_{mem}$. For instance,
$\hat{a}(0)$ should act on $\mathcal{H}_{cav}$ only but we keep
the same notation to refer to the operator $\hat{a}(0)\otimes\bigotimes_{j=1}^{N}\mathds{I}_{j}$
that acts on $\mathcal{H}_{mem}$, with identity operators represented
by $\mathds{I}$.
\end{itemize}
As illustrated in Fig.~\ref{fig:Level-structure.}, where some of
the notations are sketched out, we set

\begin{equation}
\omega_{s}+\omega_{c}=\omega_{r}.\label{eq:frequency_sum}
\end{equation}

Conventions, units and experimental values for all relevant parameters
are provided in Appendix~\ref{subsec:Experimental-parameters}. What
is more, we should recap the hypotheses that were used: 
\begin{enumerate}[label=HP~\arabic{enumi}]
\item  \emph{Dipole approximation and} \emph{Rotating-wave approximation
(RWA) } for the interaction between the fields and the atoms\label{enu:-Dipole-approximation}
\item \emph{Homogeneous coupling $g$} for all the atoms\label{enu:Homogeneous-coupling-}
\item \emph{Only one well defined cavity light mode couples to only one
atomic transition\label{enu:Only-one-well-defined-mode}}
\end{enumerate}
\noindent Let us further assume that
\begin{enumerate}[resume, label=HP~\arabic{enumi}]
\item \emph{Low excitation number (or weak excitation limit) }Almost all
atoms stay in the ground state at all times, so that, for each atom
class $k$, one takes the approximations $\hat{\sigma}_{gg}^{(k)}\approx N_{k}\mathds{I}$
and $\hat{\sigma}_{ee}^{(k)}\approx\mathds{O}$, $\hat{\sigma}_{ss}^{(k)}\approx\mathds{O}$,
$\hat{\sigma}_{es}^{(k)}\approx\mathds{O}$ (or more precisely $\frac{1}{N_{k}}\hat{\sigma}_{ee}^{(k)}=\mathcal{O}\left(\frac{1}{N_{k}}\right)$); this is in the end relevant to compute expectation values by discarding higher order terms.
Such an approximation typically assumes that each atomic frequency class
is sufficiently populated compared to the amount of excitations involved
in the system: the approximation does not refer to a single atom operator,
but to the frequency class collective operator which is the sum of
single-atom operators over the class. It is sufficient to assume that
the number of excitations is much smaller than $N$ if no class is
much more likely to interact with light than some others. In the situation
we are interested in, the number of excitations will be $1$, and
a similar approximation can be made exact regardless of the value of $N$ for the purpose of following calculations\footnote{See paragraph after Eqs.~6 of Ref.~\citep{kollath-bonig_fast_2024} as compared to the discarded terms in our equations \ref{eq:EqEvo_P} to \ref{commSHeff}.}. A consequence is that the dynamical
system for the whole set of operators $\hat{\sigma}_{\alpha,\beta}^{(k)},\quad\alpha,\beta\in\left\{ g,e,s\right\} $
given by Hamiltonian evolution is reduced to a set of equations involving
$\hat{\mathcal{E}}$ and $\hat{P}_{k},\hat{S}_{k}$ coherences only,
where the latter collective operators will behave as bosonic operators\footnote{This is sometimes called the Hollstein-Primakoff approximation in
this context.}, as we will see below.\label{enu:Low-excitation-number}
\end{enumerate}
In order to switch to a rotating frame, we introduce the following
slowly-varying operators in the Heisenberg picture:
\begin{itemize}
\item $\hat{\mathcal{E}}(t)=e^{+i\omega_{c}t}\hat{a}(t)$ for the cavity
field;
\item $\hat{P}_{k}(t)=\frac{1}{\sqrt{N_{k}}}e^{+i\omega_{c}t}\hat{\sigma}_{ge}^{(k)}(t)$
for the atomic polarisations;
\item $\hat{S}_{k}(t)=\frac{1}{\sqrt{N_{k}}}e^{+i(\omega_{c}-\omega_{r})t}\hat{\sigma}_{gs}^{(k)}(t)$
for the spin polarisations;
\item $\hat{C}_{k}(t)=\frac{1}{\sqrt{N_{k}}}e^{+i\omega_{r}t}\hat{\sigma}_{se}^{(k)}(t)\approx\mathds{O}$
with \ref{enu:Low-excitation-number}.
\end{itemize}
Those unitary transformations and hypotheses ensure that the following
commutation relations hold:

\begin{equation}
\left[\hat{\mathcal{E}}(t),\hat{\mathcal{E}}(t)^{\dagger}\right]=\mathds{I},\label{eq:E_boson}
\end{equation}

\noindent and for all $k,j$

\begin{align}
\left[\hat{P}_{k}(t),\hat{P}_{j}(t)^{\dagger}\right] & =\frac{1}{\sqrt{N_{k}N_{j}}}\left[\sum_{l=1}^{N_{k}}|g\rangle_{ll}\langle e|,\sum_{m\text{=}1}^{N_{j}}|e\rangle_{mm}\langle g|\right]\nonumber \\
 & =\frac{1}{N_{k}}\delta_{kj}\left(\hat{\sigma}_{gg}^{(k)}(t)-\hat{\sigma}_{ee}^{(k)}(t)\right)\nonumber \\
 & \approx\delta_{kj}\mathds{I},\label{eq:P_HP}
\end{align}

\noindent as well as

\begin{equation}
\left[\hat{S}_{k}(t),\hat{S}_{j}(t)^{\dagger}\right]\approx\delta_{kj}\mathds{I},\label{eq:S_HP}
\end{equation}

\noindent where the two later results make use of the low excitation
number hypothesis \ref{enu:Low-excitation-number}\footnote{See for instance Eq.~(2.52) of Ref.~\citep{hird_engineering_2021}}.
Furthermore, for every $k$, \ref{enu:Low-excitation-number} leads to
the approximations

\begin{eqnarray*}
\left[\hat{P}_{k},\hat{S}_{k}^{\dagger}\right] & = & \frac{1}{N_{k}}e^{+i\omega_{c}t}e^{-i(\omega_{c}-\omega_{r})t}\sum_{j,l}\left[|g\rangle_{jj}\langle e|,|s\rangle_{ll}\langle g|\right]\\
 & = & -\frac{1}{N_{k}}e^{+i\omega_{r}t}\sum_{j}|s\rangle_{jj}\langle e|=-\frac{1}{\sqrt{N_{k}}}\hat{C}_{k}\approx\mathds{O},
\end{eqnarray*}

\begin{eqnarray*}
\left[\hat{P}_{k},\hat{P}_{k}^{\dagger}\hat{S}_{k}\right] & = & \left[\hat{P}_{k},\hat{P}_{k}^{\dagger}\right]\hat{S}_{k}+\hat{P}_{k}^{\dagger}\left[\hat{P}_{k},\hat{S}_{k}\right]\approx\hat{S}_{k}+\mathds{O},
\end{eqnarray*}

\begin{eqnarray*}
\left[\hat{S}_{k},\hat{S}_{k}^{\dagger}\hat{P}_{k}\right] & = & \left[\hat{S}_{k},\hat{S}_{k}^{\dagger}\right]\hat{P}_{k}+\hat{S}_{k}^{\dagger}\left[\hat{S}_{k},\hat{P}_{k}\right]\approx\hat{P}_{k}+\mathds{O},
\end{eqnarray*}

\begin{align*}
\left[\hat{P}_{k},\hat{S}_{k}^{\dagger}\hat{P}_{k}\right] & =\left[\hat{P}_{k},\hat{S}_{k}^{\dagger}\right]\hat{P}_{k}+\hat{S}_{k}^{\dagger}\left[\hat{P}_{k},\hat{P}_{k}\right]\\
 & =-\frac{1}{\sqrt{N_{k}}}\hat{C}_{k}\hat{P}_{k}\approx\mathds{O},
\end{align*}

\begin{eqnarray*}
\left[\hat{S}_{k},\hat{P}_{k}^{\dagger}\hat{S}_{k}\right] & = & \left[\hat{S}_{k},\hat{P}_{k}^{\dagger}\right]\hat{S}_{k}+\hat{P}_{k}^{\dagger}\left[\hat{S}_{k},\hat{S}_{k}\right]\\
 & = & -\frac{1}{\sqrt{N_{k}}}\hat{C}_{k}^{\dagger}\hat{S}_{k}+\mathds{O}\approx\mathds{O},
\end{eqnarray*}

\noindent and

\begin{eqnarray*}
\left[\hat{S}_{k},\hat{P}_{k}^{\dagger}\hat{P}_{k}\right] & = & \left[\hat{S}_{k},\hat{P}_{k}^{\dagger}\right]\hat{P}_{k}+\hat{P}_{k}^{\dagger}\left[\hat{S}_{k},\hat{P}_{k}\right]\approx\mathds{O}.
\end{eqnarray*}

\noindent The definitions for the slowly varying operators amount
to switching to a rotating frame. In this new frame, we obtain an effective
Hamiltonian

\begin{widetext}

\begin{equation}
\hat{H}_{eff}=\hbar\sum_{k}\omega_{k}\hat{P}_{k}^{\dagger}\hat{P}_{k}-\hbar g\sqrt{N}\sum_{k}\sqrt{n_{k}}\left(\hat{\mathcal{E}}\hat{P}_{k}^{\dagger}+\hat{\mathcal{E}}^{\dagger}\hat{P}_{k}\right)-\frac{\hbar}{2}\sum_{k}\left(\Omega(t)\hat{P}_{k}^{\dagger}\hat{S}_{k}+\Omega^{*}(t)\hat{S}_{k}^{\dagger}\hat{P}_{k}\right),\label{eq:effective_Hamiltonian}
\end{equation}

\noindent where we set $\omega_{k}:=\Delta_{k}-\omega_{c}$ for the
detuning of class $k$ $|e\rangle-$level w.r.t. the cavity field frequency
$\omega_{c}$. Indeed, we show below the dynamics resulting from Heisenberg
equation $i\hbar\frac{\mathrm{d}\hat{A}}{\mathrm{d}t}=i\hbar\frac{\partial\hat{A}}{\mathrm{\partial}t}+\left[\hat{A},\hat{H}\right]$
are also given by setting $i\hbar\frac{\mathrm{d}\hat{A}}{\mathrm{d}t}:=\left[\hat{A},\hat{H}_{eff}\right]$
for any operator $\hat{A}=\hat{\mathcal{E}},\hat{P}_{k},\hat{S}_{k}$.
On the one hand, we have

\begin{align}
i\hbar\frac{\mathrm{d}\hat{\mathcal{E}}}{\mathrm{d}t}(t) & =i\hbar\left(+i\omega_{c}\right)\hat{\mathcal{E}}(t)+e^{+i\omega_{c}t}\left[\hat{a},\hat{H}\right]\nonumber \\
 & =-\hbar\omega_{c}\hat{\mathcal{E}}(t)+\hbar\omega_{c}\hat{\mathcal{E}}-e^{+i\omega_{c}t}\hbar g\sum_{k}\hat{\sigma}_{ge}^{(k)}\nonumber \\
 & =-\hbar g\sqrt{N}\sum_{k}\sqrt{n_{k}}\hat{P}_{k},\label{eq:EqEvo_E}
\end{align}

\noindent and for any index $k$, 

\begin{align}
i\hbar\frac{\mathrm{d}\hat{P}_{k}}{\mathrm{d}t}(t) & =i\hbar\left(+i\omega_{c}\right)\hat{P}_{k}(t)+\frac{1}{\sqrt{N_{k}}}e^{+i\omega_{c}t}\left[\hat{\sigma}_{ge}^{(k)},\hat{H}\right]\nonumber \\
 & =-\hbar\omega_{c}\hat{P}_{k}(t)+\hbar\Delta_{k}\frac{1}{\sqrt{N_{k}}}e^{+i\omega_{c}t}\hat{\sigma}_{ge}^{(k)}-\hbar g\frac{1}{\sqrt{N_{k}}}e^{+i\omega_{c}t}\hat{a}\left(\hat{\sigma}_{gg}^{(k)}-\hat{\sigma}_{ee}^{(k)}\right)-\frac{\hbar}{2}\Omega(t)e^{-i\omega_{r}t}\frac{1}{\sqrt{N_{k}}}e^{+i\omega_{c}t}\hat{\sigma}_{gs}^{(k)}\nonumber \\
 & \approx\hbar\omega_{k}\hat{P}_{k}(t)-\hbar g\sqrt{N}\sqrt{n_{k}}\hat{\mathcal{E}}(t)-\frac{\hbar}{2}\Omega(t)\hat{S}_{k}(t)\text{ thanks to \ref{enu:Low-excitation-number}},\label{eq:EqEvo_P}
\end{align}

\begin{align}
i\hbar\frac{\mathrm{d}\hat{S}_{k}}{\mathrm{d}t}(t) & =i\hbar\left(+i(\omega_{c}-\omega_{r})\right)\hat{S}_{k}(t)+\frac{1}{\sqrt{N_{k}}}e^{+i(\omega_{c}-\omega_{r})t}\left[\hat{\sigma}_{gs}^{(k)},\hat{H}\right]\nonumber \\
 & =-\hbar(\omega_{c}-\omega_{r})\hat{S}_{k}(t)-\hbar\omega_{s}\frac{1}{\sqrt{N_{k}}}e^{+i(\omega_{c}-\omega_{r})t}\hat{\sigma}_{gs}^{(k)}+\hbar g\frac{1}{\sqrt{N_{k}}}e^{+i(\omega_{c}-\omega_{r})t}\hat{a}\hat{\sigma}_{es}^{(k)}-\frac{\hbar}{2}\Omega^{*}(t)e^{+i\omega_{r}t}\frac{1}{\sqrt{N_{k}}}e^{+i(\omega_{c}-\omega_{r})t}\hat{\sigma}_{ge}^{(k)}\nonumber \\
 & =-\frac{\hbar}{2}\Omega^{*}(t)\hat{P}_{k}(t)\text{ thanks to Eq.~\ref{eq:frequency_sum} and \ref{enu:Low-excitation-number}.}\label{eq:EqEvo_S}
\end{align}

\noindent On the other hand, we have

\[
\left[\hat{\mathcal{E}},\hat{H}_{eff}\right]=-\hbar g\sqrt{N}\sum_{k}\sqrt{n_{k}}\hat{P}_{k},
\]

\noindent and for every $k$

\begin{equation}
\left[\hat{P}_{k},\hat{H}_{eff}\right]\approx\hbar\omega_{k}\hat{P}_{k}-\hbar g\sqrt{N}\sqrt{n_{k}}\hat{\mathcal{E}}-\frac{\hbar}{2}\Omega(t)\hat{S}_{k}\text{ with \ref{enu:Low-excitation-number}}
\label{commPHeff}
\end{equation}

\begin{equation}
\left[\hat{S}_{k},\hat{H}_{eff}\right]\approx-\frac{\hbar}{2}\Omega^{*}(t)\hat{P}_{k}\text{ with \ref{enu:Low-excitation-number}}
\label{commSHeff},
\end{equation}

\smallskip
\noindent which correspond to Eqs.~\ref{eq:EqEvo_E}, \ref{eq:EqEvo_P},
\ref{eq:EqEvo_S}.
\smallskip

\end{widetext}

Note that the number of total excitations is conserved throughout
evolution. Indeed, the total number of excitations operator

\begin{equation}
\hat{N}_{tot}=\hat{\mathcal{E}}^{\dagger}\hat{\mathcal{E}}+\sum_{k}\hat{P}_{k}^{\dagger}\hat{P}_{k}+\sum_{k}\hat{S}_{k}^{\dagger}\hat{S}_{k}\label{eq:Ntot_excitationnumber}
\end{equation}

\noindent commutes with the effective Hamiltonian, a fortiori thanks
to \ref{enu:Low-excitation-number},

\[
\left[\hat{N}_{tot},\hat{H}_{eff}\right]=\mathds{O}.
\]
\\

\paragraph{Adjunction of Langevin noise operators and input-output relations}

With the aim of including losses, we turn again to Equations \ref{eq:EqEvo_E},
\ref{eq:EqEvo_P} and \ref{eq:EqEvo_S}. To that Hamiltonian evolution,
one should add non-Hermitian behaviours from the losses and input-output
from the cavity, as well as dephasing of atomic polarisations. Related
terms will be directly inserted into the equations of motion below.

Following Refs.~\citep{gorshkov_photon_2007-2,gorshkov_photon_2007},
we introduce different noise operators \citep{lax_quantum_1966,gardiner_quantum_2004}
related to:
\begin{itemize}
\item Input-output relations \citep{caves_quantum_1982,yurke_quantum_1984,collett_squeezing_1984,gardiner_input_1985},
that represent the coupling between the cavity mode and some external
degrees of freedom with rate $\kappa$. This will let us describe
input and output photons involved in the storage and retrieval processes.
Construction of the related $\hat{\mathcal{E}}_{in}(t)$ and $\hat{\mathcal{E}}_{out}(t)$
operators for all times $t$ is discussed in Sec.~\ref{subsec:Quantum-input-output-relations};
\item Possible decay of the $\hat{P}_{k}$ coherences, phenomenologically
decribed by the coupling to some external baths $\hat{F}_{P,k}(t)$
with rate $\gamma_{P}$;
\item Possible decay of the $\hat{S}_{k}$ coherences, phenomenologically
decribed by the coupling to some external baths $\hat{F}_{S,k}(t)$
with rate $\gamma_{P}$.
\end{itemize}
The decay rates are assumed to be the same for all frequency classes.
Other sources of loss are discarded here, such as any further decay
channel of the cavity optical mode. Details for the introduction of
noise operators and quantum reservoirs are provided in Sec.~\ref{sec:Quantum-noise-and-fluctuations}
of these Appendices, see especially Eqs.~\ref{eq:typical_noise_term}
and \ref{eq:input_output_relation}. In particular, some specific
properties and hypotheses are introduced there, such as the Markovian
hypothesis (\ref{enu:Markovian-noise-hypothesis}) for the reservoirs.\\

\paragraph{Equations of motions}

\label{par:Equations-of-motions}

We end up with the following set of equations of motion, including
noise terms,

\begin{widetext}

\begin{alignat}{3}
\dot{\mathcal{\hat{\mathcal{E}}}}(t) & = & -\kappa\hat{\mathcal{E}}(t)+ig\sqrt{N}\sum_{k}\sqrt{n_{k}}\hat{P}_{k}(t)+ &  & +\sqrt{2\kappa}\hat{\mathcal{E}}_{in}(t)\nonumber \\
\dot{\hat{P}}_{k}(t) & = & -(\gamma_{P}+i\omega_{k}))\hat{P_{k}}(t)+ig\sqrt{N}\sqrt{n_{k}}\hat{\mathcal{E}}(t) & +i\frac{\Omega(t)}{2}\hat{S}_{k}(t) & +\sqrt{2\gamma_{P}}\hat{F}_{P,k}(t)\label{eq:memory_equations-1-1}\\
\dot{\hat{S}}_{k}(t) & = & -\gamma_{S}\hat{S}_{k}(t) & +i\frac{\Omega^{*}(t)}{2}\hat{P}_{k}(t) & +\sqrt{2\gamma_{S}}\hat{F}_{S,k}(t)\nonumber \\
\hat{\mathcal{E}}_{out}(t) & = & \sqrt{2\kappa}\hat{\mathcal{E}}(t)-\hat{\mathcal{E}}_{in}(t) &  & .\nonumber 
\end{alignat}

If $\gamma_{P}=\gamma_{S}=0$, a conservation law similar to \ref{eq:Ntot_excitationnumber}
still holds, now including input and output fields:

\begin{equation}
\forall t\quad\frac{\mathrm{d}}{\mathrm{d}t}\left[\hat{\mathcal{E}}^{\dagger}\hat{\mathcal{E}}+\sum_{k}\hat{P}_{k}^{\dagger}\hat{P}_{k}+\sum_{k}\hat{S}_{k}^{\dagger}\hat{S}_{k}+\int_{t}^{+\infty}\hat{\mathcal{E}}_{in}^{\dagger}(t')\hat{\mathcal{E}}_{in}(t')\mathrm{d}t'+\int_{-\infty}^{t}\hat{\mathcal{E}}_{out}^{\dagger}(t')\hat{\mathcal{E}}_{out}(t')\mathrm{d}t'\right](t)=\mathds{O}.\label{eq:conservation_input_output}
\end{equation}

\noindent If we are to follow the noise operators, the total number
operator is given by

\begin{eqnarray}
\hat{N}_{tot} & = & \quad\hat{\mathcal{E}}^{\dagger}\hat{\mathcal{E}}+\sum_{k}\hat{P}_{k}^{\dagger}\hat{P}_{k}+\sum_{k}\hat{S}_{k}^{\dagger}\hat{S}_{k}\label{eq:total_exc_number_full}\\
 &  & +\int_{t}^{+\infty}\hat{\mathcal{E}}_{in}^{\dagger}(t')\hat{\mathcal{E}}_{in}(t')\mathrm{d}t'+\int_{-\infty}^{t}\hat{\mathcal{E}}_{out}^{\dagger}(t')\hat{\mathcal{E}}_{out}(t')\mathrm{d}t'\nonumber \\
 &  & +\hat{N}_{bath}\nonumber 
\end{eqnarray}

\noindent with $\hat{N}_{bath}$ the number operator that counts for
excitations that left to the reservoirs, which we do not explicit
here but such that $\hat{N}_{tot}$ is conserved throughout evolution. 

Equivalently, we can use a continuous description of the dynamical
system 

\begin{alignat}{3}
\dot{\hat{\mathcal{E}}}(t) & = & -\kappa\hat{\mathcal{E}}(t)+igN\int n(\omega)\hat{P}_{\omega}(t)\mathrm{d}\omega &  & +\sqrt{2\kappa}\hat{\mathcal{E}}_{in}(t)\nonumber \\
\dot{\hat{P}}_{\omega}(t) & = & -(\gamma_{P}+i\omega)\hat{P}_{\omega}(t)+ig\hat{\mathcal{E}}(t) & +i\frac{\Omega(t)}{2}\hat{S}_{\omega}(t) & +\sqrt{2\gamma_{P}}\hat{F}_{P,\omega}(t)\label{eq:memory_equations-1}\\
\dot{\hat{S}}_{\omega}(t) & = & -\gamma_{S}\hat{S}_{\omega}(t) & +i\frac{\Omega^{*}(t)}{2}\hat{P}_{\omega}(t) & +\sqrt{2\gamma_{S}}\hat{F}_{S,\omega}(t)\nonumber \\
\hat{\mathcal{E}}_{out}(t) & = & \sqrt{2\kappa}\hat{\mathcal{E}}(t)-\hat{\mathcal{E}}_{in}(t), &  & .\nonumber 
\end{alignat}

\end{widetext}

\noindent where the correspondance between systems \ref{eq:memory_equations-1-1}
and \ref{eq:memory_equations-1} is described through

\[
n(\omega)\mathrm{d}\omega\equiv\sum n_{k}\delta(\omega=\omega_{k})
\]

\noindent so that

\begin{equation}
\int_{\mathbb{R}}n(\omega)\mathrm{d}\omega=\sum_{k}n_{k}=1,\label{eq:Normalisation_n}
\end{equation}

\noindent and by the fact that $\hat{P}_{\omega}$ is associated to
the atomic coherences of the $|g\rangle$-$|e\rangle$ transition
in the frequency range $[\omega,\omega+\mathrm{d}\omega]$

\[
\hat{P}_{k}=\frac{1}{\sqrt{n_{k}N}}e^{+i\omega_{c}t}\sum_{j}^{N_{k}}|g\rangle_{jj}\langle e|\equiv\sqrt{n_{k}N}\hat{P}_{\omega_{k}}\delta(\omega=\omega_{k})
\]

\noindent and similarly for $\hat{S}_{\omega}$. 

\subsubsection{Scalar set of equations}

\label{subsec:Scalar-set-of}

Equations \ref{eq:memory_equations-1-1} and \ref{eq:memory_equations-1}
involve quantum operators acting on the whole Hilbert space \ref{eq:HilbertSpace}.
Further analysis and numerical simulations will instead involve differential
equations for scalar functions. In Sec.\ref{sec:FromOperatorsToScalars},
and in particular in part \ref{par:Particular-case:syst_assumptions},
we consider particular conditions requiring that we start from an initially factorised
state between the system and the environment (\ref{enu:System-Environment-decomposition}
and \ref{enu:Initial-product-state}),
we work within the one-excitation subspace and the memory is loaded
with a pure state (\ref{enu:Pure-state-in-subspace}), noise operators
follow bosonic commutation rules (\ref{enu:Bosonic-operators}),
reservoirs (bar the input-output one) are in the ground state (\ref{enu:Single-photon-input-vacuum-noise}), and the memory is initially empty (\ref{enu:Empty-memory}). Those particular
conditions, together with assumptions made about the noise operators
(\ref{enu:Reservoir-in-the-ground-state}, \ref{enu:Markovian-noise-hypothesis},
\ref{enu:Uncorrelated-reservoir} of Appendix~\ref{sec:Quantum-reservoirs}),
ensure that we can compute all second-order correlators directly from
the integration of some scalar systems of ordinary differential equations,
namely

\begin{eqnarray}
\dot{\mathcal{E}}(t) & = & -\kappa\mathcal{E}(t)+ig\sqrt{N}\sum_{j}\sqrt{p_{j}}P_{j}(t)+\sqrt{2\kappa}\mathcal{E}_{in}(t)\nonumber \\
\dot{P_{k}}(t) & = & -(\gamma_{P}+i\omega_{k})P_{k}(t)+ig\sqrt{N}\sqrt{n_{k}}\mathcal{E}(t)+\frac{i}{2}\Omega(t)S_{k}(t)\nonumber \\
\dot{S_{k}}(t) & = & -\gamma_{S}S_{k}(t)+\frac{i}{2}\Omega^{*}(t)P_{k}(t)\nonumber \\
\mathcal{E}_{out}(t) & = & \sqrt{2\kappa}\mathcal{E}(t)-\mathcal{E}_{in}(t)\label{eq:discrete_scalar_system-2}
\end{eqnarray}

\noindent or

\noindent
\begin{eqnarray}
\dot{\mathcal{E}}(t) & = & -\kappa\mathcal{E}(t)+igN\int n(\omega)P_{\omega}(t)\mathrm{d}\omega+\sqrt{2\kappa}\mathcal{E}_{in}(t)\nonumber \\
\dot{P_{\omega}}(t) & = & -(\gamma_{P}+i\omega)P_{\omega}(t)+ig\mathcal{E}(t)+\frac{i}{2}\Omega(t)S_{\omega}(t)\nonumber \\
\dot{S_{\omega}}(t) & = & -\gamma_{S}S_{\omega}(t)+\frac{i}{2}\Omega^{*}(t)P_{\omega}(t)\nonumber \\
\mathcal{E}_{out}(t) & = & \sqrt{2\kappa}\mathcal{E}(t)-\mathcal{E}_{in}(t).\label{eq:full_scalar_system-1}
\end{eqnarray}

\noindent The former set of equations \ref{eq:discrete_scalar_system-2}
is easier use for the purpose of numerical resolution with a discretised
density (see Sec.~\ref{sec:Numerical-methods}), while the latter
\ref{eq:full_scalar_system-1} is easier to handle analytically as
we will do below. \\

Furthermore, we will argue in Appendix~\ref{subsec:PhotonStateCoherence}
that those second-order correlators are sufficient to characterise
single-photon components that may be emitted by the memory. Within
this picture, single-photon wavepackets enter and exit the quantum
memory, and $\mathcal{E}_{in}(t)$ represents the input single-photon
envelope while $\mathcal{E}_{out}(t)$ represents the output-single
photon envelope. For instance, we will start from the pure state 
\begin{equation}
|\psi_{input}\rangle=\int\mathcal{E}_{in}(t)\hat{\mathcal{E}}_{in}^{\dagger}(t)|0\rangle\mathrm{d}t\label{eq:psi_input}
\end{equation}

\noindent or equivalently

\begin{equation}
|\psi_{input}\rangle=\hat{\mathcal{E}}_{input}^{\dagger}(t)|0\rangle,\label{eq:psi_input_wavepacket}
\end{equation}

\noindent where $\hat{\mathcal{E}}_{input}^{\dagger}=\int\mathcal{E}_{in}(t)\hat{\mathcal{E}}_{in}^{\dagger}(t)\mathrm{d}t$
is the creation operator for a photon wavepacket in the mode with
temporal shape $\mathcal{E}_{in}$ and $|0\rangle$ is the vacuum
state shared by every mode in this linear setting \citep{fabre_modes_2020}.

Though our work focuses on the single-photon regime, we stress that
similar sets of equations are obtained if one takes a coherent state
as an input for the quantum memory, provided \ref{enu:Low-excitation-number}
still holds. In that situation $\mathcal{E}_{in/out}(t)$ would represent
the envelopes of input/output coherent states (see part \ref{par:Coherent-state-input}).\\

We add initial conditions (say at time $t=t_{0}$) corresponding to
an incoming single-photon state towards an empty cavity and an empty
memory,

\begin{equation}
\forall k\quad\begin{cases}
\mathcal{E}(t_{0}) & =0\\
P_{k}(t_{0}) & =0\\
S_{k}(t_{0}) & =0
\end{cases}\label{eq:initial_conditions-1}
\end{equation}

\noindent with $\mathcal{E}_{in}(t)$ representing the waveform envelope
of this incoming single-photon, with $\mathcal{E}_{in}(t<t_{0})=0$.
In particular, the conservation equation \ref{eq:conservation_input_output}
becomes

\begin{widetext}

\begin{equation}
\forall t\geq0\quad\left|\mathcal{E}(t)\right|^{2}+\sum_{k}\left|P_{k}(t)\right|^{2}+\sum_{k}\left|S_{k}(t)\right|^{2}+\int_{t_{0}}^{t}\left|\mathcal{E}_{out}(t)\right|^{2}\mathrm{d}t+\int_{t}^{+\infty}\left|\mathcal{E}_{in}(t)\right|^{2}\mathrm{d}t=1,\label{eq:conservation_scalar}
\end{equation}

\end{widetext}

\noindent where we assumed that $\gamma_{P}=\gamma_{S}=0$ and that
the input wavepacket is normalised 

\begin{equation}
\int_{t_{0}}^{+\infty}\left|\mathcal{E}_{in}(t)\right|^{2}\mathrm{d}t=\int_{-\infty}^{+\infty}\left|\mathcal{E}_{in}(t)\right|^{2}\mathrm{d}t=1.\label{eq:Ein_normalised}
\end{equation}

\subsection{Two-level structure and echoes}

We now turn to the analytical resolution of system \ref{eq:full_scalar_system-1}
in the context where $n(\omega)$ is chosen to represent an atomic
frequency comb (AFC). At first, we only focus on the inhomogeneous
$|g\rangle$-$|e\rangle$ two-level structure to introduce the notions
of AFC echoes and of impedance-matching regime. Building upon the
existing literature, we provide more general formulas to describe
such phenomenons. The full $\Lambda$-structure, that enables on-demand
readout from the memory, will be discussed in Sec.~\ref{subsec:Third-level-and}.

\subsubsection{Comb structure of the atomic density}

\begin{figure}
\begin{centering}
\includegraphics[width=1\columnwidth]{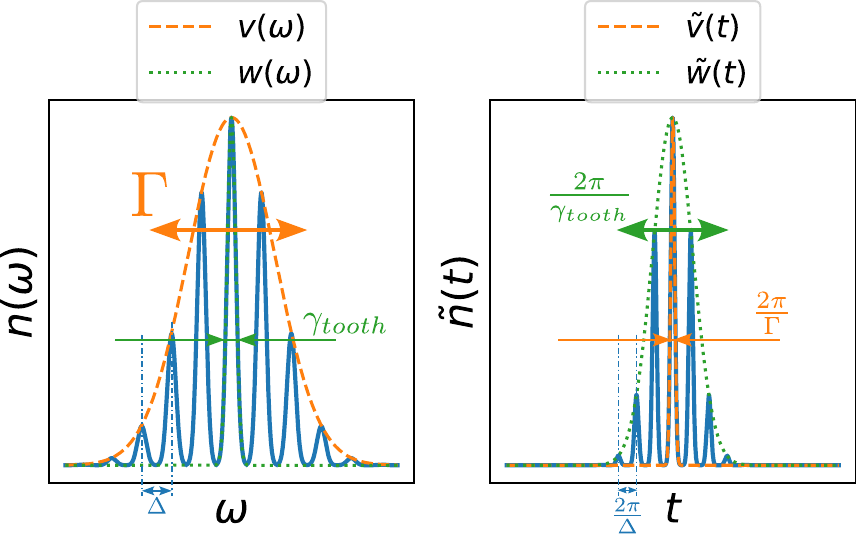}
\par\end{centering}
\caption{Comb structure $n(\omega)$ and its Fourier transform $\tilde{n}(t)$,
with the characteristic widths involved.\label{fig:Comb-structure}}
\end{figure}

We start with the description of the normalised atomic density $n(\omega)$
(see Eq.~\ref{eq:Normalisation_n}). In order to represent an AFC
structure \citep{de_riedmatten_solid-state_2008,afzelius_multimode_2009,afzelius_impedance-matched_2010}
(see Fig.~\ref{fig:Comb-structure}), we take $n(\omega)$ as a series
of well-separated peaks $w(\omega)$ (typical width $\gamma_{tooth}$)
and enveloped by a function $v(\omega)$ (typical width and variation
scale $\Gamma$), that is

\begin{align}
n(\omega) & =Z\times v(\omega)\times\sum_{n=-\infty}^{+\infty}w(\omega-n\Delta)\label{eq:n_distr_comb}\\
 & =Z\times\left(v\times\left[\mathrm{III}_{\Delta}*w\right]\right)(\omega).\nonumber 
\end{align}

\noindent A convolution product is noted $*$, $\mathrm{III}(\omega)$
stands for the infinite Dirac comb of step $\Delta$ i.e. $\mathrm{III}(\omega)=\sum_{n=-\infty}^{+\infty}\delta(\omega=n\Delta)$,
and $Z$ is a normalisation constant such that Eq.~\ref{eq:Normalisation_n}
holds. Typically, $v$ can be associated to the inhomogeneously broadened
distribution of the atomic ensemble \citep{gorshkov_photon_2007,afzelius_multimode_2009,afzelius_proposal_2013},
while $w$ gives the shape of the comb teeth that are carved within
the inhomogeneous distribution with particular optical pumping techniques
such as spectral hole burning \citep{macfarlane_coherent_1987,maniloff_power_1995,nilsson_hole-burning_2004}. 

If we assume that $\Gamma\gg\Delta>\gamma_{tooth}$ for a well-defined
comb so that the variations of $v$ are slow compared to those of
$w$, we can write

\begin{align*}
v(\omega)\times\sum_{n=-\infty}^{+\infty}w(\omega-n\Delta) & \approx\sum_{n=-\infty}^{+\infty}v(n\Delta)w(\omega-n\Delta)\\
 & \approx\frac{1}{\Delta}\int_{-\infty}^{+\infty}v(x)w(\omega-x)\mathrm{d}x,
\end{align*}

\noindent so that

\begin{align*}
\int_{-\infty}^{+\infty}n(\omega)\mathrm{d}\omega & \approx\frac{Z}{\Delta}\int_{-\infty}^{+\infty}\int_{-\infty}^{+\infty}v(x)w(\omega-x)\mathrm{d}x\mathrm{d}\omega\\
 & =\frac{Z}{\Delta}\int_{-\infty}^{+\infty}w(\omega')\mathrm{d}\omega'\int_{-\infty}^{+\infty}v(x)\mathrm{d}x.
\end{align*}

\noindent Hence, taking
\begin{equation}
Z=\frac{\Delta}{\int w(\omega)\mathrm{d}\omega\times\int v(\omega)\mathrm{d}\omega}\label{eq:normalisation_constant}
\end{equation}

\noindent leads to $\int_{-\infty}^{+\infty}n(\omega)\mathrm{d}\omega\approx1$.
To characterise the comb, we may define the comb finesse \citep{afzelius_multimode_2009}
as the ratio of the distance between comb teeth over the width of
comb teeth,

\begin{equation}
F_{AFC}:=\frac{\Delta}{\mathrm{FWHM}_{teeth}}.\label{eq:comb_finesse}
\end{equation}

\noindent It will also be useful to manipulate the Fourier transform\footnote{We will consistently use the following conventions for Fourier transforms:
$\begin{cases}
\mathrm{TF}(F)(t) & =\int_{-\infty}^{+\infty}e^{-i\omega t}F(\omega)\mathrm{d}\omega\\
\mathrm{TF}^{-1}(f)(\omega) & =\frac{1}{2\pi}\int_{-\infty}^{+\infty}e^{+i\omega t}f(t)\mathrm{d}t
\end{cases}$.} $\tilde{n}(t)$ of $n(\omega)$, which involves the Fourier transforms
$\tilde{v}(t)$ and $\tilde{w}(t)$ of $v(\omega)$ and $w(\omega)$.
Convolution representation of $n$ leads to 

\begin{eqnarray}
\tilde{n}(t) & \propto & \left(\tilde{v}*\left[\mathrm{III}_{\frac{2\pi}{\Delta}}\times\tilde{w}\right]\right)(t)\nonumber \\
 & \propto & \int_{-\infty}^{+\infty}\tilde{v}(t-t')\times\left(\sum_{n=-\infty}^{+\infty}\delta(t'=n\frac{2\pi}{\Delta})\right)\times\tilde{w}(t')\mathrm{d}t'\nonumber \\
 & \propto & \sum_{n=-\infty}^{+\infty}\int_{-\infty}^{+\infty}\tilde{v}(t-t')\delta(t'=n\frac{2\pi}{\Delta})\tilde{w}(t')\mathrm{d}t'\nonumber \\
 & \propto & \sum_{n=-\infty}^{+\infty}\tilde{v}(t-n\frac{2\pi}{\Delta})\tilde{w}(n\frac{2\pi}{\Delta})\nonumber \\
 & \approx: & Y\times\tilde{w}(t)\times\sum_{n=-\infty}^{+\infty}\tilde{v}(t-n\frac{2\pi}{\Delta}),\label{eq:ntilde_app}
\end{eqnarray}

\noindent where the last approximation can be made if $\tilde{v}$
is peaked w.r.t. characteristic variations of $\tilde{w}$, that is
still ensured by the assumption $\Gamma\gg\Delta>\gamma_{tooth}$.
Thus, $\tilde{n}(t)$ also has a comb structure, with step $\frac{2\pi}{\Delta}$
(see Fig.~\ref{fig:Comb-structure}). The normalisation constant $Y$
is computed thanks to some Fourier transform identities,

\begin{equation}
Y=Z\times\underset{\times\text{ to }*}{\underbrace{\frac{1}{2\pi}}}\times\underset{\text{Comb}}{\underbrace{\frac{2\pi}{\Delta}}}=\frac{Z}{\Delta}.\label{eq:normalisation_FT}
\end{equation}

\noindent The factor $1/2\pi$ comes from the Fourier transform of
the convolution product, while the $2\pi/\Delta$ factor stems from
the Fourier transform of the Dirac comb.

The area of the central tooth of the comb Fourier transform is defined
as

\begin{align}
D_{comb}&:=Y\tilde{w}(0)\int\tilde{v}(t)\mathrm{d}t=\frac{\tilde{w}(t=0)\int\tilde{v}(t)\mathrm{d}t}{\int w(\omega)\mathrm{d}\omega\times\int v(\omega)\mathrm{d}\omega}\nonumber\\
&=\frac{\int\tilde{v}(t)\mathrm{d}t}{\int v(\omega)\mathrm{d}\omega}.\label{eq:D_comb}
\end{align}

\noindent We used the fact for any $w$

\begin{equation}
\int_{-\infty}^{+\infty}w(\omega)\mathrm{d}w=\tilde{w}(t=0),\label{eq:FT_mean}
\end{equation}

\noindent so that $D_{comb}$ value is independent of $w$.

\noindent Similarly,

\begin{equation}
\int_{-\infty}^{+\infty}\tilde{w}(t)\mathrm{d}t=2\pi\times w(\omega=0).\label{eq:FT_mean_-1}
\end{equation}

\noindent In particular, around teeth at $t=k\frac{2\pi}{\Delta}$,
we get the peak approximation

\begin{eqnarray}
\tilde{n}(t\approx k\frac{2\pi}{\Delta}) & \approx & \frac{\tilde{w}(k\frac{2\pi}{\Delta})}{\tilde{w}(0)}\frac{\tilde{w}(0)}{\int w(\omega)\mathrm{d}\omega\times\int v(\omega)\mathrm{d}\omega}\tilde{v}(t)\nonumber\\
 & \approx & \frac{\tilde{w}(k\frac{2\pi}{\Delta})}{\tilde{w}(0)}D_{comb}\delta_{0}(t)\label{eq:peak_approximation}
\end{eqnarray}

\noindent where $\delta_{0}$ is a normalised distribution--like
Dirac peak approximation similar to the one used in Wigner-Weisskopf
theory \citep{lukin_modern_2016}. In particular, a $\frac{1}{2}$
coefficient appears when one computes a cropped overlap with a slowly
varying function $h$, 
\begin{equation}
\int_{0}^{+\infty}\delta_{0}(\omega)\times h(\omega)\mathrm{d}\omega=\frac{1}{2}h(0).\label{eq:cropped_Dirac}
\end{equation}

\begin{figure}
\begin{centering}
\includegraphics[width=1\columnwidth]{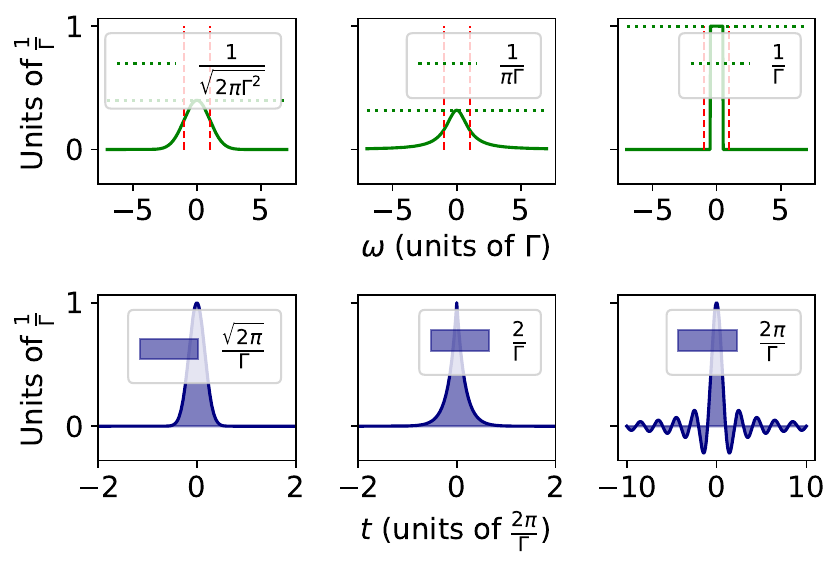}
\par\end{centering}
\caption{(Top) Comb L1-normalised envelopes, from left to right $u_{Gaussian}(\omega)=\frac{1}{\sqrt{2\pi\Gamma^{2}}}e^{-\frac{\omega{{}^2}}{2\Gamma^{2}}}$,
$u_{Lorentzian}(\omega)=\frac{1}{\pi\Gamma}\frac{1}{1+\frac{\omega^{2}}{\Gamma^{2}}}$
and $u_{Rect}(\omega)=\frac{1}{\Gamma}\mathds{1}_{\omega\in[-\frac{\Gamma}{2};+\frac{\Gamma}{2}]}$.
(Bottom) Fourier transforms of previous functions, respectively $\tilde{u}_{Gaussian}(t)=e^{-\frac{\Gamma^{2}t^{2}}{2}}$,
$\tilde{u}_{Lorentzian}(t)=e^{-\Gamma|t|}$ and $\tilde{u}_{Rect}(t)=\mathrm{sinc}\left(\frac{\Gamma t}{2}\right)$.
Integrals of the Fourier transforms are provided in the plot legends,
while vertical dashed red lines of the top figures indicate $\pm\Gamma$.
\label{fig:comb_shapes}}
\end{figure}

\noindent We plot in Fig.~\ref{fig:comb_shapes} some particular shapes
for the comb distribution, and provide in Table~\ref{tab:comb_distributions}
their properties so as to properly define the characteristic widths
$\Gamma$ and $\gamma_{tooth}$. 

\begin{table*}
\centering
\begin{tabularx}{\textwidth}{|>{\centering\arraybackslash}p{2cm}|>{\centering\arraybackslash}X|>{\centering\arraybackslash}X|>{\centering\arraybackslash}X|>{\centering\arraybackslash}X|>{\centering\arraybackslash}X|>{\centering\arraybackslash}X|>{\centering\arraybackslash}X|}
\hline
Envelope type & Frequency description $u(\omega)$ (characteristic width $\gamma$) & FWHM of $u$ & Normalisation $\int{u}(\omega)\mathrm{d}\omega$ & Fourier transform $\tilde{u}(t)$ & Fourier transform normalisation $\int\tilde{u}(t)\mathrm{d}t$ & $D_{comb}=\frac{\int\tilde{u}(t)\mathrm{d}t}{\int u(\omega)\mathrm{d}\omega}$ for $\gamma=\Gamma$ & $\mathcal{C}_{opt}:=\frac{2}{\Gamma D_{comb}}$ \\
\hline
\hline
Dirac peak & $\delta(\omega=0)$ & Undefined & $1$ & $1$ & Undefined, $2\pi$ by convention & Undefined, $2\pi$ & Undefined, $\frac{1}{\pi\Gamma}$ \\
\hline
Rectangular & $I_{\omega\in[-\frac{\gamma}{2};+\frac{\gamma}{2}]}$ & $\gamma$ & $\gamma$ & $\gamma\mathrm{sinc}\left(\frac{\gamma t}{2}\right)$ & $2\pi$ & $\frac{2\pi}{\Gamma}$ & $\frac{1}{\pi}$ \\
\hline
Gaussian & $e^{-\frac{\omega^{2}}{2\gamma^{2}}}$ & $2\sqrt{2\ln2}\gamma$ & $\sqrt{2\pi}\gamma$ & $\sqrt{2\pi}\gamma e^{-\frac{\gamma^{2}t^{2}}{2}}$ & $2\pi$ & $\frac{\sqrt{2\pi}}{\Gamma}$ & $\sqrt{\frac{2}{\pi}}$ \\
\hline
Lorentzian & $\frac{1}{1+\frac{\omega^{2}}{\gamma^{2}}}$ & $2\gamma$ & $\pi\gamma$ & $\pi\gamma e^{-\gamma|t|}$ & $2\pi$ & $\frac{2}{\Gamma}$ & $1$ \\
\hline
\end{tabularx}

\caption{Collection of functions $\omega\protect\mapsto u(\omega)$ that are
used to define comb or comb teeth envelopes ($v$ and $w$), see also
Fig.~\ref{fig:comb_shapes}. Depending on the context, the characteristic
width $\gamma\protect\geq0$ may be understood as $\Gamma$ or $\gamma_{tooth}$.
If one does not want to consider distributions centered around $\omega=0$
but rather shifted versions, it is usefull to recall that $\mathrm{TF}(\omega\protect\mapsto f(\omega-a))(t)=e^{-iat}\times\mathrm{TF}(f)(t)$.
Last column always features a $2\pi$ thanks to Eq.~\ref{eq:FT_mean_-1}.
Another consequence is that $D_{comb}$ value (Eq.~\ref{eq:D_comb})
only depends on the inhomogeneous envelope $v$, and so does $\mathcal{C}_{opt}$
(Eq.~\ref{eq:definition_Copt}). \label{tab:comb_distributions}}
\end{table*}

Finally, note that a comb structure also appears when one introduces
a discretised version of an inhomogeneous distribution for numerical
resolution, with $w(\omega)=\delta(\omega)$. This situation that
involves perfectly narrow teeth corresponds to the set of equations
\ref{eq:discrete_scalar_system-2}.

\subsubsection{Absorption and impedance-matching regime}

\label{subsec:Absorption-and-impedance-matchin}

We now look at the consequences the shape $n(\omega)$ has on the
dynamics of system \ref{eq:full_scalar_system-1}.\\

\paragraph{Integration}

We start by solving the set of equations to account for absorption
of an input pulse by the memory, building upon Refs.~\citep{afzelius_impedance-matched_2010,afzelius_multimode_2009,afzelius_proposal_2013,sangouard_analysis_2007,chaneliere_quantum_2018}. 

We consider a two-level restriction of Eq.~\ref{eq:full_scalar_system-1},
and set $\gamma_{P}=\gamma_{S}=0$, so that

\begin{alignat}{2}
\dot{\mathcal{E}}(t) & = & -\kappa\mathcal{E}(t)+igN\int n(\omega)P_{\omega}(t)\mathrm{d}\omega+\sqrt{2\kappa}\mathcal{E}_{in}(t)\nonumber \\
\dot{P_{\omega}}(t) & = & -i\omega P_{\omega}(t)+ig\mathcal{E}(t)\label{eq:2lvl_scalar_system-1}\\
\mathcal{E}_{out}(t) & = & \sqrt{2\kappa}\mathcal{E}(t)-\mathcal{E}_{in}(t).\nonumber 
\end{alignat}

Formal integration of $P_{\omega}$ with zero initial conditions \ref{eq:initial_conditions-1}
at time $t=t_{0}=0$ yields

\begin{equation}
P_{\omega}(t)=ig\int_{t_{0}}^{t}e^{-i\omega(t-t')}\mathcal{E}(t')\mathrm{d}t',\label{eq:polarisation_abs}
\end{equation}

\noindent so that for times around $t=t_{0}$ the electric field is given by

\begin{widetext}

\begin{align}
\dot{\mathcal{E}}(t\approx t_{0}) & =-\kappa\mathcal{E}(t)-g^{2}N\int n(\omega)\int_{t_{0}}^{t}e^{-i\omega(t-t')}\mathcal{E}(t')\mathrm{d}t'\mathrm{d}\omega+\sqrt{2\kappa}\mathcal{E}_{in}(t)\nonumber \\
 & =-\kappa\mathcal{E}(t)-g^{2}N\int_{t_{0}}^{t}\int e^{-i\omega(t-t')}n(\omega)\mathrm{d}\omega\mathcal{E}(t')\mathrm{d}t'+\sqrt{2\kappa}\mathcal{E}_{in}(t)\label{eq:system_abs_gen}\\
 & =-\kappa\mathcal{E}(t)-g^{2}N\int_{t_{0}}^{t}\underset{\approx:D_{\mathrm{comb}}\delta_{0}(t-t')}{\underbrace{\tilde{n}(t-t')}}\mathcal{E}(t')\mathrm{d}t'+\sqrt{2\kappa}\mathcal{E}_{in}(t)\label{eq:system_abs_gen-1}\\
 & \approx-\kappa\mathcal{E}(t)-g^{2}N\frac{D_{\mathrm{comb}}}{2}\mathcal{E}(t)+\sqrt{2\kappa}\mathcal{E}_{in}(t)\text{ }\label{eq:system_absorption}
\end{align}

\end{widetext}

\noindent where for the last equations we used relations \ref{eq:peak_approximation}
and \ref{eq:cropped_Dirac}, and the fact that if $t$ is close enough
to $t_{0}$ only the central peak of $\tilde{n}(t)$ is involved.
If $\mathcal{E}$ varies slowly enough we can set $\dot{\mathcal{E}}(t)\approx0$
\citep{lugiato_adiabatic_1984,kollath-bonig_fast_2024,afzelius_proposal_2013}.
Then from Eq.~\ref{eq:system_absorption},

\begin{align}
\mathcal{E}(t\approx t_{0}) & \approx\frac{\sqrt{2\kappa}}{\kappa+g^{2}N\frac{D_{\mathrm{comb}}}{2}}\mathcal{E}_{in}(t)\nonumber \\
 & =\frac{\sqrt{\frac{2}{\kappa}}}{1+\mathcal{C}\frac{D_{\mathrm{comb}}\Gamma}{2}}\mathcal{E}_{in}(t)\label{eq:E_approx_eq}\\
 & =:\frac{\sqrt{\frac{2}{\kappa}}}{1+\frac{\mathcal{C}}{\mathcal{C}_{opt}}}\mathcal{E}_{in}(t)\label{eq:E_approx_Copt}
\end{align}

\noindent where we define the memory cooperativity

\begin{equation}
\mathcal{C}:=\frac{g^{2}N}{\kappa\Gamma},\label{eq:definition_cooperativity}
\end{equation}

\noindent and an optimal cooperativity value

\begin{equation}
\mathcal{C}_{opt}:=\frac{2}{\Gamma D_{comb}}\label{eq:definition_Copt}
\end{equation}

\noindent which only depends on the inhomogeneous envelope shape $v$
(cf. Eq.~\ref{eq:D_comb} and Table \ref{tab:comb_distributions}).

Combining Eq.~\ref{eq:E_approx_Copt} with the input-output relation
yields

\begin{align}
\mathcal{E}_{out}(t) & =\mathcal{E}_{in}(t)\times\left(\frac{2}{1+\frac{\mathcal{C}}{\mathcal{C}_{opt}}}-1\right)\nonumber \\
 & =\mathcal{E}_{in}(t)\times\frac{1-\frac{\mathcal{C}}{\mathcal{C}_{opt}}}{1+\frac{\mathcal{C}}{\mathcal{C}_{opt}}}.\label{eq:output_matching}
\end{align}

\paragraph{Impedance-matching regime}

The situation where 
\begin{equation}
\mathcal{C}=\mathcal{C}_{opt}\label{eq:impedance_matching}
\end{equation}
 is called the impedance-matching condition \citep{afzelius_impedance-matched_2010,moiseev_efficient_2010,afzelius_proposal_2013},
and is characterised by $\mathcal{E}_{out}=0$: no light is reflected
back, and equivalently, perfect mapping of the initial input pulse
onto atomic collective excitations is achieved. In particular, we
can define the absorption efficiency of the memory

\begin{align}
\eta_{abs} & =1-\frac{\int_{t_{0}}^{t_{0}+\Theta}\left|\mathcal{E}_{out}(t)\right|^{2}\mathrm{d}t}{\int_{-\infty}^{+\infty}\left|\mathcal{E}_{in}(t)\right|^{2}\mathrm{d}t}\nonumber \\
 & =\frac{4\frac{\mathcal{C}}{\mathcal{C}_{opt}}}{\left(1+\frac{\mathcal{C}}{\mathcal{C}_{opt}}\right)^{2}}\label{eq:eta_abs}
\end{align}

\noindent where $[t_{0},t_{0}+\Theta]$ encompasses the support of
$t\mapsto\mathcal{E}_{in}(t)$, with $\Theta<\frac{2\pi}{\Delta}$.
That is, we considered a photon with a long enough frequency bandwidth
w.r.t. $\Delta$, with $\mathcal{E}_{in}(t)=0$ when $t<t_{0}$ and
$t>t_{0}+\Theta$ (as specified in Eq.~\ref{eq:Ein_normalised}, $\int_{-\infty}^{+\infty}\left|\mathcal{E}_{in}(t)\right|^{2}\mathrm{d}t=1$).
As such, $\eta_{abs}=1$ in the impedance matching condition. 

\paragraph{Fourier sampling behaviour}

\label{par:Fourier-sampling-behaviour}

Right after absorption, for $t\gtrsim t_{0}+\Theta$, Eqs.~\ref{eq:polarisation_abs}
and \ref{eq:E_approx_Copt} indicate that the collective polarisation
for frequency range $[\omega,\omega+\mathrm{d}\omega]$ is given by 

\begin{align}
P_{\omega}(t\gtrsim t_{0}+\Theta) & =\frac{ig\sqrt{\frac{2}{\kappa}}}{1+\frac{\mathcal{C}}{\mathcal{C}_{opt}}}e^{-i\omega t}\int_{-\infty}^{t}e^{+i\omega t'}\mathcal{E}_{in}(t')\mathrm{d}t'\nonumber \\
 & =\frac{i\sqrt{2\mathcal{C}}\sqrt{\frac{\Gamma}{N}}}{1+\frac{\mathcal{C}}{\mathcal{C}_{opt}}}e^{-i\omega t}\int_{-\infty}^{+\infty}e^{+i\omega t'}\mathcal{E}_{in}(t')\mathrm{d}t'\nonumber \\
 & =\frac{i2\pi\sqrt{2\mathcal{C}}\sqrt{\frac{\Gamma}{N}}}{1+\frac{\mathcal{C}}{\mathcal{C}_{opt}}}e^{-i\omega t}\mathrm{TF^{-1}}(\mathcal{E}_{in})(\omega)\nonumber \\
 & =\frac{\sqrt{\frac{\mathcal{C}}{\mathcal{C}_{opt}}}}{1+\frac{\mathcal{C}}{\mathcal{C}_{opt}}}i2\pi\sqrt{2\mathcal{C}_{opt}}\sqrt{\frac{\Gamma}{N}}e^{-i\omega t}\mathrm{TF^{-1}}(\mathcal{E}_{in})(\omega)\label{eq:polarisation_prediction}
\end{align}

\noindent i.e.,

\[
P_{\omega,opt}(t\gtrsim t_{0}+\Theta)=i\pi\sqrt{2\mathcal{C}_{opt}}\sqrt{\frac{\Gamma}{N}}e^{-i\omega t}\mathrm{TF^{-1}}(\mathcal{E}_{in})(\omega)
\]

\noindent for the optimal regime where $\ensuremath{\mathcal{C}=\mathcal{C}_{opt}}$.
That is, for a time $t\gtrsim t_{0}+\Theta$ after complete arrival
of the wavepacket $\mathcal{E}_{in}$ and before any echo emission
(see Sec.~\ref{subsec:Echoes}), we get that each frequency class
absorbs energy proportionally to the Fourier coefficient of the envelope
profile in the atomic frequency basis. In case of mismatched cooperativity
tuning, one gets a detrimental global factor that affects equally
all atoms, $\frac{1}{2}\sqrt{\eta_{abs}}$. In particular, for a comb
with a rectangular $v$ envelope and $\delta$-peaked teeth $w$,
this amounts to computing the Discrete Fourier Transform (DFT) with
respect to the atomic frequencies $\omega_{k}=-\frac{\Gamma}{2}+k\frac{\Gamma}{N-1}=-\frac{\Gamma}{2}+k\Delta$,
because $\mathcal{C}_{opt}=\frac{1}{\pi}$,

\begin{eqnarray}
P_{\omega_{k},opt,rect}(t) & = & i\sqrt{2\pi\Delta}\sqrt{\frac{N-1}{N}}\mathrm{TF^{-1}}(\mathcal{E}_{in})(\omega_{k})e^{-i\omega_{k}t}\nonumber \\
 & \underset{N\gg1}{\sim} & i\sqrt{2\pi\Delta}\mathrm{TF^{-1}}(\mathcal{E}_{in})(\omega_{k})e^{-i\omega_{k}t}.\label{eq:polarisation_FourierTransform}
\end{eqnarray}

\noindent The result from Eq.~\ref{eq:polarisation_FourierTransform}
is also mentioned in Ref.~\citep{chaneliere_quantum_2018} \footnote{See part II.A.4.}.
It is consistent with the following requirements:
\begin{itemize}
\item $\sum_{k}\left|P_{\omega_{k}}\right|^{2}=1$ for perfect absorbtion
of a single excitation;
\item $\Delta\sum\left|\mathrm{TF^{-1}}(\mathcal{E}_{in})(\omega_{k})\right|^{2}=\frac{1}{2\pi}$
by Parseval--Plancherel identity for the DFT with frequency step
$\Delta$.
\end{itemize}
\noindent We can thus provide an interpretation of absorption by the
impedance-matched inhomogeneous ensemble as some perfect signal sampling
procedure.

In the Schrödinger picture, another way to look at it is to say that
the absorption of the incoming photon leads to the creation of a collective
excitation delocalised in every atom. The Dicke--like state could
write as the collective atomic state\footnote{See for instance Refs.~\citep{dicke_coherence_1954}, \citep{duan_long-distance_2001},
\citep{afzelius_multimode_2009} Eq.~(1), or \citep{gardiner_quantum_2015}
p. 436.} 

\[
\sum_{j=1}^{N}c_{j}e^{-i\omega_{j}t}|g_{1}...g_{j-1}e_{j}g_{j+1}...g_{N}\rangle
\]

\noindent where $c_{j}$ is proportional to the Fourier coefficient
of the photon spectrum in the atomic DFT basis.

After absorption, each frequency class starts dephasing differently
from the others ($e^{-i\omega_{k}t}$ term). The resulting collective
polarisation is then suppressed so that no light gets out of the memory,
until rephasing occurs due to the regular comb structure.

\subsubsection{Echoes}

\label{subsec:Echoes}

We now build upon Refs.~\citep{afzelius_impedance-matched_2010,afzelius_multimode_2009,sangouard_analysis_2007,chaneliere_quantum_2018}
to recall that an AFC structure will feature unprompted echoes after
absorption of an input pulse \citep{dubetskii_echos_1985}.\\

\paragraph{First echo}

Let's focus on the first rephasing (around $t=t_{0}+\frac{2\pi}{\Delta}$).
To compute the field, we have to change the initial condition for
polarisation. For some time $t_{0}+\Theta\lesssim t_{abs}<t_{0}+\frac{2\pi}{\Delta}$
right after absorption, we know from Eq.~\ref{eq:polarisation_prediction}
that

\begin{widetext}

\[
P_{\omega}(t_{abs})=\frac{ig\sqrt{\frac{2}{\kappa}}}{1+\frac{\mathcal{C}}{\mathcal{C}_{opt}}}e^{-i\omega t_{abs}}\int_{-\infty}^{t_{abs}}e^{+i\omega t'}\mathcal{E}_{in}(t')\mathrm{d}t'.
\]

\noindent Taking this value as an initial condition to integrate Eq.
\ref{eq:2lvl_scalar_system-1} gives at $t\geq t_{abs}$

\begin{align*}
P_{\omega}(t) & =ig\int_{t_{abs}}^{t}e^{-i\omega(t-t')}\mathcal{E}(t')\mathrm{d}t'+e^{-i\omega(t-t_{abs})}P_{\omega}(t_{abs})\\
 & =ig\int_{t_{abs}}^{t}e^{-i\omega(t-t')}\mathcal{E}(t')\mathrm{d}t'+ig\frac{\sqrt{\frac{2}{\kappa}}}{1+\frac{\mathcal{C}}{\mathcal{C}_{opt}}}\int_{t_{0}}^{t_{abs}}e^{-i\omega(t-t')}\mathcal{E}_{in}(t')\mathrm{d}t'.
\end{align*}

\noindent Thus, as $\mathcal{E}_{in}(t\geq t_{abs})=0$, Eq.~\ref{eq:2lvl_scalar_system-1}
for $t\geq t_{abs}$ is given by

\begin{align*}
\dot{\mathcal{E}}(t) & =-\kappa\mathcal{E}(t)+igN\int n(\omega)P_{\omega}(t)\mathrm{d}\omega\\
 & =-\kappa\mathcal{E}(t)-g^{2}N\int_{t_{abs}}^{t}\int n(\omega)e^{-i\omega(t-t')}\mathrm{d}\omega\mathcal{E}(t')\mathrm{d}t'-g^{2}N\frac{\sqrt{\frac{2}{\kappa}}}{1+\frac{\mathcal{C}}{\mathcal{C}_{opt}}}\int_{t_{0}}^{t_{abs}}\int n(\omega)e^{-i\omega(t-t')}\mathrm{d}\omega\mathcal{E}_{in}(t')\mathrm{d}t'\\
 & =-\kappa\mathcal{E}(t)-g^{2}N\int_{t_{abs}}^{t}\tilde{n}(t-t')\mathcal{E}(t')\mathrm{d}t'-g^{2}N\frac{\sqrt{\frac{2}{\kappa}}}{1+\frac{\mathcal{C}}{\mathcal{C}_{opt}}}\int_{t_{0}}^{t_{abs}}\tilde{n}(t-t')\mathcal{E}_{in}(t')\mathrm{d}t'.
\end{align*}

As explained in Ref.~\citep{afzelius_multimode_2009}, the first
integral corresponds to reabsorption of the field while the second
one describes the source term, which will lead to the retrieved echo.
We now expand $\tilde{n}$ around the relevant peaks: around $t-t'=0$
in the first integral, and around $t-t'=\frac{2\pi}{\Delta}$ in the
second (see Eq.~\ref{eq:peak_approximation}). We obtain

\[
\dot{\mathcal{E}}(t)\approx-\kappa\mathcal{E}(t)-g^{2}N\frac{D_{\mathrm{comb}}}{2}\mathcal{E}(t)-g^{2}N\frac{\sqrt{\frac{2}{\kappa}}}{1+\frac{\mathcal{C}}{\mathcal{C}_{opt}}}\frac{\tilde{w}(\frac{2\pi}{\Delta})}{\tilde{w}(0)}\times D_{\mathrm{comb}}\mathcal{E}_{in}(t-\frac{2\pi}{\Delta})
\]

\noindent Neglecting again $\dot{\mathcal{E}}$ and using the input-output
relation with zero input (after some time the input pulse is over, i.e. $\mathcal{E}_{in}(t)=0$) leads to the first echo

\begin{align}
\mathcal{E}_{out}(t\approx t_{0}+\frac{2\pi}{\Delta}) & =\sqrt{2\kappa}\mathcal{E}(t\approx t_{0}+\frac{2\pi}{\Delta})\nonumber \\
 & \approx-\sqrt{2\kappa}\times\frac{1}{\kappa+g^{2}N\frac{D_{\mathrm{comb}}}{2}}\times g^{2}N\frac{\sqrt{\frac{2}{\kappa}}}{1+\frac{\mathcal{C}}{\mathcal{C}_{opt}}}\frac{\tilde{w}(\frac{2\pi}{\Delta})}{\tilde{w}(0)}\times D_{\mathrm{comb}}\times\mathcal{E}_{in}(t-\frac{2\pi}{\Delta})\nonumber \\
 & =-\frac{\tilde{w}(\frac{2\pi}{\Delta})}{\tilde{w}(0)}\times2\times\frac{g^{2}ND_{\mathrm{comb}}}{\kappa+g^{2}N\frac{D_{\mathrm{comb}}}{2}}\times\frac{1}{1+\frac{\mathcal{C}}{\mathcal{C}_{opt}}}\times\mathcal{E}_{in}(t-\frac{2\pi}{\Delta})\nonumber \\
 & =-\frac{\tilde{w}(\frac{2\pi}{\Delta})}{\tilde{w}(0)}\times\frac{4\frac{\mathcal{C}}{\mathcal{C}_{opt}}}{\left(1+\frac{\mathcal{C}}{\mathcal{C}_{opt}}\right)^{2}}\times\mathcal{E}_{in}(t-\frac{2\pi}{\Delta}).\label{eq:1st-echo-sm}
\end{align}

\end{widetext}

This last equation is consistent with Eq.~(10) of Ref.~\citep{afzelius_impedance-matched_2010}.
Note that a $\pi$-phase is included in the reemission of this first
echo. What's more, the first term of Eq.~\ref{eq:1st-echo-sm} involves
the shape of the comb teeth and corresponds to the losses due to non-recurrent
atomic dephasing, so we introduce

\begin{equation}
\eta_{F}:=\left(\frac{\tilde{w}(\frac{2\pi}{\Delta})}{\tilde{w}(0)}\right)^{2}.\label{eq:teeth_dephasing}
\end{equation}
Of course, for infinitely narrow teeth ($\tilde{w}:t\mapsto1$) $\eta_{F,\text{perfect}}=1$.
On the contrary, for Gaussian teeth ($\tilde{w}:t\mapsto\sqrt{2\pi}\gamma_{tooth}e^{-\frac{\gamma_{tooth}^{2}t^{2}}{2}}$)
$\eta_{F,\text{Gaussian}}=e^{-\gamma_{tooth}^{2}\frac{4\pi^{2}}{\Delta^{2}}}=e^{-\frac{4\pi^{2}}{8\ln(2)}\frac{1}{F_{AFC,\mathrm{Gaussian}}^2}}\approx e^{-\frac{7}{F_{AFC,\mathrm{Gaussian}}^2}}$
which is Eq.~(A21) of Ref.~\citep{afzelius_multimode_2009} and $F_{AFC}$
is the comb finesse (Eq.~\ref{eq:comb_finesse}). For rectangular
teeth ($\tilde{w}:t\mapsto\gamma\mathrm{sinc}(\frac{\gamma_{tooth}t}{2})$)
$\eta_{F,\text{Rect}}=\mathrm{sinc}^{2}\left(\frac{\gamma_{tooth}t}{2}\frac{2\pi}{\Delta}\right)=\mathrm{sinc}^{2}\left(\frac{\pi}{F_{c,\text{Rect}}}\right)$
which corresponds to the last term of Eq.~(2) in Ref.~\citep{jobez_towards_2016}.
Note that the teeth shape can be optimised to maximise the efficiencies
given a particular absorption, see Refs.~\citep{bonarota_efficiency_2010,zang_provable_2024}.\\

The second term of Eq.~\ref{eq:1st-echo-sm} involves the impedance matching
condition and we recognise $\eta_{abs}(\mathcal{C})$ (Eq.~\ref{eq:eta_abs})
so we obtain the efficiency

\begin{eqnarray}
\eta_{\text{1st-echo}} & = & \frac{\int_{t_{0}+2\pi/\Delta}^{t_{0}+2\pi/\Delta+\Theta}\left|\mathcal{E}_{out}(t)\right|^{2}\mathrm{d}t}{\int_{-\infty}^{+\infty}\left|\mathcal{E}_{in}(t)\right|^{2}\mathrm{d}t}\label{eq:1st_echo_eff}\\
 & = & \eta_{F}\times\eta_{abs}{{}^2}=\left(\frac{\tilde{w}(\frac{2\pi}{\Delta})}{\tilde{w}(0)}\right)^{2}\times\frac{16\left(\frac{\mathcal{C}}{\mathcal{C}_{opt}}\right)^{2}}{\left(1+\frac{\mathcal{C}}{\mathcal{C}_{opt}}\right)^{4}}.\nonumber 
\end{eqnarray}

\noindent In the case of perfect teeth and impedance matching, we
then get unit efficiency in the first echo: the input signal
is perfectly reemitted, up to a $\pi$-phase. This observation leads
Refs.~\citep{bonarota_efficiency_2010,bonarota_atomic_2012} to interpret
the AFC echo as a purely dispersive effect: when the peaks of the
AFC are taken infinitely narrow, the power spectrum is almost unchanged
(since $n(\omega)$ is zero almost everywhere), and the delay for
reemission appears a slow-light effect due to phases acquired along
the way.\\

\paragraph{Further echoes}

In case the input signal is not perfectly reemitted in the first echo,
some energy remains stored within the collective polarisations. This
energy is likely to be reemitted later on, when further echoes happen,
as it can be shown by recursively solving the equations around times
$k\frac{2\pi}{\Delta}$.

\subsubsection{Polarisation decay}

So far, we did not take into account the decay channel linked to $\gamma_{P}$.
In the case where $\gamma_{P}\neq0$ is small enough compared to the
other dynamical parameters\footnote{Such an approximation is for instance discussed, in the case of $\gamma_{P}$
big enough, with Fourier transform arguments in Ref.~\citep{lukin_modern_2016},
footnote~2 p.~154.}, the influence of polarisation decay can readily\textcolor{red}{{}
}be included by introducting an exponential decrease of the rephasing
efficiency\footnote{\noindent See Ref.~\citep{afzelius_proposal_2013} p. 9}
so that

\begin{equation}
\eta_{tot}(\mathcal{C})\approx\eta_{1st-echo}(\mathcal{C})\times e^{-2\gamma_{P}\frac{2\pi}{\Delta}},\label{eq:polarisation_decay}
\end{equation}

\noindent where $\frac{2\pi}{\Delta}$ is the time that the excitation
spends in the $P$ coherences.

\subsubsection{Bandwidth}

\label{subsec:Bandwidth}

The absorption bandwidth of the cavity-enhanced memory can be estimated
from the shape of the inhomogeneous distribution $v(\omega)$ and
the parameters $\kappa$, $\Gamma$, and $g\sqrt{N}$. As long as
the first AFC echo is not involved (short enough input), the AFC comb
structure does not matter. In this regard, Ref.~\citep{afzelius_proposal_2013}
provides exact results in the Lorentzian case with $\gamma_{P}=0$,
for which the bandwidth\footnote{Full resonance width at half maximum}
is found to be $4\kappa$ when $\kappa<g\sqrt{N}<\Gamma$, and $4\Gamma$
when $\kappa>g\sqrt{N}>\Gamma$ (weak coupling regime). Actually,
one can check that the same orders of magnitude hold for Gaussian
and rectangular envelopes (up to a similar but different factor than
$4$). As noted in Ref.~\citep{afzelius_proposal_2013}, taking the
input photon within the memory bandwidth readily ensures that the
assumptions $\tilde{n}(t-t')\approx D_{comb}\delta_{0}(t-t')$ and
$\dot{\mathcal{E}}\approx0$ made in Sec.~\ref{subsec:Absorption-and-impedance-matchin}\footnote{See right before and after Eq.~\ref{eq:system_absorption}.}
are met. For our study, relevant experimental parameters (see Appendix~\ref{subsec:Experimental-parameters})
correspond to the second situation, so we will seek to take small compared to $\Gamma$.
This is compatible with $\Delta<\delta\omega_{in}$ provided that
the AFC comb has a large enough number of teeth.

\subsubsection{Efficient memory}

To sum up, optimised absorption of the input pulse requires:
\begin{itemize}
\item The impedance-matching condition \ref{eq:impedance_matching};
\item $\frac{2\pi}{\delta\omega_{in}}<\frac{2\pi}{\Delta}$: the pulse
is shorter than the echo time. In the sampling picture, this means
that the sampling rate should be high enough to represent the shape
of the pulse (sort of Shannon-Nyquist theorem);
\item That the photon bandwidth shall be smaller than memory bandwidth ($\kappa$
or $\Gamma$ see above), to avoid distortions.
\end{itemize}
With the impedance-matching condition, the echo is emitted with high
efficiency if the teeth of the comb are narrow enough. If the bandwidth
of the input is too large, distortions appear.

\subsection{Third level and on-demand storage}

\label{subsec:Third-level-and}

Now, the full $\Lambda$-system structure of the atoms, including
the atomic coherences $S_{\omega}$, is considered (see Fig.~\ref{fig:Level-structure.}). Recall that in Eqs.~\ref{eq:discrete_scalar_system-2}
and \ref{eq:full_scalar_system-1}, $\Omega(t)$ represents the time-dependent
Rabi frequency for the transfer between $P$ and $S$
coherences.  As $P$ coherences are the only ones that couple to
the cavity mode and then to the output field, transferring to $S$
coherences enables long storage times.

First, we observe that if $\Omega(t)=0$, the equation for $S_{\omega}$
in system \ref{eq:full_scalar_system-1}

\[
\dot{S}_{\omega}(t)=-\gamma_{S}S_{\omega}(t)
\]

\noindent is integrated as

\begin{equation}
S_{\omega}(t)=e^{-\gamma_{S}t}S_{\omega}(0).\label{eq:polarS_decay}
\end{equation}

\noindent Consequently, this evolution does not depend on $\omega$:
no inhomogeneous phase is accumulated. If $\gamma_{S}=0$, no loss
occurs and time appears frozen: that is, the coherence will live as
long as no transfer is done. Still, we point out that this result
is a consequence of the assumption that $|s\rangle$ levels do not
exhibit any inhomogeneous broadening w.r.t. the $|g\rangle$ level.
In practice, this does not happen and dynamical decoupling methods
can be introduced to compensate for dephasing due to inhomogeneous
broadening \citep{souza_robust_2012,duranti_towards_2023}.

We can take advantage of the time-freezing property of the $S$
levels to introduce on-demand storage and readout for the quantum
memory. To this aim, transfers between $S$ and $P$ are
studied.

\subsubsection{Transfer to and from spin-levels}

\paragraph{Ideal case of closed oscillations}

First, we forget about the coupling between $P$ coherences and the
field $\mathcal{E}$, and again set $\gamma_{P}=\gamma_{S}=0$. As
a consequence, textbook results about Rabi oscillations in the semi-classical
regime apply \citep{shore_manipulating_2011,steck_quantum_nodate-1}.
Considering the situation where the driving $\Omega(t)$ has a rectangular
profile, starts at $t_{i}$, is constant of value $\Omega$ for a
period of time $T_{R}$ and gets back to $0$ afterwards, we obtain for
any time $0\leq t\leq T_{R}$

\begin{align}
\dot{P_{\omega}}(t_{i}+t) & =-i\omega P_{\omega}(t_{i}+t)+\frac{i}{2}\Omega S_{\omega}(t_{i}+t)\nonumber \\
\dot{S_{\omega}}(t_{i}+t) & =\frac{i}{2}\Omega^{*}P_{\omega}(t_{i}+t).\label{eq:transfer_dynamics}
\end{align}

\noindent Integration of Eqs.~\ref{eq:transfer_dynamics} from $t_{i}$
to $t_{i}+t$ leads to

\begin{widetext}

\begin{align*}
S_{\omega}(t_{i}+t) & =e^{-i\omega t/2}\left(\cos(\Upsilon_{\omega}t/2)+i\frac{\omega}{\Upsilon_{\omega}}\sin(\Upsilon_{\omega}t/2)\right)S_{k}(t_{i})+ie^{-i\omega t/2}\frac{\Omega}{\Upsilon_{\omega}}\sin(\Upsilon_{\omega}t/2)P_{\omega}(t_{i})\\
P_{\omega}(t_{i}+t) & =ie^{-i\omega t/2}\frac{\Omega}{\Upsilon_{\omega}}\sin(\Upsilon_{\omega}t/2)S_{\omega}(t_{i})+e^{-i\omega t/2}\left(\cos(\Upsilon_{\omega}t/2)-i\frac{\omega}{\Upsilon_{\omega}}\sin(\Upsilon_{\omega}t/2)\right)P_{\omega}(t_{i})
\end{align*}

\end{widetext}

\noindent where $\Upsilon_{\omega}=\sqrt{\Omega^{2}+\omega^{2}}$
is the generalised Rabi frequency for frequency class of detuning
$\omega$. In particular, if $S_{\omega}(t_{i})=0$ then

\[
S_{\omega}(t_{i}+T_{R})=ie^{-i\omega T_{R}/2}\frac{\Omega}{\Upsilon_{\omega}}\sin(\Upsilon_{\omega}T_{R}/2)P_{\omega}(t_{i}),
\]

\noindent which means that the excitation transfer from $P_{\omega}$
to $S_{\omega}$ is 

\begin{align*}
\left|\frac{S_{\omega}(t_{i}+T_{R})}{P_{\omega}(t_{i})}\right|^{2} & =\frac{\Omega^{2}T_{R}^{2}}{4}\mathrm{sinc}^{2}\left(\Upsilon_{\omega}T_{R}/2\right)\\
 & =\frac{\Omega^{2}T_{R}^{2}}{4}\mathrm{sinc}^{2}\left(\sqrt{\frac{\Omega^{2}T_{R}^{2}}{4}+\frac{\omega^{2}T_{R}^{2}}{4}}\right).
\end{align*}

\noindent An identical result holds for the transfer from $S_{\omega}$
to $P_{\omega}$ if $P_{\omega}(t_{i})=0$ initially.

Now, assume that $T_{R}\ll\frac{2\pi}{\Gamma}$ with $\Gamma$ the
ensemble inhomogeneous linewidth ($\omega$ is at most of the order
of $\Gamma$). Then whatever the value of $\omega$, the detuning
component $\frac{\omega^{2}T_{R}^{2}}{4}\ll1$ does not affect much
the value of population transfer. Hence,

\begin{align*}
\forall\omega\quad\left|\frac{S_{\omega}(t_{i}+T_{R})}{P_{\omega}(t_{i})}\right|^{2} & \approx_{T_{R}\ll\frac{2\pi}{\Gamma}}\left|\frac{S_{\omega=0}(t_{i}+T_{R})}{P_{\omega=0}(t_{i})}\right|^{2}\\
 & =\frac{\Omega^{2}T_{R}^{2}}{4}\mathrm{sinc}^{2}\left(\frac{\Omega T_{R}}{2}\right)\\
 & =\sin^{2}\left(\frac{\Omega T_{R}}{2}\right).
\end{align*}

\noindent In particular, for a rectangular $\pi$-pulse ($\Omega T_{R}=\pi$)
the excitation transfer is almost perfect for all frequency classes.
One can also introduce partial transfers, that we call $a\pi$-pulses
for simplicity (with $0\leq a\leq1$). By setting 
\begin{equation}
\Omega T_{R}=:a\pi=2\mathrm{Arcsin}(q),\label{eq:partial_pulse}
\end{equation}
the excitation transfer is $q^{2}$.\\

In practice in our context, it is sufficient that $T_{R}$ is chosen
much smaller than $\frac{2\pi}{\delta\omega_{in}}$ where $\delta\omega_{in}$
is the bandwidth of the input photon that was stored inside the memory.
Indeed, only the frequency classes with $\omega$ at most of the order
of $\delta\omega_{in}$ are susceptible to bear relevant amounts of
excitation (see par. \ref{par:Fourier-sampling-behaviour}), and performing
transfers on those classes only is enough to generate and retrieve
a spin wave that represents the entire photon. Numerically, we check
that a factor $10$ between the two timescales is enough.

In terms of complex coefficients, we can approximate the transfer
function between $P_{\omega}$ and $S_{\omega}$ by

\[
H_{P\leftrightarrow S}(\omega)\approx ie^{-i\omega T_{R}/2}\sin(\Omega T_{R}/2)
\]

\noindent where a frequency-class dependent phase arises. Once again,
if $T_{R}\ll\frac{2\pi}{\delta\omega_{in}}$, this inhomogeneous phase
shall not affect much the rephasing of the $P_{\omega}$ coherences
and coupling to the external field.\\

This possibility of bringing excitations back and forth between $S$
and $P$ provides a scheme for longer storage. Once the input photon
is absorbed onto $P$, a storage $\pi$-pulse is applied before the
$\frac{2\pi}{\Delta}$ delay for echo emission (see Fig.~3 in the
main text). In our rotating frame, the spin wave then does not accumulate
any phase with time. After a certain time, a readout $\pi$-pulse
is applied to retrieve the excitation: $P$ coherences phase accumulation
resumes and an echo is finally emitted, as if time between the $\pi$-pulses
had been frozen. The possibility to choose this duration enables on-demand
storage and readout from the memory.\\

\paragraph{Open system}

The dynamics of the real system are more intricate. 

To begin with, coherence losses through $\gamma_{P}$ and $\gamma_{S}$
may happen. It turns out that if the loss rates are slow enough ($\gamma_{P,S}<\Omega$,
which is the case for our strong pulses) then one stays in a regime
known as underdamped \citep{shore_manipulating_2011}. Then the previous
analysis still qualitatively holds. Some loss factors would of course
be included by hand, similarly to Eq.~\ref{eq:polarisation_decay},
\ref{eq:polarS_decay}.

Furthermore, $P$ coherences keep interacting with $\mathcal{E}$ field
during transfers. We argue that if the transfer pulses are fast enough
compared with the coupling time scales $\kappa$ and $g\sqrt{N}$
the former picture still stands, as it is observed on numerical simulations.
Moreover, if the transfer pulses are applied much before the echo,
we know that inhomogeneous phase spreading on $P_{\omega}$ coherences
(see Eq.~\ref{eq:polarisation_FourierTransform}) hamper any coupling
with light that could come out of the cavity. As such, the former
picture is quite relevant.

To this respect, we note that a naive photon stretching procedure
could involve applying longer $\pi$-pulses ($T_{R}$ comparable or
longer than $\frac{2\pi}{\delta\omega_{in}}$) so that only the central
part of the frequency spectrum would then rephase and be emitted.
A drawback of this method is that all the energy carried by discarded
frequency classes is lost, the efficiency is low and comparable to
the one of filtration of the input photon in the frequency domain.
Our more efficient shaping procedure will be presented in Section~\ref{sec:Protocol-for-photon-shaping}
of these Appendices.

\subsubsection{Multimode compatibility}

Freezing time or storing right before an echo (see Fig.~3 of the Main
Text) is useful for the use of the AFC memory in a quantum network
context, where temporal multiplexing and synchronisation of emitted
photons are crucial \citep{sangouard_quantum_2011,ortu_multimode_2022}.
The AFC can accept any time bin of arrival within a $\frac{2\pi}{\Delta}$
window. For synchronisation, a first storage $\pi$-pulse is applied
before the first possible AFC echo at time $2\pi/\Delta$, possibly
without knowing in what precise time bin the input arrived. Once the
time bin is known (e. g. by detection of a correlated photon \citep{sangouard_quantum_2011}),
synchronisation $\pi$-pulses are applied so that the AFC will immediately
follow the application of any further retrieval pulse, see Fig.~3
of the main text.

\clearpage
\newpage

\section{Quantum reservoirs}

\label{sec:Quantum-reservoirs}

\subsection{Quantum noise and fluctuations}

\label{sec:Quantum-noise-and-fluctuations}

In Sec.~\ref{sec:Modelling-a-cavity-asisted} and Ref.~\citep{gorshkov_photon_2007-2},
the Heisenberg-Langevin approach is used to include losses in the
dynamical system: decay channels of an open system are represented
by the coupling to a large assembly of bosonic oscillators. This assembly
is referred to as a quantum reservoir (or bath). From the system perspective,
the reservoir induces backaction and gives rise to noise terms in
the equations \citep{gardiner_quantum_2004}, which in turn ensure
that commutators between systems operators are preserved despite the
decay channels. In this Appendix, we clarify some statements about
quantum noise operators for the construction of Heisenberg-Langevin
equations \ref{eq:memory_equations-1-1}. To this aim, helpful derivations
may be found in Refs.~\citep{lukin_modern_2016}\footnote{See pages 148 to 153},
\citep{scully_quantum_1997}\footnote{See Part 9.1 p. 272, and 9.4 p. 283},
\citep{cohen-tannoudji_processus_2001}\footnote{See Part CIV.2 pp. 331-335 and AV pp. 386-387},
and \citep{carmichael_statistical_1999}\footnote{See Chapter 1}.

A noise term is usually included in the form of a driving term $\hat{F}_{l}(t)$
in the equation of motion of some system operator $\hat{X}_{l}$.
That gives

\begin{equation}
\frac{\mathrm{d}\hat{X}_{l}}{\mathrm{d}t}(t)=-\frac{i}{\hbar}\left[\hat{X}_{l},\hat{H}\right](t)-\gamma\hat{X}_{l}(t)+\sqrt{2\gamma}\hat{F}_{l}(t),\label{eq:typical_noise_term}
\end{equation}

\noindent where $\hat{H}$ is the Hamiltonian that governs the deterministic
evolution of the closed system, and $\gamma$ is the rate of the decay
channel associated to noise term $\hat{F}_{l}$.

Moreover, we will make use of the following hypotheses:
\begin{enumerate}[resume, label=HP~\arabic{enumi}]
\item \emph{Reservoir in the ground state hypothesis} Noise terms at time
$t$, noted $\hat{F}_{l}(t)$, are commonly defined from reservoir
operators considered at the initial time $t_{0}$ (see input-output
derivations below). Hence, we can characterise the reservoir backaction
thanks to the initial conditions and state of the reservoir. Typically,
we will assume that the reservoir is initially in the vacuum state
and bears no excitation before being coupled with the system \footnote{Making such an assumption may require some experimental tweaks, as
discussed in Ref.~\citep{gorshkov_photon_2007-2} after equation (A16).}.\label{enu:Reservoir-in-the-ground-state}
\item \emph{Markovian noise hypothesis} If one assumes that the eigenfrequencies
of the reservoir oscillators are closely spaced and thus continuously
span a broad enough spectrum, and that the coupling constant does
not vary much with frequency\footnote{For this, also see Refs.~\citep{collett_squeezing_1984,gardiner_input_1985,gardiner_quantum_2015}},
some formal integrations \emph{à la} Wigner-Weisskopf result in a
memoryless property of the reservoir. Typically, all second-order
correlators involving noise terms are delta-peaked in time: no correlation
is found between noise at different times. This property also results
in the well-known generalised Einstein relations related to some quantum
fluctuation-dissipation result.\label{enu:Markovian-noise-hypothesis}
\item \emph{Uncorrelated reservoir }A side assumption is typically that
the reservoir and the system are initially uncorrelated: their quantum
states are independent.\footnote{See discussion in Sec.~3.3 of Ref.~\citep{gardiner_quantum_2004},
or the use of this assumption in Secs. 9.1 and 16.2 of Ref.~\citep{scully_quantum_1997}
for instance.} This also explains why for \ref{enu:Reservoir-in-the-ground-state}
we can define the state of the reservoirs independently.\label{enu:Uncorrelated-reservoir}
\end{enumerate}
As recalled in Ref.~\citep{gorshkov_photon_2007-2}, a consequence
of \ref{enu:Reservoir-in-the-ground-state}, \ref{enu:Markovian-noise-hypothesis}
and \ref{enu:Uncorrelated-reservoir} is that second-order correlators
involving noise terms are all zero but the antinormal ordered one:
\[
\forall t,t'\quad\langle\hat{F_{l}}(t')\hat{F_{l}}(t)\rangle=\langle\hat{F_{l}}^{\dagger}(t')\hat{F_{l}}(t)\rangle=0,
\]
\[
\forall t,t'\quad\langle\hat{F_{l}}(t')\hat{F_{l}}^{\dagger}(t)\rangle\propto\delta(t-t').
\]

\noindent The coefficient of proportionality of the last equation
(also known as the diffusion coefficient) can be found using generalised
Einstein relations \citep{lax_quantum_1966,cohen-tannoudji_processus_2001,lukin_modern_2016}.
In our case, $\hat{X}_{l}$ operators will refer to the $\hat{P}_{k}$
and $\hat{S}_{k}$ collective operators of Appendix~\ref{sec:Modelling-a-cavity-asisted}.
It turns out that with the right normalisations, assuming that each
individual atom is coupled to an independent reservoir, using the
low-excitation hypothesis \ref{enu:Low-excitation-number}, and referring
to the generalised Einstein relations for individual operators, the
coefficient is $1$ (see Ref.~\citep{gorshkov_photon_2007-2} (Eq.~(A15) and (A16)), Appendix~of Ref.~\citep{hald_mapping_2001}).

Another consequence, related to \ref{enu:Reservoir-in-the-ground-state}
and \ref{enu:Uncorrelated-reservoir}, is that initial time second-order
correlators involving one reservoir operator and one system operator
are zero. Specifically, for every $t$
\begin{equation*}
\left\langle \hat{X}_{l}(t_{0})\hat{F}_{l}(t)\right\rangle =0
\end{equation*}

Finally, we will assume that any pair of different decay channels
is associated to two independent reservoirs, so crossed second-order
correlators are also zero (noise terms for $l\neq l'$).

\subsection{Quantum input-output relations}

\label{subsec:Quantum-input-output-relations}

A particular type of noise is involved in input-output relations \citep{gardiner_input_1985,gardiner_quantum_2004}
(see also Refs.~\citep{caves_quantum_1982,yurke_quantum_1984,collett_squeezing_1984,gardiner_quantum_2015,steck_quantum_nodate-1}).
A typical setting consists of a cavity, open on one side to coupling
to an infinite set of free space modes. For instance, consider our
cavity mode $\hat{\mathcal{E}}$ and the complete set of free space
electromagnetic modes (one-dimensional here) $\hat{c}(\omega)$ for
$\omega\in\mathbb{R}$ such that $\forall\omega,\omega'\quad\left[\hat{c}(\omega),\hat{c}^{\dagger}(\omega')\right]=\delta(\omega-\omega')$.
We assume that the coupling is introduced through the approximate
interaction Hamiltonian

\[
\hat{H}_{bath}=\hbar\int\sqrt{\frac{\kappa(\omega)}{\pi}}\left(\hat{\mathcal{E}}\hat{c}^{\dagger}(\omega)+\hat{\mathcal{E}}^{\dagger}\hat{c}(\omega)\right)\mathrm{d}\omega.
\]

In case the coupling is set aside, the free space modes are assumed
to evolve according to the Hamiltonian $\hbar\int\omega\hat{c}^{\dagger}(\omega)\hat{c}(\omega)\mathrm{d}\omega$
so that their evolution is given by 

\[
\frac{\mathrm{d}\hat{c}(\omega)}{\mathrm{d}t}(t)=-i\omega\hat{c}(\omega,t)-i\sqrt{\frac{\kappa(\omega)}{\pi}}\hat{\mathcal{E}}(t)
\]

\noindent that we integrate as
\begin{equation}
\hat{c}(\omega,t)=e^{-i\omega(t-t_{0})}\hat{c}(\omega,t_{0})-i\sqrt{\frac{\kappa(\omega)}{\pi}}\int_{t_{0}}^{t}e^{-i\omega(t-t')}\hat{\mathcal{E}}(t')\mathrm{d}t',\label{eq:bath_evolution}
\end{equation}

\noindent or

\begin{equation}
\hat{c}(\omega,t)=e^{-i\omega(t-t_{f})}\hat{c}(\omega,t_{f})+i\sqrt{\frac{\kappa(\omega)}{\pi}}\int_{t}^{t_{f}}e^{-i\omega(t-t')}\hat{\mathcal{E}}(t')\mathrm{d}t'.\label{eq:bath_evolution_back}
\end{equation}

One can define input modes from the free space modes taken at time
$t_{0}$ (understood as preceding any interaction) as 

\begin{equation}
\hat{\mathcal{E}}_{in}(t)=\frac{-i}{\sqrt{2\pi}}\int e^{-i\omega(t-t_{0})}\hat{c}(\omega,t_{0})\mathrm{d}\omega\label{eq:def_Ein}
\end{equation}

\noindent so that for all $t,t'$

\begin{eqnarray}
\left[\hat{\mathcal{E}}_{in}(t),\hat{\mathcal{E}}_{in}^{\dagger}(t')\right] & = & \frac{1}{2\pi}\int e^{-i\omega(t-t')}\mathrm{d}\omega\label{eq:Ein_commutator}\\
 & = & \delta(t-t').\nonumber 
\end{eqnarray}

Note that this Fourier-transform like operation can be seen as some
change of basis. Just as the whole set of free space modes at time
$t_{0}$ could be completely described by the collection of $\hat{c}(\omega,t_{0})$
for all $\omega$, it can be completely described by the collection
of $\hat{\mathcal{E}}_{in}(t)$ for all $t$.

Similarly, one can define output modes from the free space modes taken
at time $t_{f}>t_{0}$ (understood as following any interaction) as
\begin{equation*}
\hat{\mathcal{E}}_{out}(t)=\frac{i}{\sqrt{2\pi}}\int e^{-i\omega(t-t_{f})}\hat{c}(\omega,t_{f})\mathrm{d}\omega
\end{equation*}

\noindent so that again, for all $t,t'$
\begin{equation*}
\left[\hat{\mathcal{E}}_{out}(t),\hat{\mathcal{E}}_{out}^{\dagger}(t')\right]=\delta(t-t').
\end{equation*}

Thus, the whole set of free space modes at time $t_{f}$ is completely
described either by the collection of $\hat{c}(\omega,t_{f})$ for
all $\omega$, or by the collection of $\hat{\mathcal{E}}_{out}(t)$
for all $t$.

Then the Markovian approximation assumes that the coupling constant
between the cavity mode $\hat{\mathcal{E}}$ and any external mode
$\hat{c}(\omega)$ does not depend on the frequency $\omega$ ($\forall\omega\quad\kappa(\omega)=\kappa$)
so that $\hat{\mathcal{E}}$ evolves as 
\begin{eqnarray}
\frac{\mathrm{d}\hat{\mathcal{E}}}{\mathrm{d}t}(t) & = & -\frac{i}{\hbar}\left[\hat{\mathcal{E}},\hat{H}+\hat{H}_{bath}\right]\nonumber \\
 & = & -\frac{i}{\hbar}\left[\hat{\mathcal{E}},\hat{H}\right]-i\sqrt{\frac{\kappa}{\pi}}\int\hat{c}(\omega,t)\mathrm{d}\omega\nonumber \\
 & = & -\frac{i}{\hbar}\left[\hat{\mathcal{E}},\hat{H}\right]-\frac{\kappa}{\pi}\int\int_{t_{0}}^{t}e^{-i\omega(t-t')}\hat{\mathcal{E}}(t')\mathrm{d}t'\mathrm{d}\omega\nonumber \\
 &  & -i\sqrt{2\kappa}\frac{1}{\sqrt{2\pi}}\int e^{-i\omega(t-t_{0})}\hat{c}(\omega,t_{0})\mathrm{d}\omega\nonumber \\
 & = & -\frac{i}{\hbar}\left[\hat{\mathcal{E}},\hat{H}\right]-\kappa\hat{\mathcal{E}}(t)+\sqrt{2\kappa}\hat{\mathcal{E}}_{in}(t),\label{eq:dE_Ein}
\end{eqnarray}

\noindent where we used Eq.~\ref{eq:bath_evolution} to go from the
second to the third line. In Eq.~\ref{eq:dE_Ein}, $\hat{H}$ is the
Hamiltonian that governs all evolutions but the coupling to the external
bath. As such, we retrieve a decay at rate $\kappa$ in terms of field
intensity, and the feeding of $\hat{\mathcal{E}}(t)$ with input $\hat{\mathcal{E}}_{in}(t)$.
Similarly, one can obtain from Eq.~\ref{eq:bath_evolution_back}
\begin{equation}
\frac{\mathrm{d}\hat{\mathcal{E}}}{\mathrm{d}t}(t)=-\frac{i}{\hbar}\left[\hat{\mathcal{E}},\hat{H}\right]+\kappa\hat{\mathcal{E}}(t)-\sqrt{2\kappa}\hat{\mathcal{E}}_{out}(t).\label{eq:dE_Eout}
\end{equation}

\noindent Equating the right-hand sides of Eqs.~\ref{eq:dE_Ein} and
\ref{eq:dE_Eout} leads to the input-output relation
\begin{equation*}
2\kappa\hat{\mathcal{E}}(t)=\sqrt{2\kappa}\left(\hat{\mathcal{E}}_{in}(t)+\hat{\mathcal{E}}_{out}(t)\right)
\end{equation*}

\noindent i.e.
\begin{equation}
\hat{\mathcal{E}}_{out}(t)=\sqrt{2\kappa}\hat{\mathcal{E}}(t)-\hat{\mathcal{E}}_{in}(t).\label{eq:input_output_relation}
\end{equation}

A last thing to note is that input term at $t$ only influences $\hat{\mathcal{E}}$
at times $t'\geq t$, as expected for causality \citep{gardiner_input_1985,gardiner_quantum_2004}.
Namely,
\begin{equation*}
\left[\hat{\mathcal{E}}_{in}(t),\hat{\mathcal{E}}(t')\right]=0,\text{ if \ensuremath{t'<t}}
\end{equation*}

\noindent and
\begin{equation*}
\left[\hat{\mathcal{E}}_{out}(t),\hat{\mathcal{E}}(t')\right]=0,\text{ if \ensuremath{t'>t}}.
\end{equation*}

Finally, $\left[\hat{\mathcal{E}}_{in}(t),\hat{\mathcal{E}}^{(\dagger)}(t_{0})\right]$
and $\left[\hat{\mathcal{E}}_{out}(t),\hat{\mathcal{E}}^{(\dagger)}(t_{f})\right]$
depend on the initial conditons for the system and the set of external
free modes: see \ref{enu:Uncorrelated-reservoir} above.

\clearpage
\newpage

\section{Integration of a set of differential equations for quantum operators}

\label{sec:FromOperatorsToScalars}

In this Appendix, we explain how the scalar equations \ref{eq:discrete_scalar_system-2},
\ref{eq:full_scalar_system-1} are obtained from the equations involving
quantum operators \ref{eq:memory_equations-1-1}, \ref{eq:memory_equations-1}. 

\subsection{Integration of a vectorial ODE}

\label{subsec:Integration-of-a-vectorial-ODE}

To this end, we first recall some textbook results about the integration
of vectorial Ordinary Differential Equations (ODEs). In particular,
we stress that a formal integration of the ODE shows that any solution
is a linear function of the initial conditions.

Let $\mathscr{E}$ be a vector space of finite dimension $n$, typically
$\mathscr{E}=\mathbb{C}^{n}$. Consider the first order ODE on $\mathscr{E}$

\begin{equation}
\dot{Y}(t)=M(t)X(t)+F(t),\label{eq:non_homoegeneous_equation}
\end{equation}

\noindent with $\begin{array}{cccc}
X: & \mathbb{R} & \rightarrow & \mathscr{E}\\
 & t & \mapsto & X(t)
\end{array}$,$\begin{array}{cccc}
M: & \mathbb{R} & \rightarrow & \mathcal{L}(\mathscr{E},\mathscr{E})\\
 & t & \mapsto & M(t)
\end{array}$and $\begin{array}{cccc}
F: & \mathbb{R} & \rightarrow & \mathscr{E}\\
 & t & \mapsto & F(t)
\end{array}$. $M(t)$ are typically $n\times n$ matrices. Let $t_{0}\in\mathbb{R}$
be any distinguished initial time (we will assume that solutions of
Eq.~\ref{eq:non_homoegeneous_equation} can be regularly defined over
$\mathbb{R}$).

\subsubsection{Homogeneous equation}

First, consider the corresponding homogeneous equation

\begin{equation}
\dot{X}(t)=M(t)X(t).\label{eq:homogeneous_eq}
\end{equation}

\noindent From the linearity of the equation, it is clear that the
set of solutions of Eq.~\ref{eq:homogeneous_eq}, that we note $Sol_{H}$,
is a vector space. Let's introduce the morphism

\[
\begin{array}{cccc}
\Phi_{t_{0}}: & Sol_{H} & \rightarrow & \mathcal{E}\\
 & X & \mapsto & X(t_{0}).
\end{array}
\]

\noindent From a linear version of Cauchy-Lipschitz theorem\footnote{Also called Picard--Lindelöf theorem},
we know that $\Phi_{t_{0}}$ is a bijection. That is, every solution
of the equation \ref{eq:homogeneous_eq} is uniquely determined by
its value at time $t_{0}$. As a consequence, $Sol_{H}$ is of finite-dimension
like $\mathcal{E}$, namely of dimension $n$. We can thus introduce
a basis $(Y^{(1)}(t_{0}),\dots,Y^{(n)}(t_{0}))$ of $\mathscr{E}$
where $Y^{(1)},...,Y^{(n)}\in Sol_{H}$. We note $(Y^{(1)}(t),\dots,Y^{(n)}(t))$
the corresponding trajectories at time $t$. Another consequence of
Cauchy-Lipschitz theorem is that at any time $t$, $(Y^{(1)}(t),\dots,Y^{(n)}(t))$
is a basis of $\mathscr{E}$.  For any $t$, we can introduce the
(invertible) matrix $Q$ of the $(Y^{(1)},\dots,Y^{(n)})$ family,

\[
Q(t):=\begin{pmatrix}Y^{(1)}(t) & | & \dots & | & Y^{(n)}(t)\end{pmatrix}.
\]

\noindent Then

\begin{equation}
\dot{Q}(t)=M(t)Q(t).\label{eq:equation_fundamental_matrix}
\end{equation}

\noindent One can write any solution $X$ of the homogeneous equation
\ref{eq:homogeneous_eq} as

\begin{equation}
X(t)=Q(t).\left(Q(t_{0})^{-1}X(t_{0})\right).\label{eq:solution_homogeneous}
\end{equation}

\noindent The occurrence of $Q(t)$ in Eq.~\ref{eq:solution_homogeneous}
is a consequence of the evolution of $X$ and the property \ref{eq:equation_fundamental_matrix},
while the occurrence of $Q^{-1}(t_{0})$ expresses the change of basis
of the initial condition $X(t_{0})$ into the basis $(Y^{(1)}(t_{0}),\dots,Y^{(n)}(t_{0}))$.

\subsubsection{Non-homogeneous equation}

A more general solution for the non-homogeneous equation \ref{eq:non_homoegeneous_equation}
can be obtained thanks to the variation of parameters method. Namely,
a solution is

\begin{equation}
X(t)=Q(t)\int_{t_{0}}^{t}Q^{-1}(t')F(t')\mathrm{d}t'+Q(t)Q^{-1}(t_{0})X(t_{0})\label{eq:general_solution_fu}
\end{equation}

\noindent where $Q$ follows Eq.~\ref{eq:equation_fundamental_matrix}.
As previously mentioned, Cauchy-Lipschitz theorem ensures this solution
for initial condition $X(t_{0}$) is unique. Interestingly but not
surprisingly, the expression of $X(t)$ is a linear function of the
initial condition $X(t_{0})$ and of the time sequence of driving
terms \textbf{$F(t)$}. We shall extensively use this property in
the following discussions.

\subsection{Notation: vectors of operators}

\label{subsec:Abuse-of-notation}

Next, we give insight into a commonly used notation which involves
``vectors of operators'' to make use of the results of the previous
section.

\subsubsection{Notation}

Consider a system of differential equations for quantum operators
of the form

\begin{equation}
\begin{cases}
\frac{\mathrm{d}\hat{a}_{1}}{\mathrm{dt}} & =\sum_{j=1}^{n}M_{1j}\hat{a}_{j}\\
 & \vdots\\
\frac{\mathrm{d}\hat{a}_{n}}{\mathrm{dt}} & =\sum_{j=1}^{n}M_{nj}\hat{a}_{j}
\end{cases}.\label{eq:system_operators}
\end{equation}

\noindent We would like to use previous results from part \ref{subsec:Integration-of-a-vectorial-ODE}
to solve system \ref{eq:system_operators}, and thus write it in the
form

\begin{equation}
\frac{\mathrm{d}}{\mathrm{d}t}\begin{pmatrix}\hat{a}_{1}(t)\\
\vdots\\
\hat{a}_{n}(t)
\end{pmatrix}=M(t)\begin{pmatrix}\hat{a}_{1}(t)\\
\vdots\\
\hat{a}_{n}(t)
\end{pmatrix},\label{eq:operators_vector}
\end{equation}

\noindent where $M=[M_{ij}]_{1\leq i,j\leq n}$ and the vector coordinates
shall be quantum operators. As in Eq.~\ref{eq:equation_fundamental_matrix},
we then would like to write that system \ref{eq:operators_vector}
is solved by

\begin{equation}
X(t)=Q(t)Q^{-1}(t_{0})X(t_{0}),\label{eq:integrated_operators_vector}
\end{equation}

\noindent with $Q$ the fundamental matrix such that $\dot{Q}(t)=M(t)Q(t)$
and 

\begin{equation}
\forall t\quad X(t)=\begin{pmatrix}\hat{a}_{1}(t)\\
\vdots\\
\hat{a}_{n}(t)
\end{pmatrix}.\label{eq:vector_X_a}
\end{equation}

\noindent A way to properly introduce the notation \ref{eq:vector_X_a}
is to consider that Eq.~\ref{eq:integrated_operators_vector} means
that for all $i$

\begin{align}
\hat{a}_{i}(t) & =\sum_{l}[Q(t)Q^{-1}(t_{0})]_{il}X_{l}(t_{0})\nonumber \\
 & =\sum_{l}[Q(t)Q^{-1}(t_{0})]_{il}\hat{a}_{l}(t_{0}).\label{eq:interpretation_vector}
\end{align}

\noindent With this understanding,

\begin{align*}
\frac{\mathrm{d}\hat{a}_{i}}{\mathrm{dt}}(t) & =\sum_{l}\frac{\mathrm{d}}{\mathrm{d}t}[Q(t)Q^{-1}(t_{0})]_{il}\hat{a}_{l}(t_{0})\\
 & =\sum_{l}[\dot{Q}(t)Q^{-1}(t_{0})]_{il}\hat{a}_{l}(t_{0})\\
 & =\sum_{l}[M(t)Q(t)Q^{-1}(t_{0})]_{il}\hat{a}_{l}(t_{0})\\
 & =\sum_{l}\sum_{k}[M(t)]_{ik}[Q(t)Q^{-1}(t_{0})]_{kl}\hat{a}_{l}(t_{0})\\
 & =\sum_{k}[M(t)]_{ik}\sum_{l}[Q(t)Q^{-1}(t_{0})]_{kl}\hat{a}_{l}(t_{0})\\
 & =\sum_{k}[M(t)]_{ik}\hat{a}_{k}(t)\quad\text{from Eq.~\ref{eq:interpretation_vector}},
\end{align*}

\noindent which is merely the $i$-th line of the symbolic equation
\ref{eq:operators_vector}. Thus, as long as linear quantities are
involved, the abuse of notation consisting of a vector of operators
can be used.\\

\subsubsection{Coupled system of equations over a direct product vector space of
operators}

Previous considerations can actually be clarified further in the following
way. We note $\mathscr{\mathcal{H}}$ for the Hilbert space over which
the elements $\hat{a}_{i}(t_{0})$ of a linearly independent set of
initial conditions $(\hat{a}_{i}(t_{0}))_{i}$ act. For every $i$,
$\hat{a}_{i}(t_{0})\in\mathcal{L}(\mathcal{H})$, the vector space
of linear maps from $\mathcal{H}$ to $\mathcal{H}$. $\mathscr{E}:=\mathrm{Span}(\hat{a}_{1}(t_{0}),\dots,\hat{a}_{n}(t_{0}))$
is by definition a $n$-dimensional subspace of $\mathcal{L}(\mathcal{H})$.
Consider then the direct (Cartesian) product $\mathscr{E}_{\times}:=\underset{n\text{ times}}{\underbrace{\mathscr{E}\times\mathscr{E}\times\dots\times\mathscr{E}}}$,
of dimension $n\times n$. A basis of $\mathscr{E}_{\times}$ is given
by

\begin{widetext}

\[
\left((\hat{a}_{1}(t_{0}),\vec{0},\dots,\vec{0}),\dots,(\hat{a}_{n}(t_{0}),\vec{0},\dots,\vec{0}),\dots,(\vec{0},\dots,\vec{0},\hat{a}_{1}(t_{0})),\dots,(\vec{0},\dots,\vec{0},\hat{a}_{n}(t_{0}))\right)
\]

\end{widetext}

\noindent With a vector representation of $\mathscr{E}_{\times}$,
the system of equations \ref{eq:system_operators} with initial conditions
$(\hat{a}_{i}(t_{0}))_{i}$ involves

\begin{align*}
X(t_{0}) & =[\hat{a}_{i}(t_{0})]_{i}\\
 & \equiv\begin{pmatrix}1 & 0 & \dots & 0 & | & 0 & 1 & 0 & \dots & 0 & | & \dots & | & 0 & \dots & 0 & 1\end{pmatrix}^{T}
\end{align*}

\noindent and

\begin{align}
M(t)=
\left(
\begin{array}{c|c|c}
M_{11} \mathds{1}_n & \dots & M_{1n} \mathds{1}_n \\
\hline
\vdots & \ddots & \vdots \\
\hline
M_{n1} \mathds{1}_n & \dots & M_{nn} \mathds{1}_n \\
\end{array}
\right).
\end{align}

\noindent The trajectory for each of the $\hat{a}_{i}$ operators
is then identified as the restriction of the whole $X(t)$ trajectory
to each term of the product space: for instance $\hat{a}_{1}(t)$
is understood as the operator acting on $\mathscr{E}$ with coordinates
given the first $n$ coordinates of $X(t)$. Applying Cauchy-Lipschitz
theorem to the vector space $\mathscr{E}_{\times}$ indicates that
the solution to the system of equations is given at any time $t$
by a linear combination of the initial conditions (see Eq.~\ref{eq:solution_homogeneous}).
In particular, it shows that every restriction $\hat{a}_{i}(t)$,
as defined above, lies in the stabilised subspace $\mathscr{E}=\mathrm{Span}(\hat{a}_{1}(t_{0}),\dots,\hat{a}_{n}(t_{0}))$. 

If we now consider a non-homogeneous equation, we introduce some driving
term $F:\mathbb{R}\rightarrow V$, where $V$ is a Hilbert space (possibly
of infinite dimension), of which $\mathscr{E}_{\times}$ is a sub-vector
space. Then Eq.~\ref{eq:general_solution_fu} yields a solution for
the previous construction, that we will assume to be unique. Considering
the restrictions $\hat{a}_{i}(t)$ defined above, it appears that
at any time $t$, the operator $\hat{a}_{i}(t)$ is a linear combination
of the initial operators $(\hat{a}_{i}(t_{0}))_{i}$ and (continuously)
of the driving terms at former times $F(t')$ with $t'\leq t$. We
can then use the results of section \ref{subsec:Integration-of-a-vectorial-ODE}
to solve the equation. 

\paragraph{Example}

Consider the following system of coupled differential equations

\[
\frac{\mathrm{d}}{\mathrm{d}t}\begin{pmatrix}\hat{a}\\
\hat{b}
\end{pmatrix}=\begin{pmatrix}\kappa_{1} & g\\
g & \kappa_{2}
\end{pmatrix}\begin{pmatrix}\hat{a}\\
\hat{b}
\end{pmatrix}
\]

\noindent Let $\mathcal{H}$ be the Hilbert of quantum states over
which two initial operators at $t_{0}=0$ $\hat{a}(0)$ and $\hat{b}(0)$
are supposed to act, and consider the vector subspace of $\mathcal{L}(\mathcal{H})$,
$\mathscr{E}=\mathrm{Span}(\hat{a}(0),\hat{b}(0))$. The system of
equations is understood as a first order ODE over $\mathscr{E}\times\mathscr{E}$,
with the following vector representation in the basis $\left((\hat{a}(0),\vec{0}_{\mathscr{E}}),(\hat{b}(0),\vec{0}_{\mathscr{E}}),(\vec{0}_{\mathscr{E}},\hat{a}(0)),(\vec{0}_{\mathscr{E}},\hat{b}(0))\right)$:

\begin{align}
\left\{
	\begin{array}{rl}
	\frac{\mathrm{d}X}{\mathrm{d}t}(t) &=
		\left(
		\begin{array}{c|c}
		\kappa_1 \mathds{1} & g \mathds{1} \\
		\hline
		g \mathds{1} & \kappa_2 \mathds{1} \\
		\end{array}
		\right)
	.X(t)\\
	\\
	X(0)&=\begin{pmatrix}1 & 0 & 0 & 1 \\ \end{pmatrix}^{T}
	\end{array}
\right.
\end{align}

\noindent where $\vec{0}_{\mathscr{E}}$ is the null vector of $\mathscr{E}$.

The resolution of this equation yields a vector $X(t)$ with 4 coordinates.
The first two are the coordinates of what will be identified as $\hat{a}(t)$
in the basis $(\hat{a}(0),\hat{b}(0))$ of $\mathscr{E}$, the last
two those of $\hat{b}(t)$.

\begin{widetext}

\subsection{Integration of a linear differential system of operators}

\subsubsection{Formal integration}

With previous notations and considerations in mind, we consider an
equation of the form

\begin{equation}
\frac{\mathrm{d}X}{\mathrm{d}t}(t)=M(t)X(t)+F(t)\label{eq:equation_ops}
\end{equation}

\noindent where $X$ is thought to be a vector of operators as in
Eq.~\ref{eq:vector_X_a}. Typically, an input-output relation can
be cast into this form, where $M$ describes the coupling between
different system operators (rows of $X$) from the deterministic evolution
and $F$ gathers the different types of quantum noise in a Heisenberg-Langevin
formulation.

As explained in section \ref{subsec:Abuse-of-notation}, Eq.~\ref{eq:equation_ops}
can be formally solved using the results of \ref{subsec:Integration-of-a-vectorial-ODE}.
As such, for the operator assigned to the $k$-th row of $X$ we get

\begin{equation}
X_{k}(t)=\int_{t_{0}}^{t}\sum_{l}\underset{=:G(t,t')_{kl}}{\underbrace{[Q(t)Q^{-1}(t')]_{kl}}}F_{l}(t')\mathrm{d}t'+\sum_{l}\underset{=:G(t,t_{0})_{kl}}{\underbrace{[Q(t)Q^{-1}(t_{0})]_{kl}}}X_{l}(t_{0}).\label{eq:operator_integration}
\end{equation}

\subsubsection{Single vacuum state}

\label{subsec:Single-vacuum-state}

In the case where $X_{k}(t_{0})$ operators and $F_{l}(t)$ noise
operators are understood as annihilation operators (in the sense of
\ref{enu:System-Environment-decomposition} and Eq.~\ref{eq:vacuum_co}
below), we observe that a single vacuum state can be defined and singled
out throughout time evolution. It is a consequence of the passive
linear interferometer structure of the system \ref{eq:equation_ops}.
Indeed,

\begin{eqnarray*}
\forall k\quad X_{k}(t)|0\rangle & = & \int_{t_{0}}^{t}\sum_{l}G(t,t')_{kl}\underset{\vec{0}}{\underbrace{F_{l}(t')|0\rangle}}\mathrm{d}t'+\sum_{l}G(t,t_{0})_{kl}\underset{\vec{0}}{\underbrace{X_{l}(t_{0})|0\rangle}}\\
 & = & \vec{0}.
\end{eqnarray*}

\noindent Conservation of the excitation number $\hat{N}$ (such as
in Eq.~\ref{eq:Ntot_excitationnumber}) within the whole system can
also lead to $0\leq\langle0|\hat{X}_{k}^{\dagger}(t)\hat{X}_{k}(t)|0\rangle\leq\langle0|\hat{N}(t)|0\rangle=\langle0|\hat{N}(t_{0})|0\rangle=0$
which gives the same result.

\subsection{Application to the computation of different correlators}

\label{subsec:Application-to-the_correlators}

We now apply previous results to the determination of quantum correlators
when integrating a set of differential equations for quantum operators.
The goal is to give a clear interpretation to the paragraph ``a linear
functional of~{[}...{]}'' in Ref.~\citep{gorshkov_photon_2007}
(p. 12), and our use of scalar equations. See also Sec.~11.3 of Ref.~\citep{gardiner_quantum_2015}
for a motivation.

Note that we work within the framework of Heisenberg representation.
Any state is supposed to be constant, so that the system-environment
decomposition is set with the initial description (see Appendix~\ref{subsec:PhotonStateCoherence}).

\subsubsection{Correlators}

For a given system density matrix $\rho$, the correlator between
two operators $\hat{a}$ and $\hat{b}$ is noted $\langle\hat{a}\hat{b}\rangle$,
with

\[
\langle\hat{a}\hat{b}\rangle:=\mathrm{Tr}\left(\rho\hat{a}\hat{b}\right).
\]

The correlator between two system operators is

\begin{eqnarray}
\langle X_{k}^{(\dagger)}(t)X_{m}(t')\rangle & = & \int_{t_{0}}^{t}\int_{t_{0}}^{t'}\sum_{l}\sum_{l'}G^{(*)}(t,t'')_{kl}G(t',t''')_{ml'}\langle F_{l}^{(\dagger)}(t'')F_{l'}(t''')\rangle\mathrm{d}t''\mathrm{d}t'''\label{eq:all_time_correlator}\\
 &  & +\sum_{l}\sum_{l'}G^{(*)}(t,t_{0})_{kl}G(t',t_{0})_{ml'}\langle X_{l}^{(\dagger)}(t_{0})X_{l'}(t_{0})\rangle\nonumber \\
 &  & +\int_{t_{0}}^{t}\sum_{l}\sum_{l'}G^{(*)}(t,t'')_{kl}G(t',t_{0})_{ml'}\langle F_{l}^{(\dagger)}(t'')X_{l'}(t_{0})\rangle\mathrm{d}t''\nonumber \\
 &  & +\int_{t_{0}}^{t'}\sum_{l}\sum_{l'}G^{(*)}(t,t_{0})_{kl}G(t',t''')_{ml'}\langle X_{l}^{(\dagger)}(t_{0})F_{l'}(t''')\rangle\mathrm{d}t'''.\nonumber 
\end{eqnarray}

\noindent All then amounts to reckoning correlators of the following
type:
\begin{itemize}
\item Input-input correlators: $\langle F_{l}^{(\dagger)}(t_{1})F_{l'}(t_{2})\rangle$;
\item Input-initial condition correlators: $\langle F_{l}^{(\dagger)}(t_{1})X_{l'}(t_{0})\rangle$
or $\langle X_{l}^{(\dagger)}(t_{0})F_{l'}(t_{1})\rangle$;
\item Initial-initial conditions correlators: $\langle X_{l}^{(\dagger)}(t_{0})X_{l'}(t_{0})\rangle$.
\end{itemize}
These correlators can be computed thanks to extra assumptions on the
system state and the noise operators \citep{scully_quantum_1997,gardiner_quantum_2004,lukin_modern_2016}\textcolor{red}{.}

\subsubsection{Translation for the system of operators}
\label{sec:correspondence_notations_scalar}
In the context of the model of Appendix~\ref{sec:Modelling-a-cavity-asisted}, we highlight that this section uses the correspondence of notations

\begin{eqnarray*}
F_{1,in}^{(\dagger)}(t) & \leftrightarrow & \hat{\mathcal{E}}_{in}^{(\dagger)}(t)\\
F_{1,out}^{(\dagger)}(t) & \leftrightarrow & \hat{\mathcal{E}}_{out}^{(\dagger)}(t)\\
f_{in}(t) & \leftrightarrow & \mathcal{E}_{in}(t)\\
X_{1}^{(\dagger)}(t) & \leftrightarrow & \hat{\mathcal{E}}^{(\dagger)}(t)\\
X_{k\in[2,N+1]}^{(\dagger)}(t) & \leftrightarrow & \hat{P}_{k-1}^{(\dagger)}(t)\\
F_{k\in[2,N+1]}^{(\dagger)}(t) & \leftrightarrow & \hat{F}_{P,k-1}^{(\dagger)}(t)\\
X_{k\in[N+2,2N+1]}^{(\dagger)}(t) & \leftrightarrow & \hat{S}_{k-N-1}^{(\dagger)}(t)\\
F_{k\in[N+2,2N+1]}^{(\dagger)}(t) & \leftrightarrow & \hat{F}_{S,k-N-1}^{(\dagger)}(t),
\end{eqnarray*}

where in the rotating frame, $\mathcal{E}_{in}(t)$ ($\mathcal{E}_{out}(t)$)
for instance gives the input (output) photon envelope.

\subsubsection{First particular case: one input, vacuum noise and factorised system
and at most one excitation}

\label{par:Particular-case:syst_assumptions}

Hereinafter, we will assume all the following particular conditions \ref{enu:System-Environment-decomposition} to \ref{enu:Single-photon-input-vacuum-noise} to help with the resolution in the context of our quantum memory.
\begin{enumerate}[label=PC~1.\arabic{enumi}]
\item \emph{System-Environment decomposition }At initial time $t_{0}$,
we single out the decomposition of the Hilbert space $\mathcal{H}_{S}\otimes\mathcal{H}_{E}$
between the system $S$ and the environment $E$. As an initial condition
for the operators, we consider annihilation operators $X_{l}(t_{0})$
(that act over the system) and $F_{k}(t)$ (that act as quantum noise
operators over the environment, see Appendix~\ref{sec:Quantum-reservoirs}
for the $t$ label) such that
\begin{equation}
\begin{cases}
\forall l\quad X_{l}(t_{0})|0\rangle & =\vec{0}\\
\forall t\forall k\quad F_{k}(t)|0\rangle & =\vec{0}
\end{cases},\label{eq:vacuum_co}
\end{equation}
where the vacuum state is denoted by $|0\rangle$ and $\vec{0}$ stands
for the null vector in the Hilbert space. \label{enu:System-Environment-decomposition}
\item \emph{Initial product state }The whole state for $S$ and $E$ is
assumed to be factorised w.r.t. the former decomposition, i.e.
\begin{equation}
\rho=\rho_{S}\otimes\rho_{E},\label{eq:SE_productstate}
\end{equation}
see \ref{enu:Uncorrelated-reservoir} of Sec.~\ref{sec:Quantum-reservoirs}.\label{enu:Initial-product-state}
\item \emph{Pure state in the at-most-one-excitation subspace} The system
state $\rho_{S}$ is assumed to be pure and within the subspace of
at most one excitation, that is of the form
\begin{align}
\rho_{S} & =:\left(v_{0}|0\rangle+\sum_{l}c_{l}X_{l}^{\dagger}(t_{0})|0\rangle\right)\left(v_{0}^{*}\langle0|+\sum_{l'}c_{l'}^{*}\langle0|X_{l'}(t_{0})\right)\nonumber \\
 & =\left|v_{0}\right|^{2}|0\rangle\langle0|+v_{0}\sum_{l'}c_{l'}^{*}|0\rangle\langle0|X_{l'}(t_{0})+v_{0}^{*}\sum_{l}c_{l}X_{l}^{\dagger}(t_{0})|0\rangle\langle0|+\sum_{l}\sum_{l'}c_{l}c_{l'}^{*}X_{l}^{\dagger}(t_{0})|0\rangle\langle0|X_{l'}(t_{0}).\label{eq:Spurestate_gen}
\end{align}
In the restricted case of the one-excitation subspace, this is given by
\begin{equation}
\rho_{S}=:\left(\sum_{l}c_{l}X_{l}^{\dagger}(t_{0})|0\rangle\right)\left(\sum_{l'}c_{l'}^{*}\langle0|X_{l'}(t_{0})\right)=\sum_{l}\sum_{l'}c_{l}c_{l'}^{*}X_{l}^{\dagger}(t_{0})|0\rangle\langle0|X_{l'}(t_{0}).\label{eq:Spurestate}
\end{equation}
If one wishes to start from a mixed state, the linearity of the dynamics
can be leveraged, to study each component independently.\label{enu:Pure-state-in-subspace}
\item \emph{Bosonic operators }We assume that each $X_{l}$ corresponds
to a particular system bosonic mode initially, that is
\begin{equation}
\begin{cases}
\left[X_{l}(t_{0}),X_{l'}(t_{0})\right] & =\mathbb{O}\\
\left[X_{l}(t_{0}),X_{l'}^{\dagger}(t_{0})\right] & =\delta_{l,l'}\mathbb{I}
\end{cases}.\label{eq:mode_commutator}
\end{equation}
This is a valid approximate algebraic relation for a assembly of two-level
systems storing a few excitations compared to the number of systems
(Hollstein-Primakoff-type approximation \citep{kurucz_spectroscopic_2011}).
This particular condition is especially important for the first term
$l=1$ (see below), while for other noise terms, Markovian assumptions
leading to delta-peaked correlators may be sufficient.\label{enu:Bosonic-operators}
\item \emph{Single photon input and vacuum noise reservoirs} The environment
system $\rho_{E}$ is assumed to be vacuum for every noise component
$F_{k\neq1}$ but the first one, assumed to be a a pure state within
the subspace of at most one excitation, that is
\begin{align}
\rho_{E} & =:\left(p_{0}|0\rangle+\int_{\mathbb{R}}f(t)F_{1}(t)^{\dagger}|0\rangle\mathrm{d}t\right)\left(p_{0}^{*}\langle0|+\int_{\mathbb{R}}f^{*}(t')\langle0|F_{1}(t')\mathrm{d}t'\right)\nonumber \\
 & =\left|p_{0}\right|^{2}|0\rangle\langle0|+p_{0}^{*}\int_{\mathbb{R}}f(t)F_{1}(t)^{\dagger}|0\rangle\langle0|\mathrm{d}t+p_{0}\int_{\mathbb{R}}f^{*}(t')|0\rangle\langle0|F_{1}(t')\mathrm{d}t'+\int_{\mathbb{R}}\int_{\mathbb{R}}f(t)f^{*}(t')F_{1}(t)^{\dagger}|0\rangle\langle0|F_{1}(t')\mathrm{d}t\mathrm{d}t'.\label{eq:Epurestate_gen}
\end{align}
For a pure one-excitation (single photon) state, this is given by
\begin{equation}
\rho_{E}=:\left(\int_{\mathbb{R}}f(t)F_{1}(t)^{\dagger}|0\rangle\mathrm{d}t\right)\left(\int_{\mathbb{R}}f^{*}(t')\langle0|F_{1}(t')\mathrm{d}t'\right)=\int_{\mathbb{R}}\int_{\mathbb{R}}f(t)f^{*}(t')F_{1}(t)^{\dagger}|0\rangle\langle0|F_{1}(t')\mathrm{d}t\mathrm{d}t';\label{eq:Epurestate}
\end{equation}
Again, components within a mixed state could be studied independently
thanks to the linearity of the system.\label{enu:Single-photon-input-vacuum-noise}
\end{enumerate}
The reason for the former point \ref{enu:Single-photon-input-vacuum-noise}
is that we further regard $F_{1}\equiv F_{1,in}$ entry as some input
noise, with the related input-output relation (see Sec.~\ref{subsec:Quantum-input-output-relations})
coupling to system operator $X_{1}$ with constant $\kappa$,

\begin{equation}
F_{1,out}(t)=\sqrt{2\kappa}X_{1}(t)-F_{1,in}(t),\label{eq:inputoutput_vector}
\end{equation}

\noindent with commutation relations from Eq.~\ref{eq:Ein_commutator}

\begin{equation}
\begin{cases}
\left[F_{1,in}(t),F_{1,in}(t')\right] & =\mathbb{O}\\
\left[F_{1,in}(t),F_{1,in}^{\dagger}(t')\right] & =\delta(t-t')\mathds{1}
\end{cases}.\label{eq:noise_commutator}
\end{equation}

Taking into consideration \ref{enu:System-Environment-decomposition}
to \ref{enu:Single-photon-input-vacuum-noise}, we obtain the following
useful relations for correlators

\begin{eqnarray}
\langle F_{1,in}^{\dagger}(t_{1})F_{1,in}(t_{2})\rangle & = & \underset{=1}{\underbrace{\mathrm{Tr}_{S}(\rho_{S})}}\times\mathrm{Tr}_{E}(F_{1,in}^{\dagger}(t_{1})F_{1,in}(t_{2})\rho_{E})\nonumber \\
 & = & \mathrm{Tr}_{E}(F_{1,in}(t_{2})\rho_{E}F_{1,in}^{\dagger}(t_{1}))\nonumber \\
 & = & \int_{\mathbb{R}}\int_{\mathbb{R}}f(t)f^{*}(t')\mathrm{Tr}_{E}(F_{1,in}(t_{2})F_{1,in}(t)^{\dagger}|0\rangle\langle0|F_{1,in}(t')F_{1,in}^{\dagger}(t_{1}))\mathrm{d}t\mathrm{d}t'\nonumber \\
 & = & f(t_{2})f^{*}(t_{1})\mathrm{Tr}_{E}(|0\rangle\langle0|)\quad\text{using Eqs.~\ref{eq:vacuum_co} and \ref{eq:noise_commutator}}\nonumber \\
 & = & f(t_{2})f^{*}(t_{1}).\label{eq:in_in_correlator}
\end{eqnarray}

\noindent In addition, if $l\neq1$ or $l'\neq1$ then

\begin{eqnarray}
\langle F_{l}^{\dagger}(t_{1})F_{l'}(t_{2})\rangle & = & \mathrm{Tr}_{E}(F_{l}^{\dagger}(t_{1})F_{l'}(t_{2})\rho_{E})\nonumber \\
 & = & 0\quad\text{for vacuum of Eq.~\ref{eq:Epurestate}},\label{eq:noise_noise_correlator}
\end{eqnarray}

\noindent and similarly for the crossed terms thanks to the product
structure of $\rho$

\begin{eqnarray}
\langle F_{l}^{\dagger}(t_{1})X_{l'}(t_{0})\rangle & = & \mathrm{Tr}_{S}(\mathrm{Tr}_{E}(F_{l}^{\dagger}(t_{1})X_{l'}(t_{0})\rho_{S}\otimes\rho_{E}))\nonumber \\
 & = & \underset{\text{keep diag. only because of the trace}}{\underbrace{\mathrm{Tr}_{S}(X_{l'}(t_{0})\rho_{S})}}\times\underset{\text{keep diag. only because of the trace}}{\underbrace{\mathrm{Tr}_{E}(F_{l}^{\dagger}(t_{1})\rho_{E})}}\nonumber \\
 & = & \left(v_{0}^{*}c_{l'}\right)\times\left(p_{0}f^{*}(t_{1})\delta_{l,1}\right)\text{ for Eqs.~\ref{eq:Spurestate_gen} and \ref{eq:Epurestate_gen}}\label{eq:noise_system_correlator}\\
 & = & 0\text{ for Eqs.\,\ref{eq:Spurestate} or \ref{eq:Epurestate} where \ensuremath{v_{0}p_{0}=0}}.
\end{eqnarray}

\noindent and $\langle X_{l'}^{\dagger}(t_{0})F_{l}(t_{1})\rangle=0$.

\noindent As for the system operators we get

\begin{eqnarray}
\langle X_{l}^{\dagger}(t_{0})X_{l'}(t_{0})\rangle & = & \mathrm{Tr}_{S}(X_{l}^{\dagger}(t_{0})X_{l'}(t_{0})\rho_{S})\times\underset{=1}{\underbrace{\mathrm{Tr}_{E}(\rho_{E})}}\nonumber \\
 & = & \mathrm{Tr}_{S}(X_{l'}(t_{0})\rho_{S}X_{l}^{\dagger}(t_{0}))\nonumber \\
 & = & \sum_{m}\sum_{m'}c_{m}c_{m'}^{*}\mathrm{Tr}_{S}(X_{l'}(t_{0})X_{m}^{\dagger}(t_{0})|0\rangle\langle0|X_{m'}(t_{0})X_{l}^{\dagger}(t_{0}))\nonumber \\
 & = & c_{l'}c_{l}^{*}\mathrm{Tr}_{S}(|0\rangle\langle0|)\quad\text{using Eqs.~\ref{eq:vacuum_co} and \ref{eq:mode_commutator}}\nonumber \\
 & = & c_{l'}c_{l}^{*}.\label{eq:syst_syst_correlator}
\end{eqnarray}

We can now use the results from Eqs.~\ref{eq:in_in_correlator}, \ref{eq:noise_noise_correlator},
\ref{eq:noise_system_correlator}, \ref{eq:syst_syst_correlator}
together with the integration \ref{eq:operator_integration} to compute
correlators at any times, such as the one from Eq.~\ref{eq:all_time_correlator},

\begin{eqnarray}
\langle X_{k}^{\dagger}(t)F_{1,in}(t')\rangle & = & \int_{t_{0}}^{t}\sum_{l}G^{*}(t,t'')_{kl}\langle F_{l}^{\dagger}(t'')F_{1,in}(t')\rangle\mathrm{d}t''+\sum_{l}G^{*}(t,t_{0})_{kl}\langle X_{l}^{\dagger}(t_{0})F_{1,in}(t')\rangle\nonumber \\
 & = & \int_{t_{0}}^{t}G^{*}(t,t'')_{k1}f(t')f^{*}(t'')\mathrm{d}t''+\sum_{l}G^{*}(t,t_{0})_{kl}v_{0}c_{l}^{*}p_{0}^{*}f(t')\nonumber \\
 & = & \int_{t_{0}}^{t}G^{*}(t,t'')_{k1}f(t')f^{*}(t'')\mathrm{d}t''\text{ in the case of Eqs.\,\ref{eq:Spurestate} or \ref{eq:Epurestate} where \ensuremath{v_{0}p_{0}=0}}.\label{eq:allt_syst_noise_correlator}
\end{eqnarray}

\noindent Similarly,

\begin{eqnarray}
\langle F_{1,in}^{\dagger}(t)X_{k}(t')\rangle & = & \int_{t_{0}}^{t'}G(t',t'')_{k1}f(t'')f^{*}(t)\mathrm{d}t''+v_{0}^{*}p_{0}\sum_{l}G(t',t_{0})_{kl}c_{l}f^{*}(t)\nonumber \\
 & = & \int_{t_{0}}^{t'}G(t',t'')_{k1}f(t'')f^{*}(t)\mathrm{d}t''\text{ in the case of Eqs.\,\ref{eq:Spurestate} or \ref{eq:Epurestate} where \ensuremath{v_{0}p_{0}=0}},\label{eq:allt_syst_noise_correlator_rev-1}
\end{eqnarray}

\noindent while for the other noises

\begin{equation}
\begin{array}{ccccc}
\langle F_{l\neq1}^{\dagger}(t)X_{k}(t')\rangle & = & 0 & = & \langle X_{k}^{\dagger}(t)F_{l\neq1}(t')\rangle\end{array}.\label{eq:allt_syst_othernoise_correlator}
\end{equation}

\noindent As for two system operators, Eq.~\ref{eq:all_time_correlator}
becomes

\begin{eqnarray}
\langle X_{k}^{\dagger}(t)X_{m}(t')\rangle & = & \int_{t_{0}}^{t}\int_{t_{0}}^{t'}G^{*}(t,t'')_{k1}G(t',t''')_{m1}f(t''')f^{*}(t'')\mathrm{d}t''\mathrm{d}t'''\nonumber \\
 &  & +\sum_{l}\sum_{l'}G^{*}(t,t_{0})_{kl}G(t',t_{0})_{ml'}c_{l'}c_{l}^{*}\nonumber \\
 &  & +v_{0}^{*}p_{0}\int_{t_{0}}^{t}\sum_{l'}G^{*}(t,t'')_{k1}G(t',t_{0})_{ml'}c_{l'}f^{*}(t'')\mathrm{d}t''\nonumber \\
 &  & +v_{0}p_{0}^{*}\int_{t_{0}}^{t'}\sum_{l}G^{*}(t,t_{0})_{kl}G(t',t''')_{m1}c_{l}^{*}f(t''')\mathrm{d}t'''\label{eq:allt_syst_syst_correlator}\\
 & = & \left(\int_{t_{0}}^{t}G(t,t'')_{k1}f(t'')\mathrm{d}t''\right)^{*}\left(\int_{t_{0}}^{t'}G(t',t''')_{m1}f(t''')\mathrm{d}t'''\right)+\left(\sum_{l}G(t,t_{0})_{kl}c_{l}\right)^{*}\left(\sum_{l'}G(t',t_{0})_{ml'}c_{l'}\right)\nonumber 
\end{eqnarray}

\noindent in the case of Eqs.~\ref{eq:Spurestate} or \ref{eq:Epurestate}
where $v_{0}p_{0}=0$.

We can in turn use the results from Eqs.~\ref{eq:allt_syst_noise_correlator},
\ref{eq:allt_syst_noise_correlator_rev-1} and \ref{eq:allt_syst_syst_correlator}
to compute those involving the output operators from Eq.~\ref{eq:inputoutput_vector},

\begin{eqnarray*}
\langle F_{1,out}^{\dagger}(t_{1})F_{1,out}(t_{2})\rangle & = & \langle F_{1,in}^{\dagger}(t_{1})F_{1,in}(t_{2})\rangle-\sqrt{2\kappa}\langle F_{1,in}^{\dagger}(t_{1})X_{1}(t_{2})\rangle-\sqrt{2\kappa}\langle X_{1}^{\dagger}(t_{1})F_{1,in}(t_{2})\rangle+2\kappa\langle X_{1}^{\dagger}(t_{1})X_{1}(t_{2})\rangle\\
 & = & f(t_{2})f^{*}(t_{1})\\
 &  & -\sqrt{2\kappa}\left(\int_{t_{0}}^{t_{2}}G(t_{2},t'')_{11}f(t'')f^{*}(t_{1})\mathrm{d}t''+v_{0}^{*}p_{0}\sum_{l}G(t_{2},t_{0})_{1l}c_{l}f^{*}(t_{1})\right)\\
 &  & -\sqrt{2\kappa}\left(\int_{t_{0}}^{t_{1}}G^{*}(t_{1},t'')_{11}f(t_{2})f^{*}(t'')\mathrm{d}t''+v_{0}p_{0}^{*}\sum_{l}G^{*}(t_{1},t_{0})_{1l}c_{l}^{*}f(t_{2})\right)\\
 &  & +2\kappa\left(\int_{t_{0}}^{t_{1}}G(t_{1},t'')_{11}f(t'')\mathrm{d}t''\right)^{*}\left(\int_{t_{0}}^{t_{2}}G(t_{2},t''')_{11}f(t''')\mathrm{d}t'''\right)\\
 &  & +2\kappa\left(\sum_{l}G(t_{1},t_{0})_{1l}c_{l}\right)^{*}\left(\sum_{l'}G(t_{2},t_{0})_{1l'}c_{l'}\right)\\
 &  & +2\kappa\left(v_{0}^{*}p_{0}\int_{t_{0}}^{t_{1}}\sum_{l'}G^{*}(t_{1},t'')_{11}G(t_{2},t_{0})_{1l'}c_{l'}f^{*}(t'')\mathrm{d}t''\right.\\
 &  & \qquad\left.+v_{0}p_{0}^{*}\int_{t_{0}}^{t_{2}}\sum_{l}G^{*}(t_{1},t_{0})_{1l}G(t_{2},t''')_{11}c_{l}^{*}f(t''')\mathrm{d}t'''\right)\\
 & = & \left(f(t_{1})-\sqrt{2\kappa}\int_{t_{0}}^{t_{1}}G(t_{1},t'')_{11}f(t'')\mathrm{d}t''\right)^{*}\times\left(f(t_{2})-\sqrt{2\kappa}\int_{t_{0}}^{t_{2}}G(t_{2},t'')_{11}f(t'')\mathrm{d}t''\right)\\
 &  & -\sqrt{2\kappa}\left(v_{0}^{*}p_{0}\sum_{l}G(t_{2},t_{0})_{1l}c_{l}f^{*}(t_{1})+v_{0}p_{0}^{*}\sum_{l}G^{*}(t_{1},t_{0})_{1l}c_{l}^{*}f(t_{2})\right)\\
 &  & +2\kappa\left(\sum_{l}G(t_{1},t_{0})_{1l}c_{l}\right)^{*}\left(\sum_{l'}G(t_{2},t_{0})_{1l'}c_{l'}\right)\\
 &  & +2\kappa\left(v_{0}^{*}p_{0}\int_{t_{0}}^{t_{1}}\sum_{l'}G^{*}(t_{1},t'')_{11}G(t_{2},t_{0})_{1l'}c_{l'}f^{*}(t'')\mathrm{d}t''\right.\\
 &  & \qquad\left.+v_{0}p_{0}^{*}\int_{t_{0}}^{t_{2}}\sum_{l}G^{*}(t_{1},t_{0})_{1l}G(t_{2},t''')_{11}c_{l}^{*}f(t''')\mathrm{d}t'''\right).
\end{eqnarray*}

\noindent In the case of Eqs.~\ref{eq:Spurestate} or \ref{eq:Epurestate}
where $\ensuremath{v_{0}p_{0}=0}$, we get

\begin{eqnarray}
\langle F_{1,out}^{\dagger}(t_{1})F_{1,out}(t_{2})\rangle & = & \left(f(t_{1})-\sqrt{2\kappa}\int_{t_{0}}^{t_{1}}G(t_{1},t'')_{11}f(t'')\mathrm{d}t''\right)^{*}\times\left(f(t_{2})-\sqrt{2\kappa}\int_{t_{0}}^{t_{2}}G(t_{2},t'')_{11}f(t'')\mathrm{d}t''\right)\nonumber \\
 &  & +2\kappa\left(\sum_{l}G(t_{1},t_{0})_{1l}c_{l}\right)^{*}\left(\sum_{l'}G(t_{2},t_{0})_{1l'}c_{l'}\right).\label{eq:out_out_correlator}
\end{eqnarray}

\medskip{}
\noindent
On top of particular conditions \ref{enu:System-Environment-decomposition} to \ref{enu:Single-photon-input-vacuum-noise}, we obtain a quite simple formula when any of the two following extra conditions holds: 
\begin{enumerate}[resume, label=PC~1.\arabic{enumi}]
\item \emph{Empty memory} Either $\rho_{S}=|0\rangle\langle0|$ (that is
an initially empty system: $v_{0}=1,\forall l\quad c_{l}=0$):
\begin{equation}
\langle F_{1,out}^{\dagger}(t_{1})F_{1,out}(t_{2})\rangle_{\rho_{S}=|0\rangle\langle0|}=\left(f(t_{1})-\sqrt{2\kappa}\int_{t_{0}}^{t_{1}}G(t_{1},t'')_{11}f(t'')\mathrm{d}t''\right)^{*}\times\left(f(t_{2})-\sqrt{2\kappa}\int_{t_{0}}^{t_{2}}G(t_{2},t'')_{11}f(t'')\mathrm{d}t''\right),\label{eq:out_out_correlator_input}
\end{equation}
\label{enu:Empty-memory}
\item \emph{No input photon} Or $\rho_{E}=|0\rangle\langle0|$ (that is
no input: $p_{0}=1,\forall t\quad f(t)=0$)
\begin{equation}
\langle F_{1,out}^{\dagger}(t_{1})F_{1,out}(t_{2})\rangle_{\rho_{E}=|0\rangle\langle0|}=2\kappa\left(\sum_{l}G(t_{1},t_{0})_{1l}c_{l}\right)^{*}\left(\sum_{l'}G(t_{2},t_{0})_{1l'}c_{l'}\right).\label{eq:out_out_correlator_syst}
\end{equation}
\label{enu:No-input-photon}
\end{enumerate}

\subsubsection{Second particular case: one input, vacuum noise and factorised system
and coherent state input}

\label{par:Coherent-state-input}

In the case where the system is initially empty ($\rho_{S}=|0\rangle\langle0|$),
we can also consider the situation where the single input mode $F_{1}$
is a pure coherent state instead of a single photon. As described
in Refs.~\citep{mollow_pure-state_1975,blow_continuum_1990}, a coherent
wavepacket with temporal envelope $f$ is described by a (temporally-)multimode
coherent state

\begin{equation}
|\{f(t)\}_{t}\rangle=\exp\left[\int\left(f(t)F_{1,in}^{\dagger}(t)-f^{*}(t)F_{1,in}(t)\right)\mathrm{d}t\right]|0\rangle,\label{eq:temporal_coherent_state}
\end{equation}

\noindent which is characterised by the relation

\begin{equation}
F_{1,in}|\{f(t)\}_{t}\rangle=f(t)|\{f(t)\}_{t}\rangle.\label{eq:coherent_state_annihilation}
\end{equation}

\noindent That is, we take the input \ref{enu:Coherent-state-input} instead of \ref{enu:Single-photon-input-vacuum-noise}
\begin{enumerate}[label=PC~2.\arabic{enumi}]
\item \emph{Coherent state input}
\begin{equation}
\rho_{E}=|\{f(t)\}_{t}\rangle_{F_{1}}\langle\{f(t)\}_{t}|.\label{eq:input_temporal_coherent}
\end{equation}
\label{enu:Coherent-state-input}
\end{enumerate}

\noindent In the same time, we still assume hypotheses \ref{enu:System-Environment-decomposition}
to \ref{enu:Single-photon-input-vacuum-noise} (but for the single
photon input) as well as \ref{enu:Empty-memory}. As a result

\begin{align*}
\langle F_{1,in}^{\dagger}(t_{1})F_{1,in}(t_{2})\rangle & =\underset{=1}{\underbrace{\mathrm{Tr}_{S}(\rho_{S})}}\times\mathrm{Tr}_{E}(F_{1,in}(t_{2})|\{f(t)\}_{t}\rangle_{B_{1}}\langle\{f(t)\}_{t}|F_{1,in}^{\dagger}(t_{1}))\\
 & =\mathrm{Tr}_{E}(f(t_{2})|\{f(t)\}_{t}\rangle_{F_{1}}\langle\{f(t)\}_{t}|f^{*}(t_{1}))\\
 & =f(t_{2})f^{*}(t_{1})
\end{align*}

\noindent as in Eq.~\ref{eq:in_in_correlator} for the one-photon
state. And still if $l\neq1$ or $l'\neq1$ we get

\begin{eqnarray*}
\langle F_{l}^{\dagger}(t_{1})F_{l'}(t_{2})\rangle & = & \mathrm{Tr}_{E}(F_{l}^{\dagger}(t_{1})F_{l'}(t_{2})\rho_{E})\\
 & = & 0\quad\text{for vacuum of Eq.~\ref{eq:temporal_coherent_state}}
\end{eqnarray*}

\noindent What's more, since the system is in vacuum state (see hyp.
of Eqs.~\ref{eq:noise_system_correlator}, \ref{eq:syst_syst_correlator}),
we get

\begin{eqnarray*}
\forall l,l'\quad\langle F_{l}^{\dagger}(t_{1})X_{l'}(t_{0})\rangle & = & 0,\\
\forall l,l'\quad\langle X_{l}^{\dagger}(t_{0})X_{l'}(t_{0})\rangle & = & 0.
\end{eqnarray*}

\noindent Hence, the formula from Eq.~\ref{eq:out_out_correlator_input}
still holds for an empty system and an $F_{1}$-mode input coherent
state with temporal envelope $f$.

\subsubsection{Equivalent scalar equations for analytical or numerical resolution}

We now conclude, in the particular cases exposed before, that the
resolution of the set of equations for the operators amounts to the
resolution of a set of scalar equations.

\paragraph{System of scalar equations}

It is indeed interesting to have a look at the scalar equations equivalent
to Eq.~\ref{eq:equation_ops} ($x$ is now a vector with scalar coordinates
that evolve with time), namely

\begin{equation}
\frac{\mathrm{d}x}{\mathrm{d}t}(t)=M(t)x(t)+f(t),\label{eq:main_scalar_equation}
\end{equation}

\noindent together with the corresponding input-output like relation
for the first noise term

\begin{equation}
f_{out}(t)=\sqrt{2\kappa}x_{1}(t)-f_{in}(t),\label{eq:input_output_scalarequation}
\end{equation}

\noindent where for all times $t$ the first input is represented
by the scalar function $f_{in}$ while the other entries of $f$ are
taken to be zero

\[
\begin{cases}
f_{1}(t) & \equiv f_{in}(t)\\
f_{k\neq1}(t) & =0
\end{cases}.
\]

\noindent Moreover, the initial conditions for $x$ are defined as

\[
\forall l\quad x_{l}(t_{0})=:c_{l}.
\]

As non-commuting relations are irrelevant here, we easily obtain the
following results 

\[
x_{k}(t)=\int_{t_{0}}^{t}G(t,t')_{k1}f_{in}(t')\mathrm{d}t'+\sum_{l}G(t,t_{0})_{kl}c_{l},
\]

\begin{equation}
f_{out}(t)=f_{in}(t)-\sqrt{2\kappa}\int_{t_{0}}^{t}G(t,t')_{11}f_{in}(t')\mathrm{d}t'-\sqrt{2\kappa}\sum_{l}G(t,t_{0})_{1l}c_{l}.\label{eq:fout_integrated}
\end{equation}

\noindent And for the correlators,

\begin{eqnarray}
x_{k}^{(*)}(t)x_{m}(t') & = & \left(\int_{t_{0}}^{t}G^{(*)}(t,t'')_{k1}f_{in}^{(*)}(t'')\mathrm{d}t''\right)\left(\int_{t_{0}}^{t'}G(t',t''')_{m1}f_{in}(t''')\mathrm{d}t'''\right)\nonumber \\
 &  & +\left(\sum_{l}G^{(*)}(t,t_{0})_{kl}c_{l}^{(*)}\right)\left(\sum_{l'}G(t',t_{0})_{ml'}c_{l'}\right)\nonumber \\
 &  & +\left(\int_{t_{0}}^{t}G^{(*)}(t,t'')_{k1}f_{in}^{(*)}(t'')\mathrm{d}t''\right)\left(\sum_{l'}G(t',t_{0})_{ml'}c_{l'}\right)\nonumber \\
 &  & +\left(\sum_{l}G^{(*)}(t,t_{0})_{kl}c_{l}^{(*)}\right)\left(\int_{t_{0}}^{t'}G(t',t''')_{m1}f_{in}(t''')\mathrm{d}t'''\right)\label{eq:scalar_product_eq}
\end{eqnarray}

\paragraph{Two particular situations}

\label{par:Two-particular-situations_numerical_res}

In the two particular cases \ref{enu:Empty-memory} and \ref{enu:No-input-photon}
(the system is driven by $f_{in}$ from an initially empty system
state (that is $\forall l\quad c_{l}=x_{l}(t_{0})=0$), or the driving
term is zero $\forall t\quad f_{in}(t)=0$ (that is $\forall t\forall k\quad f_{k}(t)=0$)),
only the first two terms of the right-hand side of Eq.~\ref{eq:scalar_product_eq}
remain, that is

\begin{eqnarray*}
x_{k}^{(*)}(t)x_{m}(t') & = & \left(\int_{t_{0}}^{t}G^{(*)}(t,t'')_{k1}f_{in}^{(*)}(t'')\mathrm{d}t''\right)\left(\int_{t_{0}}^{t'}G(t',t''')_{m1}f_{in}(t''')\mathrm{d}t'''\right)\\
 &  & +\left(\sum_{l}G^{(*)}(t,t_{0})_{kl}c_{l}^{(*)}\right)\left(\sum_{l'}G(t',t_{0})_{ml'}c_{l'}\right),
\end{eqnarray*}

\noindent where either the first term (second particular case \ref{enu:No-input-photon})
or the the second line (first particular case \ref{enu:Empty-memory})
is zero. Comparing with Eq.~\ref{eq:allt_syst_syst_correlator} with
the assumptions of paragraph \ref{par:Particular-case:syst_assumptions},
we notice that any of the correlators may be extracted directly from
the resolution of scalar equations, as

\begin{equation}
\langle X_{k}^{\dagger}(t)X_{m}(t')\rangle_{\rho_{S}=|0\rangle\langle0|\text{ or }\rho_{E}=|0\rangle\langle0|}=\left.x_{k}^{*}(t)\times x_{m}(t')\right|_{f_{in}=0\text{ or }\forall l\quad c_{l}=0}.\label{eq:numerical_eff}
\end{equation}

\noindent This resolution may thus be performed numerically to find
the transfer efficiency from input to system for instance.

Identically, the results \ref{eq:out_out_correlator_input} or \ref{eq:out_out_correlator_syst}
are obtained thanks to $f_{out}$ from Eq.~\ref{eq:fout_integrated}
as long as either $f_{in}=0$ or $\forall l\quad c_{l}=0$:

\begin{equation}
\langle F_{1,out}^{\dagger}(t_{1})F_{1,out}(t_{2})\rangle_{\rho_{S}=|0\rangle\langle0|\text{ or }\rho_{E}=|0\rangle\langle0|}=\left.f_{out}^{*}(t_{1})\times f_{out}(t_{2})\right|_{f_{in}=0\text{ or }\forall l\quad c_{l}=0}.\label{eq:numerical_out}
\end{equation}
\\

As a consequence, under hypotheses \ref{enu:System-Environment-decomposition}
to \ref{enu:Empty-memory} (or \ref{enu:No-input-photon}), we can
solve our system dynamics \ref{eq:memory_equations-1-1} by solving
the scalar set of equations \ref{eq:discrete_scalar_system-2} with
the correspondence of notations of Section~\ref{sec:correspondence_notations_scalar}. This is also possible when \ref{enu:Coherent-state-input} is used instead of the input of \ref{enu:Single-photon-input-vacuum-noise}.
\end{widetext}

~

\clearpage
\newpage

\section{Protocol for photon-shaping with a cavity-assisted quantum memory}

\label{sec:Protocol-for-photon-shaping}

As explained in the main text (and recalled in Appendix~\ref{sec:Quantum-network-architecture}),
we aim to stretch and reshape the photon emitted by the cavity-assisted
quantum memory. In this Appendix, we focus on the shaping protocol
that will let us change the single photon temporal waveform of the
memory output. We suggest to proceed by taking advantage of the natural
dynamics of the system. Instead of stretching the readout pulse straightaway,
we rather consider a series of partial readouts: between each readout
pulse, time is given for partial emission from the memory with the
natural echo shape (see Eq.~\ref{eq:partial_pulse}). This leads
to a piecewise mathematical formulation of the shaping problem, that
we first expose below. We then compute the optimal weights for partial
readout for a given target shape. Achievable efficiencies are then
discussed. Most of the discussions are set within a mathematical framework
that forgets about the dynamics of the physical system, as we focus
on short control pulses and direct emission of retrieved excitations.
Numerical simulations provided in the main text (see Fig.~3) validate
the relevance of these discussions.\textcolor{red}{{} }

Note that though we introduce it in the context of an AFC quantum
memory, our shaping technique seems to be compatible with other memory
protocols based on collective coherence rephasing, such as the recently
introduced PLM-coherence technique \citep{moiseev_optical_2024}.

\begin{widetext}

\subsection{Mathematical formulation of the problem}

\label{subsec:Mathematical-formulation-of}

To start with, we give a mathematical description of the piecewise
envelope shaping problem.

Let $[a,b]$ be a real segment, $f_{target}\in\mathcal{C}_{0}([a,b],\mathbb{C})$
a continuous function. Let $N_{shape}\in\mathbb{N^{*}}$ and $h_{in}\in\mathcal{C}_{0}\left(\left[-\frac{b-a}{2N_{shape}},+\frac{b-a}{2N_{shape}}\right],\mathbb{C}\right)$
another continuous function. From a physical perspective, $[a,b]$
is meant to represent the time frame over which the shaping is performed,
with $f_{target}$ representing the target shape (in our case, the
ion photon waveform) and $h_{in}$ representing the initial photon
shape stored in and retrieved from the memory (typically, coming from
an SPDC source). The spread of $h_{in}$ profile is referred to as
a bin, and $N_{shape}$ gives the number of bins needed to cover $[a,b]$. Notations are illustrated in Fig.~\ref{fig:notations_shaping}.
\begin{figure}[hbtp]
\begin{centering}
\includegraphics[width=0.7\columnwidth]{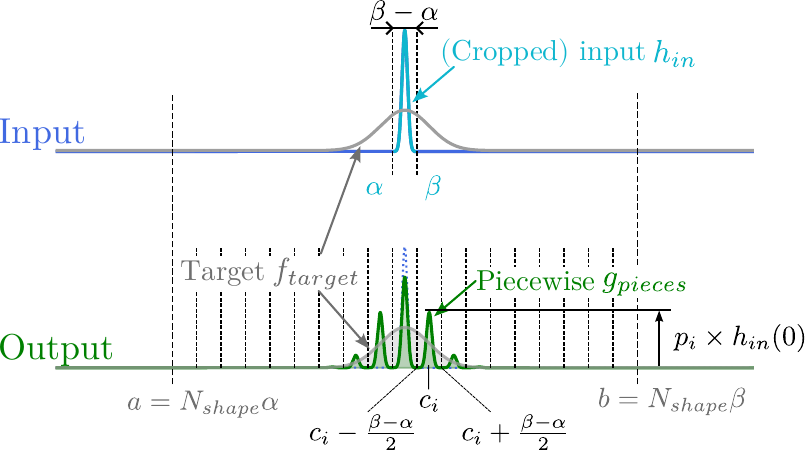}
\par\end{centering}
\caption{Notations in use for the shaping problem. A piecewise function $g_{pieces}$ is built by filling $N_{shape}$ bins (centered around the $c_i$) with $p_i$-weighted copies of an input profile $h_{in}$, so as to match a target shape $f_{target}$. $\beta - \alpha$ gives the width of the bins (see part \ref{subsec:Convergence-towards-infinity}).
\textcolor{red}{\label{fig:notations_shaping}}}
\end{figure}

We further assume that the functions are $L^{2}$-normalised, that
is 

\begin{equation}
\int_{a}^{b}\left|f_{target}\right|{{}^2}(t)\mathrm{d}t=1=\int_{-\frac{b-a}{2N_{shape}}}^{\frac{b-a}{2N_{shape}}}\left|h_{in}\right|{{}^2}(t)\mathrm{d}t.\label{eq:normalisations_profiles}
\end{equation}

\noindent The shaped photon waveform is then represented by a continuous
function $g_{pieces}\in\mathcal{C}_{0}([a,b]\rightarrow\mathbb{C})$.
Subject to the constraint of it being normalised, we will look for
a way to build a piecewise $g_{pieces}$ that maximises the overlap
with $f_{target}$. We take $g_{pieces}$ split into pieces that are
all weighted versions of profile $h_{in}$ with weights $0\leq p_{i}\leq1$
and $0\leq\theta_{i}<2\pi$ phases ($i\in[[0,N_{bins}-1]]$), that
is

\[
g_{pieces}(t):=\sum_{j=0}^{N_{shape}-1}p_{j}e^{i\theta_{j}}\mathds{1}_{\left[c_{j}-\frac{b-a}{2N_{shape}},c_{j}+\frac{b-a}{2N_{shape}}\right]}(t)\times h_{in}(t-c_{j}).
\]

\noindent We introduced the bins centers positions 

\[
\forall i\in[[0,N_{shape}-1]]\quad c_{i}:=\frac{b-a}{2N_{shape}}+i\frac{b-a}{N_{shape}}=\left(i+\frac{1}{2}\right)\frac{b-a}{2N_{shape}},
\]

\noindent and $\mathds{1}$ stands for the indicatrix function. 

We then want to optimise the objective functional

\begin{eqnarray}
R_{f_{target}}(p,\theta,h_{in}) & = & \left|\int_{a}^{b}f_{target}(t)^{*}g_{pieces}(t)\mathrm{d}t\right|=\left|\sum_{j=0}^{N_{shape}-1}p_{j}e^{i\theta_{j}}{\int_{c_{j}-\frac{b-a}{2N_{shape}}}^{c_{j}+\frac{b-a}{2N_{shape}}}f_{target}(t)^{*}h_{in}(t-c_{j})\mathrm{d}t}\right|\label{eq:objective_overlap}
\end{eqnarray}

\noindent subject to the normalisation constraint

\begin{equation}
\sum_{i=0}^{N_{shape}-1}p_{i}{{}^2}=1.\label{eq:normalisation_constraint}
\end{equation}

\noindent The normalisation constraint \ref{eq:normalisation_constraint}
ensures that

\begin{eqnarray*}
1 & = & \int_{a}^{b}\left|g_{pieces}\right|{{}^2}(t)\mathrm{d}t=\sum_{j=0}^{N_{shape}-1}p_{j}{{}^2}\int_{c_{j}-\frac{b-a}{2N_{shape}}}^{c_{j}+\frac{b-a}{2N_{shape}}}\left|h_{in}\right|{{}^2}(t-c_{j})\mathrm{d}t=\sum_{j=0}^{N_{shape}-1}p_{j}{{}^2}\times1
\end{eqnarray*}

\noindent since $\ensuremath{h_{in}}$ is normalised (Eq.~\ref{eq:normalisations_profiles}).
For convenience, we introduce the overlap $J_{j}^{f_{target},h_{in}}$ between the input translated to bin $j$ and the target restricted to that bin,
\begin{equation}
J_{j}^{f_{target},h_{in}} := \int_{c_{j}-\frac{b-a}{2N_{shape}}}^{c_{j}+\frac{b-a}{2N_{shape}}}f_{target}(t)^{*}h_{in}(t-c_{j})\mathrm{d}t.
\end{equation}

\end{widetext}

\subsection{Optimal weights for the piecewise target}

For a given arbitrary $h_{in}$, the optimal weights $(p_{i})$ can
be found in the following way. $R_{f_{target}}(p,h_{in})$ with given
$h_{in}$ is bounded by the Cauchy-Schwarz inequality on $\mathbb{C}^{N_{bins}}$

\begin{eqnarray}
R_{f_{target}} & = & \left|\sum_{j=0}^{N_{shape}-1}p_{j}e^{i\theta_{j}}J_{j}^{f_{target},h_{in}}\right|\label{eq:opt_p_bound_C}\\
 & \leq & \sqrt{\underset{=1\text{ by constraint}}{\underbrace{\left(\sum_{j=0}^{N_{shape}-1}p_{j}^{2}\right)}}\left(\sum_{j=0}^{N_{shape}-1}\left|J_{j}^{f_{target},h_{in}}\right|^{2}\right)},\nonumber 
\end{eqnarray}

\noindent and the two sides are equal if and only if the two vectors
$\left(\left[p_{j}e^{i\theta_{j}}\right]^{*}\right)_{j\in[[0,N_{shape}-1]]}$
and $(J_{j}^{f_{target},h_{in}})_{j\in[[0,N_{shape}-1]]}$ are linearly
dependant. Thus, the optimal normalised $(p_{j}^{opt})_{j}$ and $(\theta_{j}^{opt})_{j}$
are such that

\begin{equation}
\forall j\quad p_{j}^{opt}=\frac{\left|J_{j}^{f_{target},h_{in}}\right|}{\sqrt{\sum_{j=0}^{N_{shape}-1}\left|J_{j}^{f_{target},h_{in}}\right|^{2}}},\label{eq:arb_h_opt_pi}
\end{equation}

\noindent and

\begin{equation}
\forall j\quad\theta_{j}^{opt}=-\mathrm{Arg}\left(J_{j}^{f_{target},h_{in}}\right).\label{eq:arb_h_opt_thetai}
\end{equation}

In this case, the value for the overlap \ref{eq:objective_overlap}
is

\begin{eqnarray}
R_{f_{target}} & = & \left|\sum_{j=0}^{N_{shape}-1}\left[p_{j}^{opt}\right]^{*}e^{-i\theta_{j}^{opt}}J_{j}^{f_{target},h_{in}}\right|\nonumber \\
 & = & \frac{\sum_{j=0}^{N_{shape}-1}\left|J_{j}^{f_{target},h_{in}}\right|^{2}}{\sqrt{\sum_{j=0}^{N_{shape}-1}\left|J_{j}^{f_{target},h_{in}}\right|^{2}}}\nonumber \\
 & = & \sqrt{\sum_{j=0}^{N_{shape}-1}\left|J_{j}^{f_{target},h_{in}}\right|^{2}}.\label{eq:arb_h_opt_R}
\end{eqnarray}

Note that in the model, we fixed the position of the bin centers $c_{i}$.
This position is actually not important for the limit $N_{shape}\rightarrow\infty$,
but could be optimised for finite-size shaping.\textcolor{red}{{} }In
particular, for small values of $N_{shape}$, one should change the
bin centers for even and odd values of $N_{shape}$. However, for
most situations where $N_{shape}>5$ the difference is not so big,
as verified numerically. We thus keep that hypothesis, even if the
position might require to be fine tuned manually.

\subsection{Translation into partial readouts on the memory}

The $(p_{i}^{2})_{i}$ are the absolute weights for the distribution
of emission probabilities with $\sum_{i=0}^{N_{shape}-1}p_{i}{{}^2}=1$.
As the pieces will be emitted sequentially, it is important to translate
the $(p_{i})_{i}$ into relative weights $(q_{i})_{i}$ that indicate
how much of the remaining amplitude is to be emitted at step $i$.
Then those weights shall be converted into control parameters depending
on what kind of impulsions are used to control emission from the memory.

\subsubsection{Relative amplitudes chain}

We look for a succession of impulsions to retrieve a proportion $q_{i}^{2}$
of the remaining amplitude for every step $i$, and keep $(1-q_{i}^{2})$.

By induction on $N_{shape}\geq1$ we can show that for any $(q_{i})_{0\leq i\leq N_{bins}-1}$
such that $q_{N_{shape}-1}=1$ (requirement that the memory is empty
after last pulse), we get

\[
\sum_{i=0}^{N_{shape}-1}q_{i}^{2}\times\prod_{j<i}(1-q_{j}^{2})=1.
\]

\noindent This gives the absolute amplitudes retrieved at every step,
the $(p_{i})_{i}$, as a function of the relative amplitudes $(q_{i})_{i}$,
namely
\[
\begin{cases}
p_{0}^{2} & =q_{0}^{2}\\
\forall i>0\quad p_{i}^{2} & =q_{i}^{2}\prod_{j<i}(1-q_{j}^{2})
\end{cases}.
\]

\noindent If the $(p_{i})_{i}$ are given and assumed non-zero, we
can compute the $(q_{i})_{i}$ recursively as

\[
\begin{cases}
q_{0}^{2} & =p_{0}^{2}\\
q_{1}^{2} & =\frac{p_{1}^{2}}{(1-q_{0}^{2})}\\
 & \vdots\\
q_{i}^{2} & =\frac{p_{i}^{2}}{\prod_{j<i}(1-q_{j}^{2})}
\end{cases},
\]

\noindent and obtain by induction an explicit conversion formula

\[
\begin{cases}
q_{0}^{2} & =p_{0}^{2}\\
\forall i>0\quad q_{i}^{2} & =\frac{p_{i}^{2}}{1-\sum_{j<i}p_{j}^{2}}
\end{cases}.
\]

\noindent Note that for every $i$, $q_{i}>p_{i}$ as expected. If
there is an index $i$ such that $q_{i}=1$, then all the next weights
will be zero.

\subsubsection{Chain of $\pi$-pulses}

The conversion of relative amplitudes $q_{i}$ into control amplitudes
$\Omega_{i}$ for partial pulses shone onto the atoms depends on the
type of control pulses. For rectangular pulses, the formula was given
in Section \ref{sec:Modelling-a-cavity-asisted}, and corresponds
to Eq.~\ref{eq:partial_pulse}. Furthermore, the optimal phase $\theta_{i}^{opt}$
(see Eq.~\ref{eq:arb_h_opt_thetai}) can be added to each control
pulse. When dealing with real wavepackets no phase is required.

\subsection{Suboptimal efficiency for infinite stretching}

\label{subsec:Suboptimal-efficiency-for}

\subsubsection{Insight from photon shaping}

\begin{figure}
\begin{centering}
\includegraphics[width=1\columnwidth]{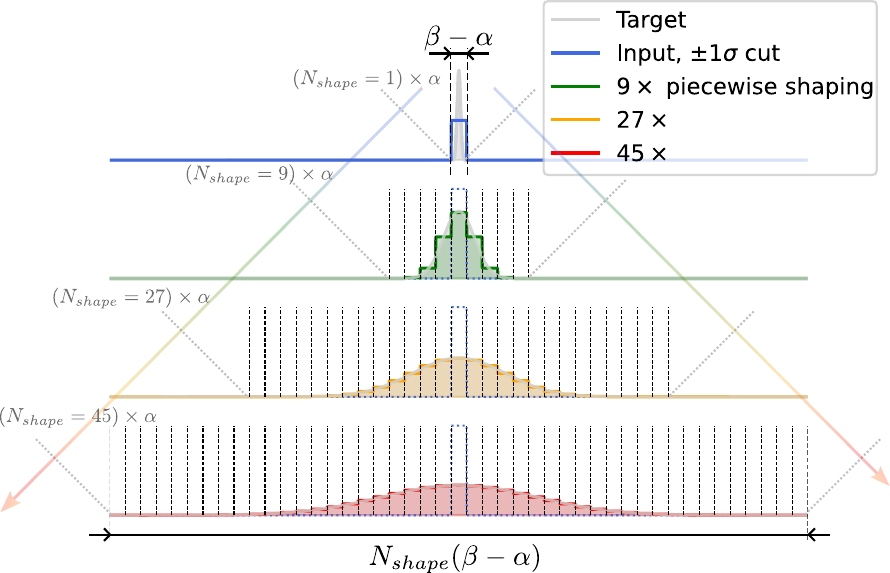}
\par\end{centering}
\caption{Instead of reducing the size of the pieces (rectangular here) used
to build the piecewise approximations, we rather keep the input pieces
constant and study how well they can be used to approximate a L\texttwosuperior -normalised
function $f_{target}$ with growing width (resp. $N_{shape}=9$, $27$
and $45$ stretching factors for green, orange and red curves w.r.t.
the top grey Gaussian).\textcolor{red}{\label{fig:constantBINstretchedENV}}}
\end{figure}

The previous description started from the definition of $f_{target}$
over $[a,b]$ and described the construction of a piecewise function
$g_{pieces}$ from bins that are $N_{shape}$ subdivisions of $[a,b]$.
The bins are filled by copies of $h_{in}$, the width of which is
constrained by the bin size $\frac{b-a}{N_{shape}}$. From now on,
we will rather take the perspective of a photon stretching procedure,
where $h_{in}$ has a fixed temporal width over $[\alpha,\beta]$.
The piecewise function $g_{pieces}$ is then made from $N_{shape}$
weighted copies of $h_{in}$. The aim is to set the weights to maximise
the overlap with $f_{target}$ whose width can be increased (see Fig.~\ref{fig:constantBINstretchedENV}).

\subsubsection{Convergence towards infinity}

\label{subsec:Convergence-towards-infinity}

We use our shaping scheme by carving a longer envelope with successive
echoes. It turns out that when looking at the limit of a target much
longer than the input, one gets an asymptotic value for the overlap.
We hereby provide a proof for this limit, restricting to a case were
we assume that $f_{target}$ is real-valued and that the input $h_{in}$
is real-valued and has a constant sign, let's say positive\textbf{.
}Thanks to this assumption, we can use the generalised mean value
theorem to obtain that for every $j$, there exists a $\tau_{j}\in\left[c_{j}-\frac{b-a}{2N_{shape}};c_{j}+\frac{b-a}{2N_{shape}}\right]$
such that

\begin{eqnarray*}
J_{j}^{f_{target},h_{in}} & = & \int_{c_{j}-\frac{b-a}{2N_{shape}}}^{c_{j}+\frac{b-a}{2N_{shape}}}f_{target}(t)h_{in}(t-c_{j})\mathrm{d}t\\
 & = & f_{target}(\tau_{j})\times\int_{c_{i}-\frac{b-a}{2N_{shape}}}^{c_{i}+\frac{b-a}{2N_{shape}}}h_{in}(t-c_{j})\mathrm{d}t\\
 & = & f_{target}(\tau_{j})\times\int_{-\frac{b-a}{2N_{shape}}}^{+\frac{b-a}{2N_{shape}}}h_{in}(t)\mathrm{d}t\\
 & = & f_{target}(\tau_{j})\times\frac{b-a}{N_{shape}}\\
 &  & \qquad\qquad\times\left(\frac{N_{shape}}{b-a}\int_{-\frac{b-a}{2N_{shape}}}^{+\frac{b-a}{2N_{shape}}}h_{in}(t)\mathrm{d}t\right)\\
 & =: & f_{target}(\tau_{j})\times\frac{b-a}{N_{shape}}\times\hat{h}_{in}(0)
\end{eqnarray*}

\noindent where we denote $\hat{h}_{in}(0)$ the mean value of $h_{in}$
over its support.

Then the optimal overlap from Eq.~\ref{eq:arb_h_opt_R} involves
a set of $(\tau_{j})_{j\in[0,N_{shape}-1]}$ where $\forall j\quad\tau_{j}\in\left[c_{j}-\frac{b-a}{2N_{shape}};c_{j}+\frac{b-a}{2N_{shape}}\right]$
such that

\begin{widetext}

\begin{eqnarray}
R_{f_{target}}^{opt}(p^{opt},h_{in})^{2} & = & \sum_{j=0}^{N_{shape}-1}\left|J_{j}^{f_{target},h_{in}}\right|^{2}\nonumber \\
 & = & \hat{h}_{in}(0)^{2}\times\left(\frac{b-a}{N_{shape}}\right)^{2}\times\sum_{j=0}^{N_{shape}-1}\left|f_{target}(\tau_{j})\right|^{2}\label{eq:R_opt_hf}\\
 & = & \hat{h}_{in}(0)^{2}\times\frac{b-a}{N_{shape}}\times\sum_{j=0}^{N-1}\int_{c_{j}-\frac{b-a}{2N_{shape}}}^{c_{j}+\frac{b-a}{2N_{shape}}}\left|f_{target}(\tau_{j})\right|^{2}\mathrm{d}t.\nonumber 
\end{eqnarray}

\end{widetext}

Now, as illustrated in Fig.~\ref{fig:constantBINstretchedENV}, we
keep the width of $h_{in}$ constant, while the profile $f_{target}$
is stretched. As such, we take $a=N_{shape}\alpha$ and $b=N_{shape}\beta$
and $N_{shape}$ bins centered at $c_{i}$. We further define $c_{i}^{(0)}=\frac{c_{i}}{N_{shape}}$.
Besides we define a reference profile $\psi:[\alpha,\beta]\rightarrow\mathbb{R}$
such that $\int_{\alpha}^{\beta}\psi^{2}=1$ and define a sequences
of growing target profiles

\begin{align*}
f_{target,N_{shape}}: & \left[N_{shape}\alpha;N_{shape}\beta\right]\rightarrow\mathbb{R}\\
 & t\mapsto\frac{1}{\sqrt{N_{shape}}}\psi\left(\frac{t}{N_{shape}}\right).
\end{align*}
By means of a change of variables, one can check that $\forall N_{bins}\quad\int_{a=N_{shape}\alpha}^{b=N_{shape}\beta}f_{target,N_{shape}}^{2}=1$.
We write $R_{N_{shape}}$ for the optimal overlap $R_{f_{target,N_{shape}}}^{opt}(p^{opt},h_{in})$
of Eq.~\ref{eq:R_opt_hf}. 

Let $\varepsilon>0$ and as $\psi^{2}$ is uniformly continuous over
$[\alpha;\beta],$ let $\delta>0$ such that $\forall x,y\quad\left|x-y\right|\leq\delta\Rightarrow\left|\psi^{2}(x)-\psi^{2}(y)\right|\leq\varepsilon$.
Then 

\begin{widetext}

\begin{eqnarray*}
 &  & \left|\sum_{i=0}^{N_{shape}-1}\int_{N_{shape}c_{i}^{(0)}-\frac{\beta-\alpha}{2}}^{N_{shape}c_{i}^{(0)}+\frac{\beta-\alpha}{2}}f_{target,N_{shape}}(\tau_{i})^{2}\mathrm{d}t-\int_{a=N_{shape}\alpha}^{b=N_{shape}\beta}f_{target,N_{shape}}(t)^{2}\mathrm{d}t\right|\\
 & = & \frac{1}{N_{shape}}\left|\sum_{i=0}^{N_{shape}-1}\int_{N_{shape}c_{i}^{(0)}-\frac{\beta-\alpha}{2}}^{N_{shape}c_{i}^{(0)}+\frac{\beta-\alpha}{2}}\psi\left(\frac{\tau_{i}}{N_{shape}}\right)^{2}\mathrm{d}t-\int_{a=N_{bins}\alpha}^{b=N_{bins}\beta}\psi\left(\frac{t}{N_{shape}}\right)^{2}\mathrm{d}t\right|\\
 & \leq & \frac{1}{N_{shape}}\sum_{i=0}^{N_{shape}-1}\int_{N_{shape}c_{i}^{(0)}-\frac{\beta-\alpha}{2}}^{N_{shape}c_{i}^{(0)}+\frac{\beta-\alpha}{2}}\left|\psi\left(\frac{\tau_{i}}{N_{shape}}\right)^{2}-\psi\left(\frac{t}{N_{shape}}\right)^{2}\right|\mathrm{d}t\\
 & \leq & \frac{1}{N_{shape}}\sum_{i=0}^{N_{shape}-1}\int_{N_{shape}c_{i}^{(0)}-\frac{\beta-\alpha}{2}}^{N_{shape}c_{i}^{(0)}+\frac{\beta-\alpha}{2}}\varepsilon\mathrm{d}t\\
 & = & \frac{1}{N_{shape}}\times N_{shape}\times(\beta-\alpha)\times\varepsilon,
\end{eqnarray*}

\noindent where the first inequality stems from triangular inequality,
and the last inequality holds if $N_{shape}$ is big enough (in that case
$\left|\frac{\tau_{i}}{N_{shape}}-\frac{t}{N_{shape}}\right|\leq\frac{\beta-\alpha}{N_{shape}}\leq\delta$).
As such, if $N_{shape}$ is big enough one gets

\[
\left|\frac{1}{\hat{h}_{in}(0)^{2}(\beta-\alpha)}R_{N_{shape}}^{2}-\int_{\alpha}^{\beta}\psi(t)^{2}\mathrm{d}t\right|=\left|\sum_{i=0}^{N_{shape}-1}\int_{c_{i}^{(0)}-\frac{\beta-\alpha}{2N_{shape}}}^{c_{i}^{(0)}+\frac{\beta-\alpha}{2N_{shape}}}f_{target}(\tau_{i})^{2}\mathrm{d}t-\int_{a=N_{shape}\alpha}^{b=N_{shape}\beta}f_{target,N_{shape}}(t)^{2}\mathrm{d}t\right|\leq\varepsilon.
\]

\noindent That is, as we consider positive quantities

\begin{equation}
R_{N_{shape}}\xrightarrow[N_{shape}\rightarrow+\infty]{}\sqrt{\beta-\alpha}\times\underset{=\hat{h}_{in}(0)}{\underbrace{\left(\frac{1}{\beta-\alpha}\int_{\alpha}^{\beta}h_{in}(t)\mathrm{d}t\right)}}\times\underset{=1}{\underbrace{\sqrt{\int_{\alpha}^{\beta}\psi(t)^{2}\mathrm{d}t}}}=\frac{1}{\sqrt{\beta-\alpha}}\int_{\alpha}^{\beta}h_{in}(t)\mathrm{d}t.\label{eq:overlap_limit_RN}
\end{equation}

\end{widetext}

\noindent To sum up, when $N_{shape}\rightarrow\infty$, the overlap
$R_{N_{shape}}$ has a limit which only depends on the memory input $h_{in}$ (also assumed to be the unshaped output) and not on $f_{target}$. From the left-hand side of equality
\ref{eq:overlap_limit_RN}, we see that, given a bin size $(\beta-\alpha)$, the
limit of the overlap is optimal when the mean value of the input profile
$h_{in}$ is maximised. From Cauchy-Schwarz inequality, this corresponds
to $h_{in}:t\mapsto\frac{1}{\sqrt{\beta-\alpha}}$, and then $R_{N_{shape}}\rightarrow1$. However, the shape $h_{in}$ is usually given before the shaping. As it is studied in paragraph~\ref{optimisation_cropped}, the optimisation is rather to be done on the choice of $\alpha$ and $\beta$.

Moreover, it turns out that the convergence \ref{eq:overlap_limit_RN} is quite
fast. As expected for Riemann-sum-like results, it actually goes as
$\frac{1}{N_{shape}^{2}}$. We observe numerically that for targets
about $10$ times longer than the input, the overlap is already less
than $1$~\% away from the theoretical limit.

As a side remark, we conjecture that a similar result stands when
a real-valued $h_{in}$ does not keep a constant sign, by focusing
on different slices of its natural domain where it does. Moreover,
we point out that our proof was inspired by demonstrations of Riemann-Lebesgue
lemma and Fejér's theorem \citep{fejer_lebesguessche_1910,young_introduction_1988},
that might also provide a more direct way to the result. From this
perspective, $g_{pieces}$, a weighted regular juxtaposition of $h_{in}$
kernels, appears as a modulated periodic function, that somehow samples
the wider $f_{target}$ as a comb of identity approximations when $N_{shape}\rightarrow\infty$.

\begin{widetext}

\subsubsection{Optimisation of the limit: cropped-echo technique}
\label{optimisation_cropped}

We now allow $\int_{\alpha}^{\beta}h_{in}(t)^{2}\mathrm{d}t\leq1$
by defining $h_{in}$ as a L\texttwosuperior -normalised function
over a broader support $[\alpha';\beta']$, and cutting it between
$\alpha$ and $\beta$ (that is $\alpha'\leq\alpha<\beta\leq\beta'$)
so as to maximise the quantity \textbf{$\frac{1}{\sqrt{\beta-\alpha}}\int_{\alpha}^{\beta}h_{in}(t)\mathrm{d}t$}.
This is what we call the cropped-echo technique. For instance, let
us take $\alpha'\rightarrow-\infty$ and $\beta'\rightarrow+\infty$
and consider a L\texttwosuperior -normalised Gaussian over $\mathbb{R}$
with standard deviation $\sigma$ (that is fixed). Then we choose $\alpha=-M\sigma$ and $\beta=M\sigma$ with a varying width
ratio $M$ to define $h_{in}$ as the restriction of Gaussian to $[\alpha,\beta]$ (thus not normalised), as illustrated in Fig.~\ref{fig:shaping_and_cutoff}. The resulting overlap
limit \ref{eq:overlap_limit_RN} related to the cut of this Gaussian is

\begin{eqnarray}
\frac{1}{\sqrt{2M\sigma}}\int_{-M\sigma}^{M\sigma}\frac{1}{(\pi\sigma^{2})^{1/4}}e^{-t^{2}/2\sigma^{2}}\mathrm{d}t & = & \frac{1}{\pi^{1/4}\sqrt{2M}\sigma}\times\sqrt{2}\sigma\int_{-M/\sqrt{2}}^{M/\sqrt{2}}e^{-u^{2}}\mathrm{d}u\text{ with the change of variables \ensuremath{u=\frac{t}{\sigma\sqrt{2}}}}\nonumber \\
 & = & \frac{1}{\pi^{1/4}\sqrt{M}}\times\sqrt{\pi}\times\frac{2}{\sqrt{\pi}}\int_{0}^{M/\sqrt{2}}e^{-u^{2}}\mathrm{d}u\nonumber \\
 & = & \frac{\pi^{1/4}}{\sqrt{M}}\mathrm{erf}\left(\frac{M}{\sqrt{2}}\right),\label{eq:Merf_limit_demo}
\end{eqnarray}

\noindent which is represented in Fig.~\ref{fig:ErfPrediction}.
It is seen that the overlap is optimal around $M\approx1.4$, reaching
about $0.94$. For an exponential input, $h_{in}:t\mapsto\sqrt{2}e^{-t}\mathds{1}_{x\geq0}$,
and truncation between $\alpha=0$ and $\beta=M$, we obtain the truncation
value limit \ref{eq:overlap_limit_RN} $\sqrt{\frac{2}{M}}(1-e^{-M})$.
The optimal value is about $0.9$, found for $M\approx1.26$.

\begin{figure*}
\begin{centering}
\includegraphics[width=0.33\textwidth]{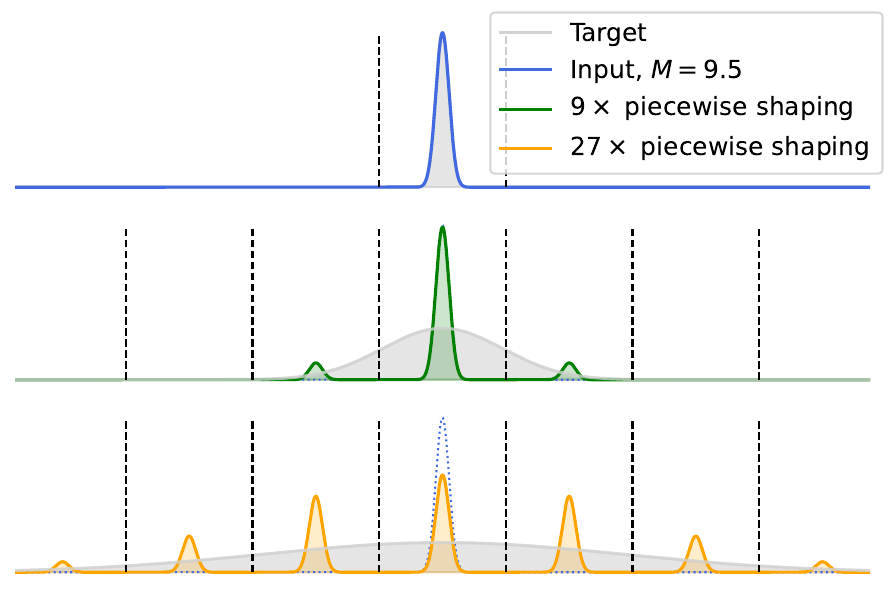}\includegraphics[width=0.33\textwidth]{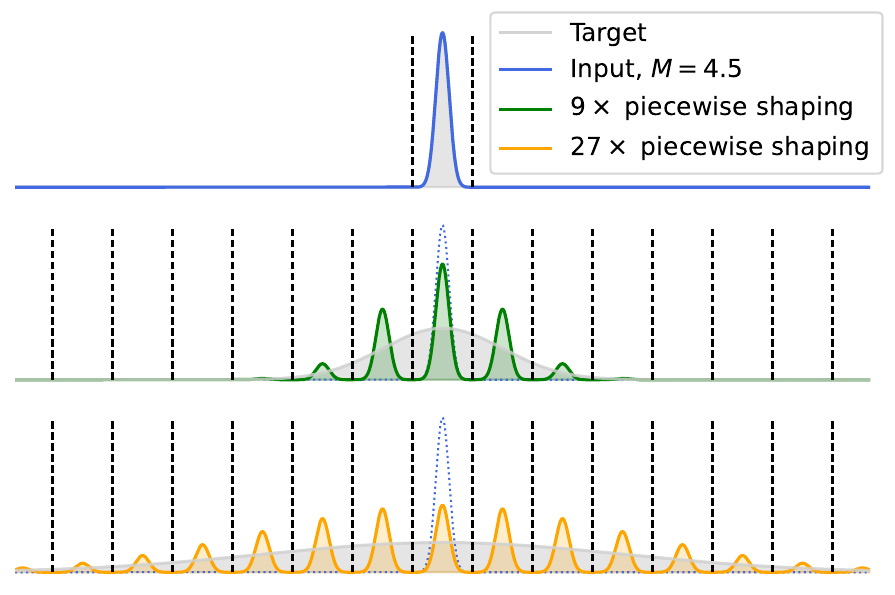}\includegraphics[width=0.33\textwidth]{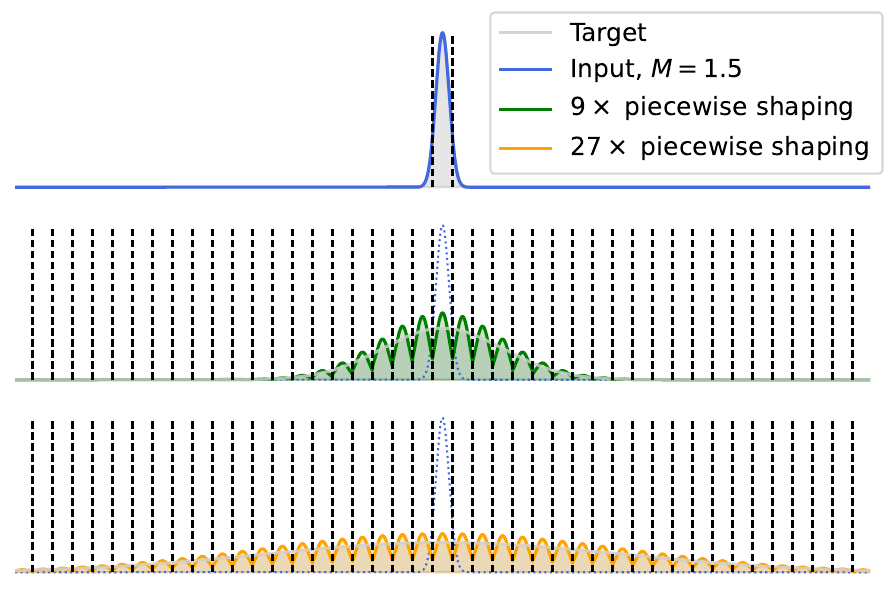}
\par\end{centering}
\caption{For a Gaussian input profile (in blue on top), we apply different
cutoff factors ($M=9.5$, $4.5$ and $1.5$ from left to right, bins
are delimited by the dashed vertical black lines) to define the piece
$h_{in}$ used to build the piecewise function $g_{pieces}$ (in green
and in orange) that is used to match the target $f_{target}$ (in
grey) using optimal coefficients from Eq.~\ref{eq:arb_h_opt_pi}.
This $f_{target}$ is a L\texttwosuperior -normalised Gaussian function
chosen $9$ times (middle) or $27$ times (bottom) as wide as the
initial uncropped Gaussian input. \label{fig:shaping_and_cutoff}}
\end{figure*}

Crucially, we note that the cropped-echo technique gives nice results
on AFC simulations, as illustrated in the main text. In this section,
the mathematical formulation dealt with independent time bins of weighted
copies of $h_{in}$. When implementing that on an AFC quantum memory,
the late cropped-part of an echo readout stays in the memory and may
interfere with later partial readouts and thus change the shape of
the resulting echoes. Such interferences could be observed, but did
not happen to be detrimental to the efficiencies.

\smallskip{}
Finally, we recall that in Fig.~3 of the main text, we evaluate the
efficiency of the shaping process in two steps. First, the loss of
energy is considered, as it matters for the whole network entanglement
generation rate (see Appendix~\ref{sec:Quantum-network-architecture}).
In our mathematical context, loss of energy due to the memory (see
Eq.~\ref{eq:1st_echo_eff}) can also be taken into account with some
undernormalised $h_{in}$. The losses due to the left and right crops
are $\int_{-\infty}^{-M\sigma}\left|h_{in}(t)\right|^{2}\mathrm{d}t$
and $\int_{M\sigma}^{+\infty}\left|h_{in}(t)\right|^{2}\mathrm{d}t$.
Second, the overlap with the target shape is taken renormalised with
respect to the losses, that is by considering

\[
\int\frac{\mathcal{E}_{out}}{\sqrt{\eta_{out}}}\cdot f_{target}^{*},
\]

\noindent where $\eta_{out}$ gives the total energy of the shaped
echo (integral bounds defined accordingly). In our context, $\mathcal{E}_{out}$
corresponds to $g_{pieces}$ and $\eta_{out}$ is given by 
\begin{equation}
\int_{a}^{b}\left|g_{pieces}\right|{{}^2}(t)\mathrm{d}t=\sum_{j=0}^{N_{shape}-1}p_{j}{{}^2}\int_{c_{j}-M\sigma}^{c_{j}+M\sigma}\left|h_{in}\right|{{}^2}(t-c_{j})\mathrm{d}t=\int_{-M\sigma}^{M\sigma}\left|h_{in}\right|{{}^2}(t)\mathrm{d}t,\label{eq:remaining_energy_gpieces}
\end{equation}

\noindent so that the renormalised asymptotic overlap \ref{eq:overlap_limit_RN}
in the Gaussian case is

\begin{eqnarray}
\frac{1}{\sqrt{\int_{-M\sigma}^{M\sigma}\left|\frac{1}{(\pi\sigma^{2})^{1/4}}e^{-t^{2}/2\sigma^{2}}\right|^{2}\mathrm{d}t}}\times\frac{\pi^{1/4}}{\sqrt{M}}\mathrm{erf}\left(\frac{M}{\sqrt{2}}\right) & = & (\pi\sigma^{2})^{1/4}\frac{1}{\sqrt{\sigma}\pi^{1/4}\sqrt{\mathrm{erf}\left(M\right)}}\times\frac{\pi^{1/4}}{\sqrt{M}}\mathrm{erf}\left(\frac{M}{\sqrt{2}}\right)\nonumber \\
 & = & \pi^{1/4}\times\frac{\mathrm{erf}\left(\frac{M}{\sqrt{2}}\right)}{\sqrt{M\mathrm{erf}\left(M\right)}}.\label{eq:renormalised_overlap}
\end{eqnarray}

\noindent This renormalised overlap matters for the fidelity of the
entanglement generated by the whole network (again, see Appendix~\ref{sec:Quantum-network-architecture}).
Both quantities \ref{eq:remaining_energy_gpieces} and \ref{eq:renormalised_overlap}
are plotted in Fig.~\ref{fig:ErfPrediction}. The former increases
monotonously with $M$, while latter seems to decrease monotonously
with $M$, and has the limit $1$ when $M$ goes to $0$. A nice trade-off
the two turns out to be given by the optimal value of the non-renormalised
overlap \ref{eq:Merf_limit_demo}. For $M\approx1.4$, the remaining
energy is about $0.95$ so that the renormalised overlap is almost
$0.97$. Realistic AFC simulation outputs are consistent but slightly
different from these values, as shown in Fig.~3 of the main text.

\begin{figure}
\begin{centering}
\includegraphics[width=0.7\columnwidth]{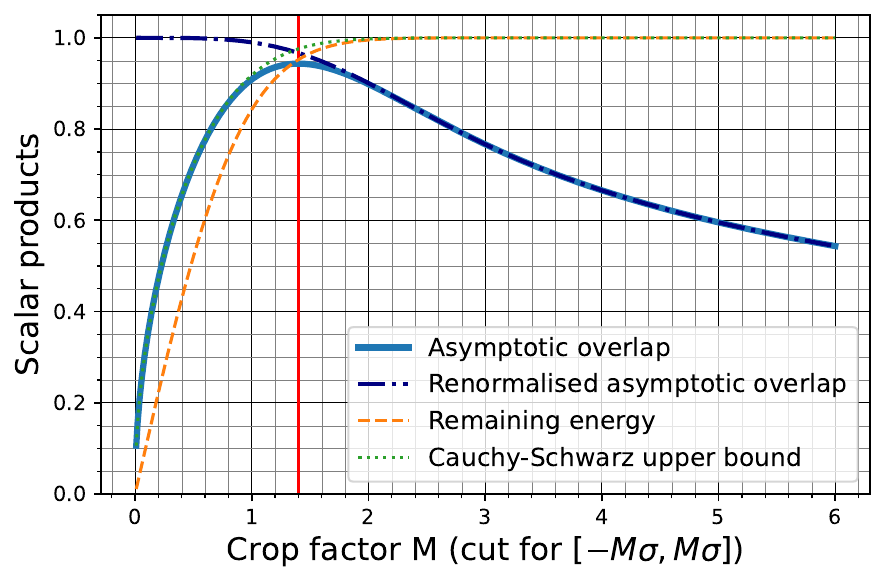}
\par\end{centering}
\caption{For Gaussian input profiles, the predicted asymptotic overlaps (Eq.~\ref{eq:Merf_limit_demo}
in blue, and renormalised \ref{eq:renormalised_overlap} in deep blue,
dash-dotted) are plotted as a function of the crop factor $M$. The
non-renormalised overlap features an optimum slightly below $M\approx1.4$
(vertical red line), about $94\%$. The Cauchy-Schwarz upper-bound
$\left|\frac{1}{\beta-\alpha}\protect\int_{\alpha}^{\beta}h_{in}(t)\times1\mathrm{d}t\right|\protect\leq\sqrt{\protect\int_{\alpha}^{\beta}h_{in}(t)^{2}\mathrm{d}t}$
is given by the green dotted line. The remaining energy after crop
$\protect\int_{\alpha}^{\beta}h_{in}(t)^{2}\mathrm{d}t$ is given
by the orange dashed line. At the former optimal point, one still
finds $95\%$ of the initial energy in the cut profile.\label{fig:ErfPrediction}}
\end{figure}

\end{widetext}

~

\clearpage
\newpage

\section{Coherence of pure single-photon state output}

\label{subsec:PhotonStateCoherence}

We showed in Section~\ref{subsec:Absorption-and-impedance-matchin}
that, in the impedance-matching regime, the full device efficiency
of the cavity-assisted AFC can reach $\eta_{1st-echo}=1$ (cf. Eq.~\ref{eq:1st_echo_eff}).
In Section~\ref{sec:Protocol-for-photon-shaping}, we argued that
a purposefully chosen sequence of partial readouts results in a reshaping
of the photon wavepacket, with a high efficiency. On top of that efficiency
and that shape, we are interested in the properties of the output
photon. Namely, we would like to make sure that purity is preserved,
i.e. the quantum state of light emitted by the memory, after a pure
single-photon state was absorbed, can be described by a pure single-photon
state, at least conditioned on success of reemission. Such an outcome is expected since getting the post-selected state does not involve tracing over an environment of any kind. As such, we give a technique to retrieve the coherent state description from the resolutions of the memory Heisenberg equations, as done in Appendix~\ref{sec:FromOperatorsToScalars} and Section~\ref{subsec:Application-to-the_correlators}.

\subsection{Need for a quantum state description}

To start with, we underline that for our model, derivations could
have been performed in the Schrödinger picture, by considering the
evolution of state coordinates with time. For similar systems of equations,
especially leveraging the one-excitation limit, it is for instance
done in Refs.~\citep{gardiner_quantum_2015,motzoi_precise_2018,kiilerich_input-output_2019,morin_deterministic_2019,bernad_analytical_2024}.
However, we chose so far to solve a set of Heisenberg-Langevin equations
in the Heisenberg picture, as it seems to be the natural framework
for introducing input-output relations as well as effects of quantum
noise and damping. In the context of this Letter, we also seek to
compute interferometric quantities that would involve both the photon
coming from the memory and from an ion node in the full network of
Fig.~1 in the main text. For the ion, and for the Hong-Ou-Mandel experiment
presented in Sec.~\ref{sec:Ion-Photon-state-and}, a description in
terms of quantum states rather than quantum operators was used in
the Literature \citep{meraner_indistinguishable_2020,krutyanskiy_entanglement_2023}.
Hence, we shall find a way of translating our previous derivations
that dealt with quantum operators into a Schrödinger--like description
of the output photon in terms of quantum state.\textcolor{red}{{} }

\subsection{Extracting a quantum state description from the Heisenberg picture}

In the Heisenberg picture, the state of the whole system is not meant
to change: only operators evolve with time. Still, the interaction
between the input photon and the memory can be regarded as a scattering
process~\citep{lehmann_zur_1955,legero_characterization_2006,fan_input-output_2010,raimond_atoms_2013}.
Describing the output state then amounts to factorising the Hilbert
space in another way and to performing a change of basis. We start
at time $t_{0}$ with the decomposition between the memory (cavity
and atoms), the light environment (modes that carry a possible incoming
photon) and the other reservoirs (baths for quantum noise) as

\begin{equation}
\mathcal{H}=\mathcal{H}_{mem}(t_{0})\otimes\mathcal{H}_{light-env}(t_{0})\otimes\mathcal{H}_{baths}(t_{0}).\label{eq:factorisation_init}
\end{equation}

\noindent In the Schrödinger picture, evolution happens from time
$t_{0}$ to $t_{f}$ within the zero and one-excitation eigenspaces
of the total number operator $\hat{N}_{tot}$ (cf. Eq.~\ref{eq:Ntot_excitationnumber}),
while in the Heisenberg picture a convenient basis to describe the
(initial at time $t_{0}$) state of the system is

\begin{widetext}

\begin{equation}
\mathcal{B}_{t_{0}}:=\left\{ \hat{\mathcal{E}}^{\dagger}(t_{0})|0\rangle\right\} \cup\left\{ \hat{P}_{k}^{\dagger}(t_{0})|0\rangle,\hat{S}_{k}^{\dagger}(t_{0})|0\rangle\right\} _{1\leq k\leq N}\cup\left\{ \hat{\mathcal{E}}_{in}^{\dagger}(t)|0\rangle\right\} _{t\in\mathbb{R}}\cup\left\{ \hat{F}_{P,k}^{\dagger}(t)|0\rangle\right\} _{1\leq k\leq N,t\in\mathbb{R}}\cup\left\{ \hat{F}_{S,k}^{\dagger}(t)|0\rangle\right\} _{1\leq k\leq N,t\in\mathbb{R}},\label{eq:base_t0}
\end{equation}

\noindent where $|0\rangle$ is again the vacuum state shared by all
the subsystems. Note that we include all the set of quantum noise operators
to take into account the excitations that would be within the reservoirs
at time $t_{0}$. As detailed in Sec.~\ref{sec:Quantum-reservoirs},
the label $t$ in the noise operators does not refer to any time evolution
but rather to a change of basis in the description of the bath modes.
For example, the light environment is either fully described by the
set of free space modes at time $t_{0}$ $\hat{c}_{light-env}(\omega,t_{0})^{\dagger}$
for all frequencies $\omega$, or by the set of input modes $\hat{\mathcal{E}}_{in}(t)=-\frac{i}{\sqrt{2\pi}}\int e^{-i\omega(t-t_{0})}\hat{c}_{light-env}(\omega,t_{0})\mathrm{d}\omega$
for all times $t$. The input state is defined within this basis,
and typically

\begin{eqnarray}
\rho_{ph,input} & = & |0\rangle_{mem}\langle0|\otimes\left(\iint\mathcal{E}_{in}(t)\mathcal{E}_{in}^{*}(t')\hat{\mathcal{E}}_{in}^{\dagger}(t)|0\rangle_{light-env}\langle0|\hat{\mathcal{E}}_{in}(t')\mathrm{d}t\mathrm{d}t'\right)\otimes|0\rangle_{baths}\langle0|\nonumber \\
 & = & \iint\mathcal{E}_{in}(t)\mathcal{E}_{in}^{*}(t')\hat{\mathcal{E}}_{in}^{\dagger}(t)|0\rangle\langle0|\hat{\mathcal{E}}_{in}(t')\mathrm{d}t\mathrm{d}t'=:|\psi_{input}\rangle\langle\psi_{input}|,\label{eq:rho_phinput}
\end{eqnarray}

\noindent as discussed for Eq.~\ref{eq:psi_input} and paragraph \ref{par:Particular-case:syst_assumptions}.

In order to represent the whole system standing at $t_{f}$ after
interaction, we rather use the more suitable basis

\begin{equation}
\mathcal{B}_{t_{f}}:=\left\{ \hat{\mathcal{E}}^{\dagger}(t_{f})|0\rangle\right\} \cup\left\{ \hat{P}_{k}^{\dagger}(t_{f})|0\rangle,\hat{S}_{k}^{\dagger}(t_{f})|0\rangle\right\} _{1\leq k\leq N}\cup\left\{ \hat{\mathcal{E}}_{out}^{\dagger}(t)|0\rangle\right\} _{t\in\mathbb{R}}\cup\mathcal{B}_{bath,t_{f}},\label{eq:base_tf}
\end{equation}

\end{widetext}

\noindent where we include the memory operators at time $t_{f}$,
the light environment operators $\hat{\mathcal{E}}_{out}^{\dagger}(t)$
for all labels $t$ (or equivalently mode operators $\hat{c}_{light-env}^{\dagger}(\omega,t_{f})$
for all frequencies $\omega$) and a suitable basis build from other
baths creation operators at time $t_{f}$. This change in representations
amounts to a formal refactorisation of the Hilbert space \ref{eq:factorisation_init}
as

\begin{equation}
\mathcal{H}=\mathcal{H}_{mem}(t_{f})\otimes\mathcal{H}_{light-env}(t_{f})\otimes\mathcal{H}_{baths}(t_{f}).\label{eq:factorisation_final}
\end{equation}

Once the stance of decomposition from Eq.~\ref{eq:base_tf} is taken,
the state of the output photon alone is obtained by tracing out other
degrees of freedom (the memory and the other reservoirs). From basis
$\mathcal{B}_{t_{f}}$, it corresponds to tracing out along all basis
vectors except the set $\left\{ \hat{\mathcal{E}}_{out}^{\dagger}(t)|0\rangle\right\} _{t\in\mathbb{R}}$.
It is then clear that a suitable description for the photon state
is

\begin{eqnarray}
\rho_{ph,output} & := & \mathrm{Tr}_{\mathcal{B}_{t_{f}}(mem,baths)}(\rho_{ph,input})\label{eq:rho_phoutput}\\
 & = & (1-\eta_{out})|0\rangle\langle0|+\eta_{out}\rho_{out},\nonumber 
\end{eqnarray}

\noindent where $\eta_{out}$ is the efficiency that embraces all
possible retrieval times and and $\rho_{out}$ is a normalised density
matrix that is described in the subspace generated by the output states
$\left\{ \hat{\mathcal{E}}_{out}^{\dagger}(t)|0\rangle\right\} _{t\in\mathbb{R}}$
only. The vacuum component in the mixed state \ref{eq:rho_phoutput}
is a consequence of the fact that we work in the one-excitation subspace:
the excitation is always present in the whole system, be that in the
output field, or in the other parts (memory or other reservoirs).
The latter are the ones that contribute to the vacuum component. As
such, $\rho_{out}$ is obtained from the order-two correlators of
output field operators of Sec.~\ref{subsec:Application-to-the_correlators},
namely

\begin{widetext}

\begin{eqnarray}
\eta_{out}\rho_{out} & = & \iint\langle0|\mathcal{\hat{E}}_{out}(t')\rho_{ph,output}\hat{\mathcal{E}}_{out}^{\dagger}(t)|0\rangle\mathcal{\hat{E}}_{out}^{\dagger}(t)|0\rangle\langle0|\mathcal{\hat{E}}_{out}(t')\mathrm{d}t\mathrm{d}t'\nonumber \\
 & = & \iint\mathrm{Tr}\left(\mathcal{\hat{E}}_{out}^{\dagger}(t)\hat{\mathcal{E}}_{out}(t')\rho_{ph,input}\right)\mathcal{\hat{E}}_{out}^{\dagger}(t)|0\rangle\langle0|\mathcal{\hat{E}}_{out}(t')\mathrm{d}t\mathrm{d}t'\nonumber \\
 & = & \iint\mathcal{E}_{out}^{*}(t)\mathcal{E}_{out}(t')\mathcal{\hat{E}}_{out}^{\dagger}(t)|0\rangle\langle0|\mathcal{\hat{E}}_{out}(t')\mathrm{d}t\mathrm{d}t'\label{eq:eq_rho_out}
\end{eqnarray}

\end{widetext}

\noindent where we used the fact that we work within the one-excitation
subspace to go from the first equality to the second one, which in
turn corresponds to Eq.~\ref{eq:numerical_out} of Sec.~\ref{sec:FromOperatorsToScalars}
and uses the hypothesis of an initially empty memory as in Eq.~\ref{eq:rho_phinput}.
Hence, $\rho_{out}$ has the structure of a pure state in the output
flying qubit basis so we write

\begin{equation}
\rho_{ph,output}=(1-\eta_{out})|0\rangle\langle0|+\eta_{out}|\psi_{output}\rangle\langle\psi_{output}|,\label{eq:rho_ph_output_complet}
\end{equation}

\noindent and $|\psi_{out}\rangle$ has the temporal waveform $t\mapsto\mathcal{\mathcal{E}}_{out}(t)$
over output modes $\hat{\mathcal{E}}_{out}^{\dagger}$, i.e.

\begin{equation}
|\psi_{output}\rangle=\frac{1}{\sqrt{\eta_{out}}}\int\mathcal{E}_{out}(t)\hat{\mathcal{E}}_{out}^{\dagger}(t)|0\rangle\mathrm{d}t.\label{eq:psi_phout}
\end{equation}

Finally, Eq.~\ref{eq:psi_phout} shows that $\eta_{out}$ is obtained
from the scalar equations by computing

\begin{equation}
\eta_{out}=\int\left|\mathcal{E}_{out}(t)\right|^{2}\mathrm{d}t.\label{eq:eta_out}
\end{equation}

\noindent According to this result, the whole conditional output mode
is pure. In particular, the coherence spans over the first reflection
and the later echoes. Similarly, the shaping process that involves
partial control pulses keeps the purity, as long as the control pulses
are coherent with one another. Moreover, the dephasing losses (rates
$\gamma_{P}$ and $\gamma_{S}$) only affect the efficiency of the
whole process and do not break the coherence of the output photon.
On the contrary, they rather increase the value of $\eta_{out}$.\\

\clearpage
\newpage

\section{Experimental parameters}

\label{subsec:Experimental-parameters}

In this Section, we explain how the memory parameters used in the
main text were chosen. More precisely, we provide a way of translating
our model parameters such as $g\sqrt{N}$,$\kappa$ and $\Gamma$
into quantities that can be assessed and measured experimentally\footnote{With a similar spirit to the one of Appendix~I of Ref.~\citep{morin_deterministic_2019}.}.
Some connections with the notations of some references that introduce
similar models are also discussed. We summarise our choices in Table
\ref{tab:Parameters-choices.}.

\begin{table*}
\begin{centering}
\begin{tabularx}{1\textwidth}{|>{\raggedleft\arraybackslash}X|>{\centering\arraybackslash}X|>{\centering\arraybackslash}X|>{\centering\arraybackslash}X|}
\hline 
\multirow{1}{=}{\textbf{Parameters}} & \textbf{Designation} & \multirow{1}{=}{\textbf{Value}} & \textbf{Reference}\tabularnewline
\hline 
\hline 
Cavity decay rate & $\kappa$ & $2\pi\times55\mathrm{MHz}$ & Eq.~\ref{eq:kappa_exp}\tabularnewline
\hline 
Inhomogeneous width & $\Gamma$ & $2\pi\times4$ MHz & Eq.~\ref{eq:Gamma_exp}\tabularnewline
\hline 
Collective coupling & $g\sqrt{N}$ & \begin{cellvarwidth}[t]
\centering
$2\pi\times8.4$ MHz

(corresponds to $\tilde{d}=0.45$ in free space)
\end{cellvarwidth} & Eq.~\ref{eq:g2N_cav}\tabularnewline
\hline 
Number of teeth & $N_{teeth}$ & $67$ & Eq.~\ref{eq:N_teeth_exp}\tabularnewline
\hline 
Comb step & $\Delta$ & $2\pi\times61$ kHz & Eq.~\ref{eq:Delta_exp}\tabularnewline
\hline 
Tooth width & $\gamma_{tooth}$ & $2\pi\times1$ kHz & Eq.~\ref{eq:gamma_tooth_exp}\tabularnewline
\hline 
Optical transition decay & $\gamma_{P}$ & $0$ & Eq.~\ref{eq:gammaPS_exp}\tabularnewline
\hline 
Spin transition decay & $\gamma_{S}$ & $0$ & Eq.~\ref{eq:gammaPS_exp}\tabularnewline
\hline 
\end{tabularx}
\par\end{centering}
\caption{Parameters chosen to feed the model, based on achievable experimental
values.\label{tab:Parameters-choices.}}

\end{table*}

\subsection{Frequency units}

First, we stress that by convention it is understood that parameters
such as $\kappa$, $\Gamma$ and $g\sqrt{N}$ are given in $\mathrm{rad.s^{-1}}$.
As such, the value in $\mathrm{rad.s^{-1}}$ is equal to the value
in $\mathrm{Hz}$ multiplied by $2\pi$. We will consistently highlight
the latter factor, by writing for instance $\Gamma=2\pi\times4$ MHz
to refer to the value in $\mathrm{rad.s}^{-1}$ ($25$~$\mathrm{rad.s}^{-1}$
approximately).

Depending on the references, some values can be found in both units
and one should make sure that the right conversion is made. For instance
in Ref.~\citep{duranti_efficient_2024}, the comb frequencies are
given in $\mathrm{MHz}$, with echo time expressed as $\frac{1}{\Delta}$
where we recall that $\Delta$ is the distance between two comb teeth.
We rather consider all frequencies in $\mathrm{rad.s^{-1}}$ so that
the memory recurrence time is given by $\frac{2\pi}{\Delta}$.

Note that even if $\kappa$ refers to some decay rate of the cavity,
for which an expression in $\mathrm{Hz}$ would seem natural, it is
the value in $\mathrm{rad.s^{-1}}$ which is used in the relation
with cavity finesse (see Eq.~\ref{eq:cavity_finesse}), as mentioned
in Ref.~\citep{afzelius_proposal_2013}\footnote{See right before Eq.~(1) of Ref.~\citep{afzelius_proposal_2013}.}.

\subsection{Cavity finesse and decay rate}

For a one-sided cavity (left mirror with reflectivity $R_{1}\lesssim1$
and perfect right mirror with $R_{2}=1$) of length $L_{cav}$, the
cavity finesse is defined as 

\begin{align}
\mathcal{F}_{cav} & :=\frac{\Delta\omega_{FSR}}{\Delta\omega_{FWHM}}\label{eq:cavity_finesse}\\
 & =\frac{\pi}{2}\frac{1}{\mathrm{Arcsin}\left(\frac{1-\sqrt{R_{1}}}{2R_{1}^{1/4}}\right)}\approx\frac{\pi R_{1}^{1/4}}{1-\sqrt{R_{1}}},\nonumber 
\end{align}

\noindent where $\Delta\omega_{FSR}=\frac{\pi c}{L_{cav}}$ is the
Fabry-Perot resonator free-spectral range \citep{siegman_lasers_1986},
and $\Delta\omega_{FWHM}$ the FWHM of its transmission peaks. The
field intensity decays with a rate $-\frac{\ln(R_{1})}{2L_{cav}/c}$
per cavity round-trip, so that we define the field amplitude decay
rate $\kappa$ as half that quantity

\begin{equation}
\kappa\approx\frac{c}{4L_{cav}}(1-R_{1}).\label{eq:kappa_cav}
\end{equation}

\noindent As such, the ``lifetime'' of a photon within the cavity
is $\frac{1}{2\kappa}$, and the ``the number of passes a photon
would make through an empty cavity before leaking out'' is $\frac{1}{2\kappa}\times\frac{c}{L_{cav}}\approx\frac{\mathcal{F}}{\pi}$
\citep{gorshkov_photon_2007-2}. Expanding $\mathcal{F}_{cav}$ and
$\kappa$ in terms of $\ln(R_{1})$ for small values of $1-R_{1}$
leads to the approximation

\begin{equation}
\mathcal{F}_{cav}\approx\frac{\pi c}{2L_{cav}\kappa}=\frac{\Delta\omega_{FSR}}{2\kappa},\label{eq:Fcav_VS_kappa}
\end{equation}

\noindent which is sometimes taken as a definition.

In Refs.~\citep{duranti_efficient_2024}, the cavity is chosen such
that $R_{1}=0.4$ and $R_{2}=0.97$. This yields $\mathcal{F}_{cav}\approx6.5$
as mentioned in the text. In addition, the cavity has a length $L_{cav}=208$~mm,
so that 
\begin{equation}
\kappa=3.4\times10^{8}\text{ }\mathrm{rad.s^{-1}}=2\pi\times55\text{ MHz}.\label{eq:kappa_exp}
\end{equation}
This last value can also be obtained from the values of the free spectral
range $\Delta\omega_{FSR}=2\pi\times7.2\times10^{8}$~MHz and the
cavity linewidth $\gamma_{cav}=\frac{\Delta\omega_{FSR}}{\mathcal{F}_{cav}}=2\pi\times111$~MHz. 

Note that in the experiment of Refs.~\citep{duranti_towards_2023,duranti_efficient_2024}
the vacuum chamber with the crystal is mounted between the two cavity
mirrors, which leads to losses due to reflections on the chamber windows.
These losses impact every round-trip. Note that we did not include
these intra-cavity losses in our model.

\subsection{Comb structure and optical depth}

\subsubsection{Probing the transmission profile}

A standard way to evaluate the atomic density within a crystal is
to perform a transmission experiment. Typically, one shines a probe
continuous laser at frequency $\xi$ through the crystal (without any
cavity), and the measured transmission curve $T(\xi)$ can be used
to extract the optical density of the crystal as a function of $\xi$
(see for instance Fig.~1.c) in Ref.~\citep{duranti_efficient_2024}).

To describe such an experiment without any cavity, Maxwell-Bloch or
related Heisenberg-Langevin equations in free space must be introduced
instead of system \ref{eq:memory_equations-1-1}, as it is done in
Refs.~\citep{gorshkov_photon_2007,sangouard_analysis_2007,afzelius_multimode_2009,sekatski_photon-pair_2011}
for instance. In the envelope approximation \citep{lukin_modern_2016},
one can write propagation along an axis with $z$ coordinate as

\begin{equation}
\begin{cases}
\partial_{t}\mathcal{E}(z,t)+c\partial_{z}\mathcal{E}(z,t) & =igN\int n(\omega)\sigma_{\omega}(z,t)\mathrm{d}\omega\\
\hfill\partial_{t}\sigma_{\omega}(z,t) & =-i\omega\sigma_{\omega}(z,t)+ig\mathcal{E}(z,t),
\label{eq:free_space_system}
\end{cases}
\end{equation}

\noindent where $\int n(\omega)\mathrm{d\omega}=1$ gives the atomic
distribution normalisation, $N$ the total number of atoms that interact
with the field, and $\sigma_{\omega}$ the collective polarisation
for frequency classes around detuning $\omega$. Note that no damping
is included here.

To solve this quickly, we will use rough notations of Fourier transforms
and distribution theory, with a convention where the plane waves involve
$e^{-i\xi t}$: $f(t)=\int\hat{f}(\xi)e^{-i\xi t}\mathrm{d}\xi$ ($\xi$
is understood as the angular frequency in $\mathrm{rad.s}^{-1}$).
Taking the Fourier transform $t\leftrightarrow\xi$ of the equations
yields

\[
\begin{cases}
-i\xi\hat{\mathcal{E}}(z,\xi)+c\partial_{z}\mathcal{\hat{\mathcal{E}}}(z,\xi) & =igN\int n(\omega)\hat{\sigma}_{\omega}(z,\xi)\mathrm{d}\omega\\
-i\xi\hat{\sigma}_{\omega}(z,\xi) & =-i\omega\hat{\sigma}_{\omega}(z,\xi)+ig\mathcal{\hat{\mathcal{E}}}(z,\xi),
\end{cases}
\]

\noindent which is

\[
\begin{cases}
\hat{\sigma}_{\omega}(z,\xi) & =\frac{g}{\omega-\xi}\hat{\mathcal{E}}(z,\xi)\\
\partial_{z}\mathcal{\hat{\mathcal{E}}}(z,\xi) & =-i\frac{\xi}{c}\hat{\mathcal{E}}(z,\xi)+i\frac{g^{2}N}{c}\left(\int_{\mathbb{R}}\frac{n(\omega)}{\omega-\xi}\mathrm{d}\omega\right)\hat{\mathcal{E}}(z,\xi).
\end{cases}
\]

\noindent The last equation is solved with an exponential; the imaginary
part of the argument yields the decay rate of the field amplitude
as a function of $z$. In particular, one gets for the field intensity

\begin{equation}
\partial_{z}\left|\mathcal{\hat{\mathcal{E}}}(z,\xi)\right|^{2}=+i\frac{2g^{2}N}{c}\mathrm{Im}\left(\int_{\mathbb{R}}\frac{n(\omega)}{\omega-\xi}\mathrm{d}\omega\right)\left|\mathcal{\hat{\mathcal{E}}}(z,\xi)\right|^{2}\label{eq:prop_equation}
\end{equation}

\noindent The integral can be computed using Kramers-Kronig relations
(or Sokhotski-Plemelj theorem) stemming from complex analysis. If
$n$ is a continuous function on the real line, one gets for the imaginary
part (below, we don't need the dispersive part of the equation) \footnote{See for instance Ref.~\citep{steck_quantum_nodate-1} p. 590} 

\begin{equation}
n(\xi)=\frac{1}{i\pi}\mathrm{v.p.}\int_{\mathbb{R}}\frac{n(\omega)}{\omega-\xi}\mathrm{d}\omega\label{eq:real_Kramers_Kronig}
\end{equation}

\noindent with v.p. denoting the Cauchy principal value of the integral.
Thus Eq.~\ref{eq:prop_equation} becomes a Beer Lambert--like equation
\citep{crisp_propagation_1970}

\begin{equation}
\partial_{z}\left|\mathcal{\hat{\mathcal{E}}}(z,\xi)\right|^{2}=-\frac{2\pi g^{2}N}{c}n(\xi)\left|\mathcal{\hat{\mathcal{E}}}(z,\xi)\right|^{2},\label{eq:absorption_differential_equation}
\end{equation}

\noindent which is solved as

\[
\left|\mathcal{\hat{\mathcal{E}}}(z,\xi)\right|^{2}=e^{-\frac{2\pi g^{2}N}{c}n(\xi)z}\left|\mathcal{\hat{\mathcal{E}}}(0,\xi)\right|^{2}.
\]

\noindent So the single pass transmission at frequency $\xi$ is given by

\[
T(\xi)=\frac{\left|\mathcal{\hat{\mathcal{E}}}(z=L_{crys},\xi)\right|^{2}}{\left|\mathcal{\hat{\mathcal{E}}}(z=0,\xi)\right|^{2}}=e^{-\frac{2\pi g^{2}N}{c}n(\xi)L_{crys}},
\]

\noindent where we note $L_{crys}$ for the crystal sample length
along $z$-axis. 

The optical depth is defined with the natural logarithm of the transmittance
for laser detuning $\omega$ (formerly $\xi$)\footnote{Similarly, the absorbance (or optical density) could be defined with
the decimal logarithm.},

\begin{equation}
\mathrm{OD}(\omega):=-\ln T(\omega)=\frac{2\pi g^{2}N}{c}n(\omega)L_{crys}.\label{eq:optical_depth_definition}
\end{equation}

\noindent So, in this undamped case, the optical depth is found to
be proportional to the atomic distribution $n$. As such, it is understood
that if one shines a continuous laser at frequency $\omega$ through
the crystal, the measured absorbance curve $T(\omega)$ can be used
to extract the optical density as a function of $\omega$. We can
also define the absorption coefficient $\alpha(\omega)$ such that
$\mathrm{OD}(\omega)=\alpha(\omega)L_{crys}$: $\alpha(\omega)=\frac{2\pi g^{2}N}{c}n(\omega)$,
as used in Refs.~\citep{bonarota_efficiency_2010,chaneliere_efficient_2010,bonarota_atomic_2012}\footnote{See for instance that Eq.~\ref{eq:absorption_differential_equation}
is consistent with Eq.~(1) of Ref.~\citep{bonarota_atomic_2012}
or Eq.~(6) of Ref.~\citep{bonarota_efficiency_2010}.}

\subsubsection{Mean optical depth $\tilde{d}$}

What is more, we can solve system \ref{eq:free_space_system} with
the same tricks as for the cavity case, first by integrating

\[
\sigma_{\omega}(t)=e^{-i\omega(t-t_{0})}\sigma_{\omega}(t_{0})+ig\int_{t_{0}}^{t}e^{-i\omega(t-t')}\mathcal{E}(z,t')\mathrm{d}t'
\]

\noindent and inserting that result into the equation for $\mathcal{E}(z,t)$
where we neglect temporal variations, which yields

\begin{eqnarray*}
\frac{\partial\mathcal{E}}{\partial z}(z,t) & = & -\frac{g^{2}N}{c}\int_{t_{0}}^{t}\tilde{n}(t-t')\mathcal{E}(z,t')\mathrm{d}t'\\
 &  & +i\frac{gN}{c}\int e^{-i\omega(t-t_{0})}\sigma_{\omega}(t_{0})\mathrm{d}\omega.
\end{eqnarray*}

\noindent Absorption only involves the central peak of $\tilde{n}(t)$
(as for Eq.~\ref{eq:system_abs_gen-1}) and if we assume that the
atomic ensemble is initially empty ($\sigma_{\omega}(t_{0})=0$)
we get

\[
\frac{\partial\mathcal{E}}{\partial z}(z,t\approx t_{0})\approx-\frac{g^{2}N}{c}\frac{D_{comb}}{2}\mathcal{E}(z,t).
\]

The mean absorption coefficient per unit length of a light pulse in
the AFC structure can be introduced as 

\[
\frac{\tilde{\alpha}}{2}=\frac{g^{2}ND_{comb}}{2c},
\]

\noindent and the mean optical density as 

\begin{equation}
\tilde{d}=\tilde{\alpha}L_{crys}=\frac{g^{2}ND_{comb}}{c}L_{crys}.\label{eq:mean_d}
\end{equation}

\noindent$\tilde{d}$ corresponds to the absorption power of the
inhomogeneous ensemble without the teeth. 

\smallskip{}
Now recall from Eqs.~\ref{eq:n_distr_comb} and \ref{eq:normalisation_constant}
that we wrote the AFC atomic distribution as 
\begin{equation}
n(\omega)\approx\frac{\Delta}{\int w(\omega)\mathrm{d}\omega\times\int v(\omega)\mathrm{d}\omega}\times v(\omega)\times\sum_{n=-\infty}^{+\infty}w(\omega-n\Delta),\nonumber
\end{equation}
with $w(0)=v(0)=1$ (see Table~\ref{tab:comb_distributions}).
Consider the maximal value $d_{max}=\mathrm{OD}(0)$ reached by the central peak of the optical density, that we also call the peak optical depth. From Eq.~\ref{eq:optical_depth_definition}, it is linked to $n(0)=\frac{\Delta}{\int w(\omega)\mathrm{d}\omega\times\int v(\omega)\mathrm{d}\omega}=\frac{\Delta}{\int w(\omega)\mathrm{d}\omega}\times\frac{D_{comb}}{2\pi}=F_{AFC}\times\frac{\mathrm{FWHM}_{teeth}}{\int w(\omega)\mathrm{d}\omega}\times\frac{D_{comb}}{2\pi}$,
with a factor $\frac{2\pi g^{2}N}{c}L_{crys}$. As such,

\[
\frac{d_{max}}{\tilde{d}}=\frac{2\pi n(0)}{D_{comb}}=F_{AFC}\times\frac{\mathrm{FWHM}_{teeth}}{\int w(\omega)\mathrm{d}\omega}
\]

\noindent has a value that only depends on the comb finesse 
and the geometry of the teeth (the last factor is $1$, $2\sqrt{\frac{\ln2}{\pi}}\approx0.94$
and $\frac{2}{\pi}\approx0.64$ for rectangular, Gaussian and Lorentzian
teeth). Since $F_{AFC}=\frac{\Delta}{\gamma_{tooth}}$ for an
AFC with rectangular teeth and a rectangular envelope,  $\tilde{d}$
does appear as the usual optical depth ``averaged over comb teeth''
\citep{afzelius_impedance-matched_2010,duranti_efficient_2024}.

\subsubsection{Comb structure and inhomogeneous width $\Gamma$}

When measuring an AFC transmission profile (such as the one of Fig.~1.c)
of Ref.~\citep{duranti_efficient_2024}), the width parameters of
$w$ ($\gamma_{tooth}$) and $v$ ($\Gamma$) can be estimated directly
from the profile, as well as the comb step $\Delta$, so that the
comb finesse $F_{AFC}$ (Eq.~\ref{eq:comb_finesse}) and $D_{comb}$
parameter (Eq.~\ref{eq:D_comb}) are reckoned. 

In particular, in Ref.~\citep{duranti_efficient_2024}, one finds
a rectangular AFC envelope with
\begin{equation}
\Gamma=2\pi\times4\text{ MHz}.\label{eq:Gamma_exp}
\end{equation}
Actually, the maximum inhomogeneous width within which the AFC can
be efficiently carved is limited by the width of the $^{1}D_{2}$
$\pm\frac{1}{2}$ to $\pm\frac{3}{2}$ transition for $\mathrm{Pr}_{3+}:\mathrm{Y}_{2}\mathrm{Si}\mathrm{O}_{5}$,
namely $2\pi\times4.6$ MHz \citep{duranti_towards_2023}.

\subsubsection{Retrieving the collective coupling constant $g\sqrt{N}$}

\label{subsec:Retrieving-the-collective_g2N}

We highlight the fact that the values of $\tilde{d}$ and $d_{max}$ were obtained so far for the free-space case. When dealing with a Fabry-Perot style return-trip modelling as in Ref.~\citep{afzelius_impedance-matched_2010}, $\tilde{d}$
actually appears as the relevant quantity to retrieve the impedance-matching
condition in the cavity case:
\[
\sqrt{R_{1}}=\sqrt{R_{2}}e^{-\tilde{d}}.
\]

\noindent Referring to Eqs.~\ref{eq:Fcav_VS_kappa} and \ref{eq:kappa_cav}
when $R_{2}=1$, that leads to $\frac{2L_{cav}}{c}\kappa=\tilde{d}$,
or

\begin{equation}
\mathcal{F}_{cav}\times\tilde{d}=\pi,\label{eq:imp_match_Fd}
\end{equation}
which is consistent with Ref.~\citep{gorshkov_photon_2007-2}\footnote{See end of paragraph III of Ref.~\citep{gorshkov_photon_2007-2}.}.

On the contrary, the value of $g\sqrt{N}$ that
we want to use in our model should refer to the situation where the
light mode is confined within the cavity. Referring to Eqs.~\ref{eq:g_cav_def}
and \ref{eq:quantisation_volume}, we know that $g$ scales as the
inverse square root of the quantisation volume. This volume is the
product of the mode cross section $A$ (that is usually taken smaller
than the crystal cross-section\footnote{Which is a crucial assumption for $g^{2}N$ not to depend on $A$,
as mentioned in Ref.~\citep{gorshkov_photon_2007-2}.}) and the quantisation length (so far in this Section, it was implicitly
$L_{crys}$). Once the crystal is embedded in a cavity\footnote{Note that in Refs.~\citep{afzelius_demonstration_2010,moiseev_efficient_2010},
$L_{crys}=L_{cav}$ seems to be assumed.} of length $L_{cav}\geq L_{crys}$, the value of $g\sqrt{N}$ that
should be considered in the equations (such as \ref{eq:memory_equations-1-1})
is to be rescaled from the free space value $\left[g\sqrt{N}\right]_{free}$ (subscript introduced here) used in Eq.~\ref{eq:mean_d}, namely

\begin{equation}
g\sqrt{N}=\left[g\sqrt{N}\right]_{free}\times\sqrt{\frac{L_{crys}}{L_{cav}}}\leq\left[g\sqrt{N}\right]_{free}.
\end{equation}

\noindent This is given by
\begin{equation}
    g\sqrt{N} = \sqrt{\frac{\tilde{d}c}{D_{comb}L_{cav}}}.
    \label{eq:gsqrtN_from_d}
\end{equation}

\noindent As such, Eq.~\ref{eq:imp_match_Fd} is also given by

\[
\frac{2L_{cav}}{c}\kappa=\frac{g^{2}ND_{comb}}{c}L_{cav},
\]

\noindent or equivalently

\[
\frac{g^{2}N}{\kappa\Gamma}=\frac{2}{D_{comb}\Gamma}.
\]

\noindent This last equation is nothing more than our impedance-matching
condition of Eq.~\ref{eq:impedance_matching}. Hence, $\tilde{d}$
is the value to use for the impedance-matching condition in the sense
of Ref.~\citep{duranti_efficient_2024}, while $g\sqrt{N}$
will be fed to our model.

\smallskip{}
Typically, in Refs.~\citep{duranti_towards_2023,lago-rivera_telecom-heralded_2021},
we find a maximum reachable optical depth in the crystal $\left[d_{max}\right]_{free}=6$, which corresponds to $\left[\alpha\right]_{free}=20\text{ }\mathrm{cm^{-1}=2\times10^{3}}~\mathrm{m^{-1}}$ for a crystal length $L_{crys}=3$~mm \citep{duranti_towards_2023, duranti_efficient_2024}. A background absorption $d_{0}=0.006$ is also mentioned\footnote{This background absorption is linked with ions at the pedestal of the peaks (see p.~33 of Ref.~\citep{duranti_towards_2023}), due to imperfect optical pumping in the comb preparation \citep{seri_quantum_2017}.}. In the experiments \citep{duranti_towards_2023, duranti_efficient_2024}, the AFC peaks are carved with a peak optical depth around $3$ and a comb finesse around $6$ so that the mean optical depth is then set to $\left[\tilde{d}\right]_{free}=0.45$
(or $0.4$). For $\Gamma=2\pi\times4$~MHz, this is consistent with the impedance-matching condition \ref{eq:impedance_matching} for a comb with a rectangular envelope and the cavity $\kappa$ Eq.~\ref{eq:kappa_exp}, which yields the value

\begin{equation}
\left[g\sqrt{N}\right]_{cav}=2\pi\times8.4\text{ MHz},\label{eq:g2N_cav}
\end{equation} 
that we consider for the simulations. Equivalently, with $L_{cav}=208$~mm mentioned at Eq.~\ref{eq:kappa_exp}, we obtain $\tilde{d} = 0.48$.

\subsubsection{Number of teeth and multimode capacity}

The number of teeth can easily be read on the transmission profile,
and is usually chosen either to fix the comb step $\Delta\sim\frac{\Gamma}{N_{teeth}-1}$
and thus the echo time given $\Gamma$ \citep{duranti_towards_2023},
or to enhance the multimode capacity of the memory \citep{lago-rivera_telecom-heralded_2021}.

Indeed, we should point out that the number of modes that can be stored
in the memory is also related to the number of teeth $N_{teeth}$.
Basically, the maximum number of modes $N_{modes}$ is the ratio between
the comb recurrence time $\frac{2\pi}{\Delta}\approx\frac{2\pi}{\Gamma}\times N_{teeth}$
and the duration of the mode wavepacket $\frac{2\pi}{\delta\omega_{in}}$,
that is: $N_{modes}\sim N_{teeth}\frac{\delta\omega_{in}}{\Gamma}$.
The incoming photon maximum bandwidth is limited by $\Gamma$
in our regime where $\kappa\gg\Gamma$ (see part \ref{subsec:Bandwidth}).
This gives the maximal number of modes one can suitably store: $N_{modes}\lesssim N_{teeth}$.
In case the bandwidth is limited by $\kappa$, when $\Gamma\gg\kappa,$
$N_{teeth}\gtrsim N_{modes}\frac{\Gamma}{\kappa}$, so in order to
store at least $10$ modes we should have at least about $100$ to
$1000$ teeth in the AFC comb. More discussion about an AFC memory
multimode capacity can be found in Ref.~\citep{ortu_multimode_2022}.

\noindent For our study, we will take

\begin{equation}
N_{teeth}=67\label{eq:N_teeth_exp}
\end{equation}

\noindent which corresponds to

\begin{equation}
\Delta\sim\frac{\Gamma}{N_{teeth}-1}=2\pi\times61\text{ kHz}\label{eq:Delta_exp}
\end{equation}

\noindent and is similar to the $25$ \textmu s echo experiment of
Fig.~S4 in Ref.~\citep{lago-rivera_telecom-heralded_2021}.

\subsubsection{Tooth width}

We can also note that $\frac{\Gamma}{N_{teeth}}$ is ultimately limited
by the homogeneous linewidth of one atom $\gamma_{h}$, that is $N_{teeth}\leq\frac{\Gamma}{\gamma_{h}}$,
since a tooth cannot be narrower than $\gamma_{h}$. Hence, the number
of modes in our regime one can hope to store is ultimately limited
by $N_{modes}\lesssim\frac{\Gamma}{\gamma_{h}}$.

In Refs.~\citep{duranti_towards_2023,duranti_efficient_2024}, combs
with a finesse $F_{AFC}\approx6$ are prepared ($5.8$ or $6.5$),
which corresponds to a tooth width $\gamma_{tooth}$ around $2\pi\times80$
kHz.

We will rather take a tooth width of the order of the homogeneous
linewidth for our simulations, which is said to be of the order of
$2\pi\times2$~kHz in Ref.~\citep{duranti_towards_2023}. In our
case, we consider Gaussian teeth for the numerical simulations, of
FWHM $2\sqrt{2\ln(2)}\gamma_{tooth}$ with

\begin{equation}
\gamma_{tooth}=2\pi\times1\text{ kHz}.\label{eq:gamma_tooth_exp}
\end{equation}

\subsection{Polarisation decay}

In our model, we saw that non-infinitely narrow comb teeth result
in only partial rephasing of the echo (see Eq.~\ref{eq:1st_echo_eff}):
energy will still be in the atoms but cannot couple back with the
cavity field due to the dephasing between the different frequency
classes that form each tooth. Instead, $\gamma_{P}$ and $\gamma_{S}$
stand for irreversible loss of coherence through some bath decay channel,
or some effective loss of coherence\footnote{Typically see Ref.~\citep{duranti_towards_2023}, pp. 24-25 (homogeneous
linewidth), p. 35 (loss because of holes dephasing), and p.84 (decay
results).}. We can for instance use $\gamma_{S}$ to phenomenologically reproduce
the effect of the inhomogeneous broadening of the spin $|g\rangle$-$|s\rangle$
transition.

In our simulations, we decide to set

\begin{equation}
\gamma_{P}=\gamma_{S}=0\text{ kHz}.\label{eq:gammaPS_exp}
\end{equation}

\noindent If we were to include decay due to Lorentzian inhomogeneous broadening on the spin transition, a phenomenological approach would be to take a non-zero $\gamma_{S}$. Actually, the spin broadening is experimentally observed to be Gaussian (see Ref.~\citep{duranti_towards_2023} with $T_{eff}^{2}\approx90$ \textmu s). Taking $\gamma_{S}$ from $10$ to $20$~kHz would then be an incomplete approach, to be replaced by the introduction of frequency classes for the $\hat{S}$ transitions.

\subsection{Input photon width}

In a quantum repeater setting, the photons that are stored in the
memory are usually produced by specifically designed SPDC sources.
In Refs.~\citep{lago-rivera_telecom-heralded_2021,duranti_towards_2023},
we find photon bandwidths around $2\pi\times1$ to $2$~MHz, with
a double-exponential temporal structure. For the simulations, we typically
take Gaussian-shaped wavepackets, with an intensity FWHM $1/(2\times1.5\times10^{6})\approx330$~ns,
which corresponds to an intensity standard deviation equal to $1/(4\sqrt{2\ln(2)}\times1.5\times10^{6})=140$~ns and a frequency bandwidth FWHM equal to $\frac{2\ln{2}}{\pi\times 330 \times 10^{-9}} = 1.3$~MHz.

\subsection{Control $\pi$-pulses laser power}

From Eq.~\ref{eq:Rabi_frequency}, we know that the Rabi frequency
is proportional to the electric field amplitude $E_{r}$, thus proportional
to the square root of the field intensity $I=\left|E_{r}\right|{{}^2}$.
Ref.~\citep{nilsson_hole-burning_2004} indicates that $I_{0}=250$~W/cm\texttwosuperior{}
leads to Rabi frequency $\Omega_{0}=2\pi\times1.6$~MHz for the 3/2-3/2
transition (this is consistent with the control beams of Ref.~\citep{duranti_towards_2023}).
If we take rectangular control pulses of time width $\tau_{e}=0.07$~\textmu s
so as to almost cover the bandwidth of a $1.5$~MHz photon stored
in the memory we get $\Omega_{e}=\pi/\tau_{e}=2\pi\times7.1$ MHz.
So we would need $I_{e}=I_{0}\times\frac{\Omega_{e}^{2}}{\Omega_{0}^{2}}\approx250*(7.1/1.6){{}^2}=5$~kW/cm\texttwosuperior .
Switching to a 50-\textmu m-wide beam waist (smaller zone), one gets
about $100$~mW of power required.

\subsection{Filtration of the memory output}

The filtration mentioned in the main text (right boxes of Fig.~3)
is performed with a box filter function in the Fourier space, similarly
to the pit used in the experiments \citep{duranti_towards_2023,duranti_efficient_2024}.
The width of the box filter, $2\pi\times0.15$ MHz, is chosen such
that the value is small enough for the filtered wavepackets to appear
smoothed, but big enough not to loose too much energy from the cropped-echo
wavepacket.

\newpage
\clearpage

\section{Numerical methods}

\label{sec:Numerical-methods}

\subsection{Dynamical system}

The discretised set of scalar equations \ref{eq:discrete_scalar_system-2} is solved numerically. Typically, we use a Runge-Kutta solver of order
8 to integrate the differential equations. Despite such a scheme not
being symplectic, energy conservation (see Eq.~\ref{eq:conservation_scalar})
was observed as expected. Integration is performed in the time domain,
so we make sure that the time steps involved are short enough compared
with any time scale of the system dynamics (in particular, the width
of the input wavepacket). The number of frequency classes for the
discretised set is also chosen so as to resolve the particular shape
of the AFC comb structure given by $n(\omega)$. We typically take
$21$ frequency classes per Gaussian tooth.

\subsection{Hong-Ou-Mandel interferences}

Computing the multiparameter integrals in the expressions
of HOM visibilities of next Appendix (see \ref{eq:visi_mixed_mixed} with $T<+\infty$)
can be numerically involved. Once numerical approximations of the
wavepackets are known, the integrals' values are estimated by means of
quasi-Monte-Carlo methods. Typically, we take for each integral between
$1000$ and $10000$ statistically independent samples, each time
with $1024$ evaluations of the functions.

\newpage
\clearpage

\section{Ion-Photon state and Hong-Ou-Mandel visibility}

\label{sec:Ion-Photon-state-and}

In this Appendix, we sum up results extracted from a theoretical model
of Refs.~\citep{meraner_indistinguishable_2020,krutyanskiy_entanglement_2023}
that describes photon states emitted by trapped ions as observed experimentally.
We also discuss the link that can be found between the visibility
and the wavepacket overlap for photons involved in a HOM experiment.

\subsection{Ion emission and Raman scattering}

First, we recall some results from Refs.~\citep{krutyanskiy_light-matter_2019,meraner_indistinguishable_2020,krutyanskiy_entanglement_2023}, where experiments dealt with a single $^{40}$Ca$^{+}$ atom trapped in the focus of
an optical cavity. 

The ion was modeled as a three-level $\Lambda$
system with a ground state $|S\rangle$, an excited state $|P\rangle$,
and a metastable state $|D\rangle$. After the ion's motional state is Doppler cooled, the ion is prepared in $|S\rangle$ by optical
pumping. Single photons are then generated by driving the $|S\rangle$-$|P\rangle$ transition off-resonantly with a Raman laser pulse. This triggers the emission of a polarized photon into a vacuum cavity
mode. The emission is due to a cavity-mediated Raman process, and the cavity frequency is detuned from resonance with the $|P\rangle$-$|D\rangle$ transition \citep{keller_continuous_2004}.
As reported in Refs.~\citep{meraner_indistinguishable_2020,krutyanskiy_entanglement_2023},
the observed results were accurately reproduced by theoretical predictions
from a model using a master equation approach. A photon waveform obtained
from this model, as detected on photon-count histograms, is typically
asymmetric with a FWHM of about $11$~\textmu s, with a fast rising
front and a longer decreasing tail \citep{krutyanskiy_entanglement_2023}. 

The quality of the photons can be assessed by Hong Ou Mandel--like
interference, which quantifies how well two photons bunch when combined
on a balanced beam splitter (see next Section). In the case of pure and indistinguishable
waveforms, the photons bunch perfectly, resulting in a maximal interference
visibility. Perfect bunching was not observed experimentally: the visibility of
the interference decreased when the duration of the coincidence detection
window increased\footnote{See Fig.~3(b) of Ref.~\citep{krutyanskiy_entanglement_2023}.}.

The primary source of imperfections was identified as coming from
spontaneous emission. Following decays from $|P\rangle$ to
$|S\rangle$ during the Raman laser pulse, a cavity photon can still
be emitted while the Raman laser is on. Since every spontaneously
scattered photon carries away the information that the cavity photon
has not yet been emitted, each cavity photon is thus a continuous
temporal mixture of shifted pure wavepackets (of $\sim6$~\textmu s
FWHM). As a result, those photons do not bunch perfectly. As mentioned
in the Supplemental Material of Refs.~\citep{meraner_indistinguishable_2020,krutyanskiy_entanglement_2023},
the continuous mixture can be written as

\begin{eqnarray}
\rho_{emit} & = & P_{0}|0\rangle\langle0|+P_{ns}|\Psi(\cdot)\rangle\langle\Psi(\cdot)|\label{eq:ion_mixture}\\
 &  & +\ensuremath{\int_{0}^{+\infty}P(s)|\Psi(\cdot-s)\rangle\langle\Psi(\cdot-s)|\mathrm{d}s}\nonumber \\
 & =: & P_{0}|0\rangle\langle0|+\ensuremath{\int_{0}^{+\infty}\bar{P}(s)|\Psi(\cdot-s)\rangle\langle\Psi(\cdot-s)|\mathrm{d}s,}\nonumber 
\end{eqnarray}

\noindent where $P_{0}$ gives the proportion of vacuum in the mixture,
$|\Psi(\cdot)\rangle$ is the pure-state component that is emitted
by the ion when no scattering happens (with probability $P_{ns}$),
and $|\Psi(\cdot-s)\rangle$ are the shifted components emitted conditioned
on a last scattering event happening at time $s$ (with probability
$P(s)$). The $\Psi(\cdot)$ functions give the temporal wavepacket profiles of each component and are such that, given a time $s$, $\Psi(t-s)=0$ if $t\leq s$.
They can be computed by solving a master equation
representing the ion-photon system up to a time $t$. We take $t$
big enough (after the start of the emission) such that we can substitute the $t$ dependence by a $\cdot$
in the shifted $\Psi$ and use $+\infty$ in the integral upper bound. $\bar{P}(s)$
is a probability distribution ``$\bar{P}:t\mapsto\delta_{0}(t)+P(t)$'',
such that, for any test function $\varphi$, $\int_{0}^{+\infty}\delta_{0}(s)\varphi(s)\mathrm{d}s=\varphi(0)$
and $\int_{0}^{+\infty}\delta_{0}(s)\mathrm{d}s=1$. 
\noindent
Since $\mathrm{Tr}(\rho_{emitt})=1$,
we have $P_{0}+P_{ns}+\int_{0}^{+\infty}P(s)\mathrm{d}s=1$, i.e.,
$P_{0}+\int_{0}^{+\infty}\bar{P}(s)\mathrm{d}s=1$.

\subsection{HOM visibility}

Here, we consider a general setting where we perform a HOM--like
experiment with two photons described by a mixture of pure wavepackets.
That is, we use the representation of Eq.~\ref{eq:ion_mixture},
where $P(s)$ now represents the general weight of the mixture and
$P_{0}$ accounts for the probability of not emitting or of losing
the photon. We are interested in finding a relation between the visibility
of the HOM interference and the wavepackets' overlap. For the purpose
of our study and the discussions of Fig.~4 in the main text, we consider
in particular the situation where:
\begin{itemize}
\item The first photon is emitted by an ion from the experiments~\citep{krutyanskiy_entanglement_2023}
(node B only). To reconstruct the continuous mixed state, we
retrieve the wavepackets by numerically solving the master
equations developed for Ref.~\citep{krutyanskiy_entanglement_2023}\footnote{For simplicity we assume that both polarisations share the same wavepacket.}.
\item The second photon is emitted by the AFC memory described in Appendix~\ref{sec:Modelling-a-cavity-asisted}, with or
without shaping. As such, discussions from Sec.~\ref{subsec:PhotonStateCoherence}
indicate that it is described by a mixture of one vacuum component
and one pure single-photon component, the waveform of which is retrieved
from our numerical simulations.
\end{itemize}
For the interference experiment, we assume that the two photons are
indistinguishable w.r.t. to all other degrees of freedom (thanks to
quantum frequency conversion, for instance). In the next paragraph,
we will use subscripts $a$ and $b$ to refer to the first and the
second photons.

\subsubsection{HOM dip}

\begin{figure}
\begin{centering}
\includegraphics[width=0.7\columnwidth]{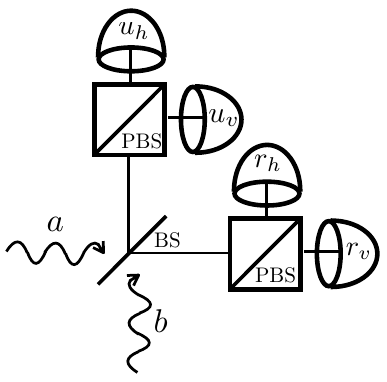}
\par\end{centering}
\caption{\label{fig:HOM-setup-retrieved}For a HOM--like experiment
on two input photons $a$ and $b$, the setup involves one beam splitter
(BS), two polarising beam splitters (PBS) and four non-photon-number-resolving
photodetectors ($u_{h}$, $u_{v}$, $r_{h}$, and $r_{v}$).}
\end{figure}

We consider the setup presented in Fig.~\ref{fig:HOM-setup-retrieved},
with two photons combined on a 50:50 beam splitter. Each output path of
the beam splitter contains a polarising beam splitter followed by
two (non photon-number resolving) detectors. We are interested in
two-click coincidence events on opposite outputs of the beam splitter,
for the pairs $\{u_{h},r_{h}\},\{u_{v},r_{v}\},\{u_{v},r_{h}\},\{u_{h},r_{v}\}$.
The rate of coincidence corresponding to a click at detector $u_{h}$ at
time $t_{1}$ and a click at detector $r_{v}$ at time $t_{2}$ is
labelled $\mathrm{det}_{h,v}(t_{1},t_{2})$, and similarly for the
other pairs ($\mathrm{det}_{h,h}(t_{1},t_{2})$, $\mathrm{det}_{v,v}(t_{1},t_{2})$
and $\mathrm{det}_{v,h}(t_{1},t_{2})$). We define the probability
to detect two clicks delayed by at most $T$ as 
\begin{equation}
\mathrm{Det}_{h,v}(T):=\iint_{|t_{1}-t_{2}|\leq T}\mathrm{det}_{h,v}(t_{1},t_{2})\mathrm{d}t_{1}\mathrm{d}t_{2},\label{eq:Det_hv_T}
\end{equation}
and similarly for the other coincidences.

\begin{widetext}

\subsubsection{Asymptotic visibilities}

\paragraph{Mixed photon vs mixed photon}

Assuming that all detectors have the same efficiency, the asymptotic
visibility ($T\rightarrow+\infty$) of the HOM experiment is linked
to the two-click coincidences:

\begin{eqnarray}
V_{HOM}^{\infty} & = & 1-\frac{\mathrm{Det}_{h,h}(\infty)+\mathrm{Det}_{v,v}(\infty)}{\mathrm{Det}_{v,h}(\infty)+\mathrm{Det}_{h,v}(\infty)}\nonumber \\
 & = & 1-\frac{2\iint_{\mathbb{R}^{2}}\iint_{0}^{+\infty}\bar{P}_{a}(s_{a})\bar{P}_{b}(s_{b})\left|\Psi_{a}(t_{1}-s_{a})\Psi_{b}(t_{2}-s_{b})-\Psi_{a}(t_{2}-s_{a})\Psi_{b}(t_{1}-s_{b})\right|^{2}\mathrm{d}s_{a}\mathrm{d}s_{b}\mathrm{d}t_{1}\mathrm{d}t_{2}}{2\iint_{\mathbb{R}^{2}}\begin{array}{c}
\left(\int_{0}^{+\infty}\bar{P}_{a}(s_{a})\left|\Psi_{a}(t_{2}-s_{a})\right|^{2}\mathrm{d}s_{a}\int_{0}^{+\infty}\bar{P}_{b}(s_{b})\left|\Psi_{b}(t_{1}-s_{b})\right|^{2}\mathrm{d}s_{b}\right.\\
\left.+\int_{0}^{+\infty}\bar{P}_{a}(s_{a})\left|\Psi_{a}(t_{1}-s_{a})\right|^{2}\mathrm{d}s_{a}\int_{0}^{+\infty}\bar{P}_{b}(s_{b})\left|\Psi_{b}(t_{2}-s_{b})\right|^{2}\mathrm{d}s_{b}\right)
\end{array}\mathrm{d}t_{1}\mathrm{d}t_{2}}\nonumber \\
 & = & 1-\frac{2\iint_{0}^{+\infty}\bar{P}_{a}(s_{a})\bar{P}_{b}(s_{b})\iint_{\mathbb{R}^{2}}\left(\begin{array}{c}
\left|\Psi_{a}(t_{1})\right|^{2}\left|\Psi_{b}(t_{2})\right|^{2}+\left|\Psi_{a}(t_{2})\right|^{2}\left|\Psi_{b}(t_{1})\right|^{2}\\
-2\mathrm{Re}\left(\Psi_{a}(t_{1}-s_{a})\Psi_{b}(t_{2}-s_{b})\Psi_{a}^{*}(t_{2}-s_{a})\Psi_{b}^{*}(t_{1}-s_{b})\right)
\end{array}\right)\mathrm{d}t_{1}\mathrm{d}t_{2}\mathrm{d}s_{a}\mathrm{d}s_{b}}{4\iint_{0}^{+\infty}\bar{P}_{a}(s_{a})\bar{P}_{b}(s_{b})\mathrm{d}s_{a}\mathrm{d}s_{b}\int_{\mathbb{R}}\left|\Psi_{a}(t)\right|^{2}\mathrm{d}t\int_{\mathbb{R}}\left|\Psi_{b}(t)\right|^{2}\mathrm{d}t}\nonumber \\
 & = & 1-\frac{\int_{0}^{+\infty}\bar{P}_{a}(s_{a})\mathrm{d}s_{a}\int_{0}^{+\infty}\bar{P}_{b}(s_{b})\mathrm{d}s_{b}}{\iint_{0}^{+\infty}\bar{P}_{a}(s_{a})\bar{P}_{b}(s_{b})\mathrm{d}s_{a}\mathrm{d}s_{b}}+\frac{\iint_{0}^{+\infty}\bar{P}_{a}(s_{a})\bar{P}_{b}(s_{b})\left|\langle\Psi_{b}(\cdot-s_{b})|\Psi_{a}(\cdot-s_{a})\rangle\right|^{2}\mathrm{d}s_{a}\mathrm{d}s_{b}}{\iint_{0}^{+\infty}\bar{P}_{a}(s_{a})\bar{P}_{b}(s_{b})\mathrm{d}s_{a}\mathrm{d}s_{b}}\nonumber \\
 & = & \frac{1}{(1-P_{0,a})(1-P_{0,b})}\iint_{0}^{+\infty}\bar{P}_{a}(s_{a})\bar{P}_{b}(s_{b})\left|\langle\Psi_{b}(\cdot-s_{b})|\Psi_{a}(\cdot-s_{a})\rangle\right|^{2}\mathrm{d}s_{a}\mathrm{d}s_{b}\label{eq:visi_mixed_mixed}\\
 & = & \frac{1}{(1-P_{0,a})(1-P_{0,b})}\left(P_{ns,a}P_{ns,b}\left|\langle\Psi_{b}(\cdot)|\Psi_{a}(\cdot)\rangle\right|^{2}+P_{ns,a}\int_{0}^{+\infty}P_{b}(s_{b})\left|\langle\Psi_{b}(\cdot-s_{b})|\Psi_{a}(\cdot)\rangle\right|^{2}\mathrm{d}s_{b}\right.\nonumber \\
 &  & \qquad\left.+P_{ns,b}\int_{0}^{+\infty}P_{a}(s_{a})\left|\langle\Psi_{b}(\cdot)|\Psi_{a}(\cdot-s_{a})\rangle\right|^{2}\mathrm{d}s_{a}+\iint_{0}^{+\infty}P_{a}(s_{a})P_{b}(s_{b})\left|\langle\Psi_{b}(\cdot-s_{b})|\Psi_{a}(\cdot-s_{a})\rangle\right|^{2}\mathrm{d}s_{a}\mathrm{d}s_{b}\right).\nonumber 
\end{eqnarray}

\noindent We have also assumed that for each emitter, the wavepackets for
both polarisations are identical, and have used the fact that $\int_{\mathbb{R}}\left|\Psi_{a}(t)\right|^{2}\mathrm{d}t=1=\int_{\mathbb{R}}\left|\Psi_{b}(t)\right|^{2}\mathrm{d}t$. 

\noindent If neither $a$ nor $b$ has a vacuum component, the visibility
is

\begin{eqnarray}
V_{HOM}^{\infty} & = & \iint_{0}^{+\infty}\bar{P}_{a}(s_{a})\bar{P}_{b}(s_{b})\left|\langle\Psi_{b}(\cdot-s_{b})|\Psi_{a}(\cdot-s_{a})\rangle\right|^{2}\mathrm{d}s_{a}\mathrm{d}s_{b}.\label{eq:visi_mixed_novacuum}
\end{eqnarray}

\paragraph{Pure photon vs mixed photon}

In the case where photon $a$ is mixed and photon $b$ is either pure
or lost, we take $P_{ns,b}=1-P_{0,b}$ and $\forall s\quad P_{b}(s)=0$,
so that

\begin{eqnarray}
V_{HOM}^{\infty,pure-mixed} & = & \frac{1}{1-P_{0,a}}\int_{0}^{+\infty}\bar{P}_{a}(s_{a})\left|\langle\Psi_{b}(\cdot)|\Psi_{a}(\cdot-s_{a})\rangle\right|^{2}\mathrm{d}s_{a},\label{eq:visi_pure_mixed}
\end{eqnarray}

\noindent which is the fidelity between the two photon states.

\paragraph{Pure photon vs pure photon}

In the case where both photons $a$ and $b$ are pure without any
vacuum component, we take $P_{0,a}=P_{0,b}=0$, $\forall s\quad P_{a}(s)=0=P_{b}(s)$
and $P_{ns,a}=P_{ns,b}=1$ so that

\begin{equation}
V_{HOM}^{\infty,pure-pure}=\left|\langle\Psi_{b}(\cdot)|\Psi_{a}(\cdot)\rangle\right|^{2},\label{eq:visi_pure_pure}
\end{equation}

\noindent which is again nothing more than the overlap between the
two photon states, or their fidelity.

\end{widetext}

\subsubsection{Visibilities}

Above, we considered all possible coincidences, with a window length
$T\rightarrow+\infty$. Instead, one could look at finite $T$ and
compute the integrals for $t_{1},t_{2}$ such that $\left|t_{1}-t_{2}\right|\leq T$
(rather than over $\mathds{R}^{2}$). This computation is more involved
and typically requires numerical methods to estimate the integrals
(see Appendix~\ref{sec:Numerical-methods}). It is observed that the visibility
decreases with $T$ \citep{meraner_indistinguishable_2020} and goes
to $1$ in the pure-vs-pure case when $T\rightarrow0$.

\clearpage
\newpage

\end{document}